\documentclass[final]{article}
\usepackage{amsmath}
\usepackage[colorlinks = true,urlcolor = blue,hyperfootnotes=false]{hyperref}
\usepackage{graphicx}
\usepackage{wrapfig}
\usepackage{xcolor}
\usepackage{parskip}
\usepackage{float}
\definecolor{lightBlue}{rgb}{0.67, 0.9, 0.93}
\usepackage[letterpaper,portrait,margin=1in,footnotesep=1.5em]{geometry}  
\usepackage{setspace}
\usepackage{amssymb}
\usepackage[style=base,font={footnotesize},format=hang,indention=0cm]{caption}
\usepackage{tikz}
\newcommand{\xlink}{  
    \tikz[x=1ex, y=1.2ex, baseline=0ex]{
        \node at (0.1,0.1) {~};  
        \begin{scope}[x=1.3ex, y=1.3ex]
            \clip (-0.1,-0.1) 
                --++ (-0, 1.2) 
                --++ (0.6, 0) 
                --++ (0, -0.6) 
                --++ (0.6, 0) 
                --++ (0, -1);
            \path[draw, 
                line width = 1, 
                rounded corners=0.5, blue] 
                (0,0) rectangle (1,1);
        \end{scope}
        \path[draw, line width = 1, blue] (0.5, 0.5) 
            -- (1.3, 1.3);
        \path[draw, line width = 1, blue] (0.8, 1.3) 
            -- (1.3, 1.3) -- (1.3, 0.8);
    }
}
\newenvironment{enumerateTight}
    { \begin{enumerate}
        \setlength{\itemsep}{0pt}
        \setlength{\parskip}{0pt}
        \setlength{\parsep}{0pt}
    }
    { \end{enumerate}
}
\newcommand{\comment}[1]{}
\usepackage[bottom,hang,flushmargin]{footmisc}
\newif\ifforReview

\begin{document}
\sloppy   

\begingroup  
  \centering
  \LARGE The Problematic Hoover Index\\[1em]
  \large Robin W. Spencer\footnote{The author is a retired biochemist and data analyst, email: spencerrw@alum.mit.edu} \par
\endgroup
\vspace{1cm}
\begin{abstract} The Hoover index H, derived from the distribution of population density, has a long history in population geography.  But it is prone to misinterpretation and serious measurement artifacts, some of which have been recognized for years.  Here its problems, old and new, are put in a common framework and quantitatively dissected with thought-experiments, experimental reaggregation over a thousand-fold range, modeling, and simulation.  H has long been interpreted as urbanization, which simple examples show is incorrect.  Values of H near zero are taken as an ideal or primordial state, but the mechanism and distributions of settlement growth show that H can be expected in the range 0.4-0.7 and remain nearly constant across continents, centuries of time, and wide ranges of population density.  Technically, the measurement of H is confounded by choices of aggregation scale, inclusion of unpopulated areas, and truncated input data, which produce wide swings in its value.  Spatially isotropic modeling and simulation reproduce its trends with aggregation scale and time, nullifying many frontier- or nation-specific interpretations.  I suggest methods to recognize and address its major artifacts and point out several trends with valid interpretations. 
\end{abstract}
\vspace{0.15cm}
\parbox{1\textwidth}{\centering
    {\fontfamily{cmss}\selectfont 
    
    heterogeneity, distribution, pareto, MAUP, aggregation, simulation
    }
}
\setstretch{1.125}
\setlength{\parskip}{0.5\baselineskip}

\subsection*{Introduction}

    Social scientists who study complex systems make use of numeric descriptors in order to grapple with that complexity (Farrell et al 2025).  Some descriptors, such as prices in an open economy, are aggregated by thousands of independent actors with local knowledge, and as such are valid indicators (Hayek 1945).  But other metrics are created by individuals and narrowly demonstrated.  They may become ``scientized", dressed up with abstruse mathematics (Hayek 1975), and interpreted well beyond their original context.  A poor metric can mislead a field for years.

    No one would dispute the value of a numeric index that measures the geographic distribution of people.  Such an index could reveal issues of fairness, give context to historic and current events, and inform policy.  The Hoover index H has played this role in the same way that the Gini coefficient G has long been used to measure distributions of wealth.  However H has significant problems in the way that it is measured and in the way that it is interpreted.  While some of its problems have been noted since 1934 (Gehlke \& Biehl 1934, Fotheringham \& Wong 1991, Openshaw 1984, Wong 2004, Andresen 2021, Ye 2022) in most cases the problems are anecdotal or their origins go unexplored.  Thought-experiments, experimental rescaling, and simulation bring known and new Hoover problems into a common framework and suggest root causes and diagnostic tests.  Most literature results with H are shown to be due to artifacts and uninterpretable -- because non-geographic random number models reproduce the same trends in time and aggregation.  What remains are several valid uses for H: it can be used to identify regions that have self-similar distributions of settlement, though they are neither highly dense nor sparse; it can identify the scale (in km or miles) at which a large region begins to become homogeneous; and at particular sampling scales, H can detect the unusual spatial correlations of isolated cities.
    
    While H and G are closely related derivations of the same Lorenz curve, I assume for these discussions that H is exclusively used to measure relative population density and G to measure wealth.  Wealth is generally measured at the indivisible, individual level and need not be aggregated, so G need not share the ``denominator problems"\footnote{It is useful to distinguish artifacts that occur in the numerator (population) or the denominator (sampled area) of population density.} that plague spatial metrics. Thus the problems with H do not necessarily imply problems with G.

\subsection*{Roadmap}

    The Hoover index, also known as the Ricci, Pietra, or Robin Hood index, has a simple geometric interpretation from the Lorenz curve and is easily calculated as $H =(1/{2n})\sum_{i}(|x_{i} - \bar{x}| / \bar{x}) $  (Costa \& Pérez-Duarte 2019) where $x_{i}$ are the population densities of the sampled sub-regions (or individual wealth for economic applications).  This simplicity conceals a large number and diversity of factors and analyst choices that affect H, which can be organized and connected in a logical roadmap (Figure 1).  Here the input data are in green; these are the observables for which we seek understanding and context through geographic comparison, historic tracking, and model building.  In light red are the consequences on H: it generally takes on a narrow range of values over all times and places. Population growth alone can move H up or down.  It can be wildly swayed by the inclusion or exclusion of empty regions, which is appropriate or not in different contexts.  At particular sampling scales, spatial correlation will increase H.  And H has two classes of artifacts which are basically errors: one related to population growth and another to the choice of sample block scale and zoning.  The latter is called out in the literature as the modifiable areal unit problem, or MAUP (Openshaw 1984 and others above).
    
    Most of the following discussions begin with population density and then pursue downstream consequences on H.  But we also seek fundamentals that drive H, and that entails tracking backwards from density to two universal and ancient observations, namely the sizes of cities and the distances that people migrate.  These factors suggest why the natural range of H is 0.4 - 0.7 (not zero) and why it varies only slightly in any given study.  Everything else is an artifact or secondary effect.

     \begin{figure}[htbp]
        \centering
        \captionsetup{width=10cm}
        \includegraphics[width=12cm]{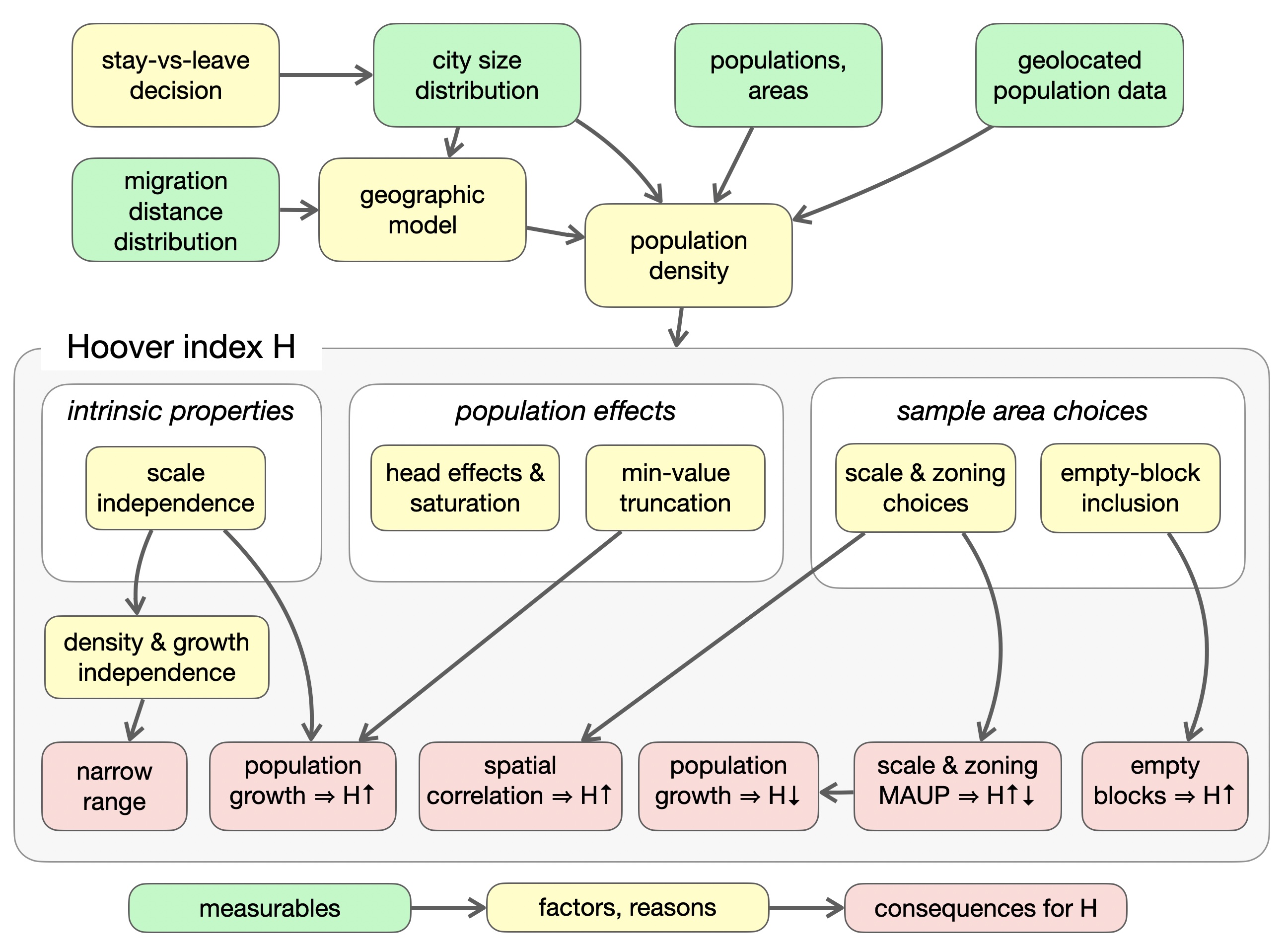}
        \caption*{Figure 1: Roadmap}
     \end{figure}

\pagebreak
     \begin{figure}[htbp]
        \centering
        \captionsetup{width=14cm}
        \includegraphics[width=16cm]{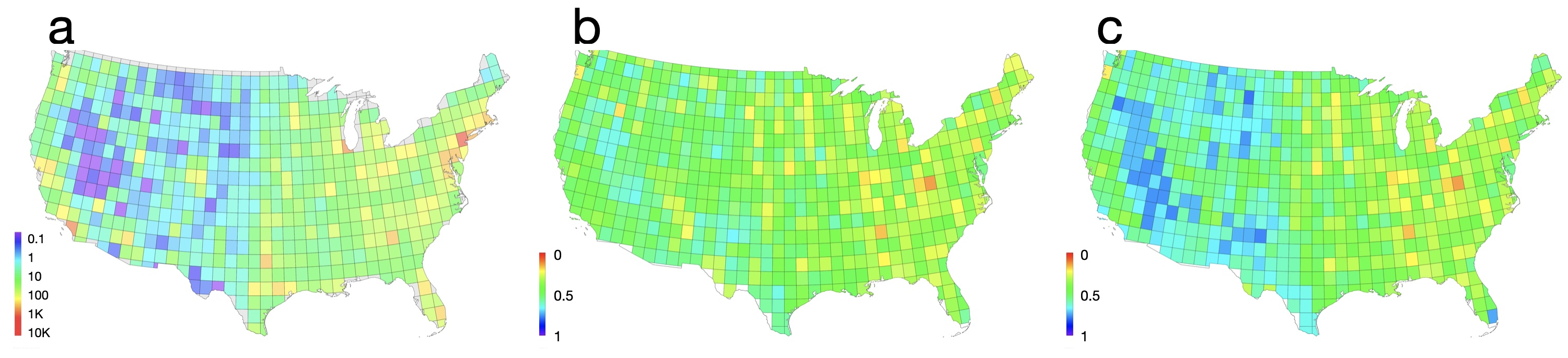}
        \caption*{Figure 2: Maps based on 2020 census data for 30,323 zipcodes in the 48 contiguous US states (a) Population density for 1.2x1.2\textdegree blocks in persons/sq km, (b) H for 1.2x1.2\textdegree blocks based on densities of each block's 16 interior 0.3x0.3\textdegree blocks, excluding interior blocks without data. (c) As (b) but including empty blocks.}
     \end{figure}

    \begin{wrapfigure}[16]{r}{8cm}
        \vspace{-12pt}
        \centering
        \captionsetup{width=6cm}
        \includegraphics[width=7cm]{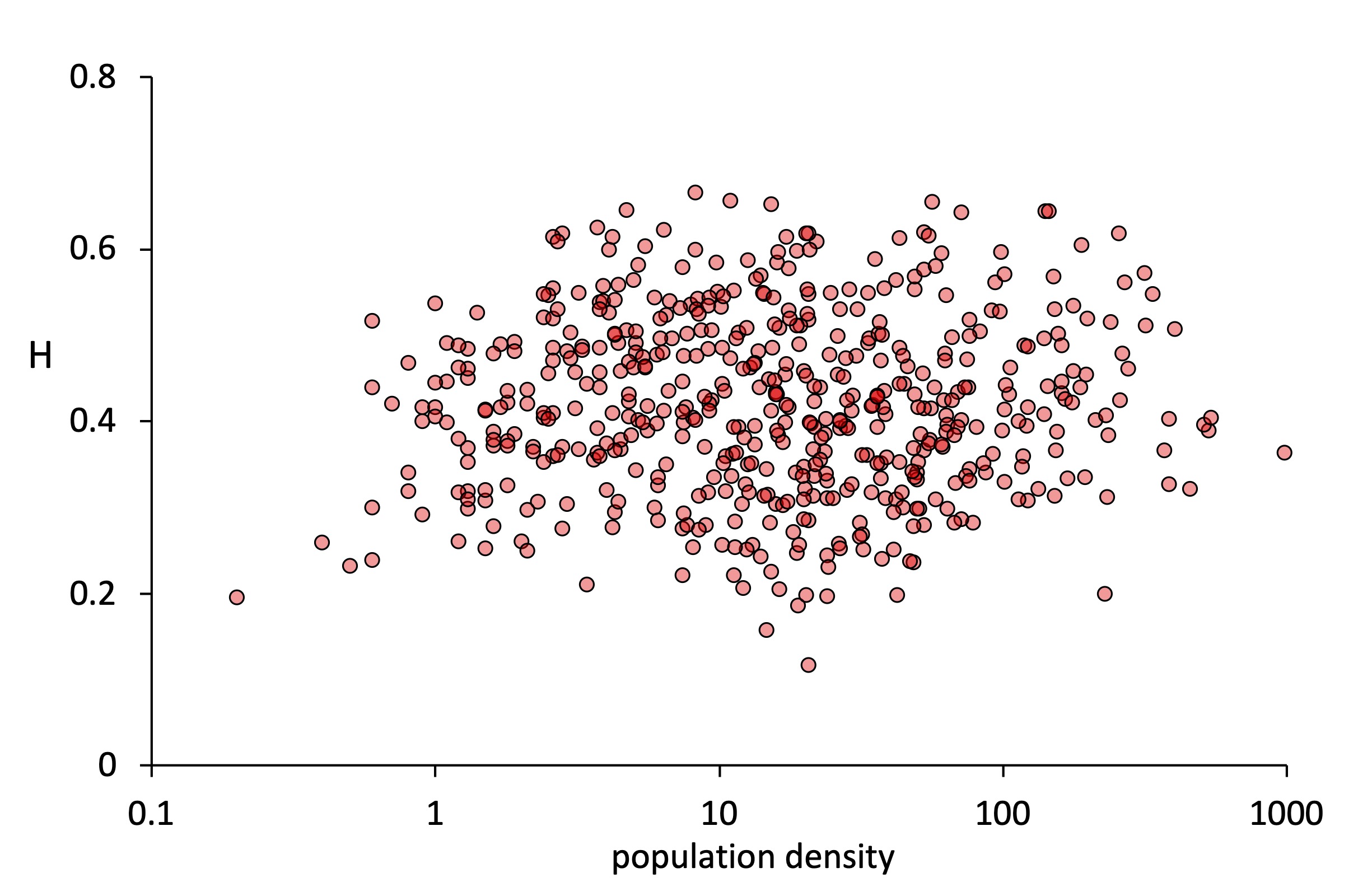} 
        \caption*{Figure 3: H versus population density (people per square kilometer) for the sampled blocks of Figure 2b.}
    \end{wrapfigure}

    The three maps of Figure 2, all from the same dataset, illustrate the magnitude of the issues.  The population density by 1.2\textdegree \:block (a) has a 2700-fold range from greater New York City to near-empty Nevada deserts, and yet the value of H calculated for the same blocks (excluding empty sub-blocks) (b) is $H = 0.42 \pm 0.10, n = 536$, a narrow range and with no significant variation for dense major cities or the sparse north prairies. Figure 3 shows that there is no correlation of H with population density over three orders of magnitude.  In map (c) the calculation of H includes empty sub-blocks which now highlights sparse western regions -- interestingly with H $\rightarrow$ 1, the opposite direction to what an interpretation of H as urbanization would imply.

\subsection*{H is Scale Independent}

    The Hoover index has an intrinsic property, true by its construction, that affects all interpretations: it is scale-independent.  The axes of the Lorenz plot are dimensionless, normalized on [0..1], and are typically cumulative fractions of area and population (or population and wealth for the Gini coefficient).  Thus H is independent of the underlying units of measurement; whether population is counted in individuals or millions of people, oe area in hectares or square miles, makes no difference to the Lorenz curve or value of H.  While this is common knowledge, another consequence is underappreciated: H is also independent of population growth and population density.
    
     Consider a region (e.g. state or county) with sub-regions (e.g. counties or townships) with population densities of 4, 2, 1, 1 (in any units, perhaps people per square mile); this has H = 0.25.  Now suppose that over time the populations grow in-place ten-fold to 40, 20, 10, 10.  H remains unchanged at 0.25.  Surely a ten-fold increase in population density would make this region decidedly more urban than it was, but H does not reflect that.  Likewise, if a given percent of people leave every sub-region, H also remains the same. 
     
     H is variable under aggregation: adjacent regions with the same H may have a very different value if considered together.  For example regions with subregions of densities 2, 1, 1 and 10, 5, 5 each has H = 0.17, but considered together have H = 0.33.  Thus H for countries cannot be compared to H for states, nor states to counties -- or for that matter, entities to other entities of different sizes: California is roughly fifty Delawares in both population and area, with all the aggregation artifacts that implies if their H values were compared.
     
     Some might complain that the numeric examples above are toy experiments.  But it is precisely because they are trivial -- with results intrinsic to the construction of H -- that it is difficult to see how H can be said to be a \emph{concentration} index: it is a \emph{heterogeneity} index that does not measure urbanization or concentration and it cannot be aggregated.
    
    \begin{wrapfigure}[27]{r}{0.5\textwidth}
        \centering
        \captionsetup{width=7cm}
        \includegraphics[width=6cm]{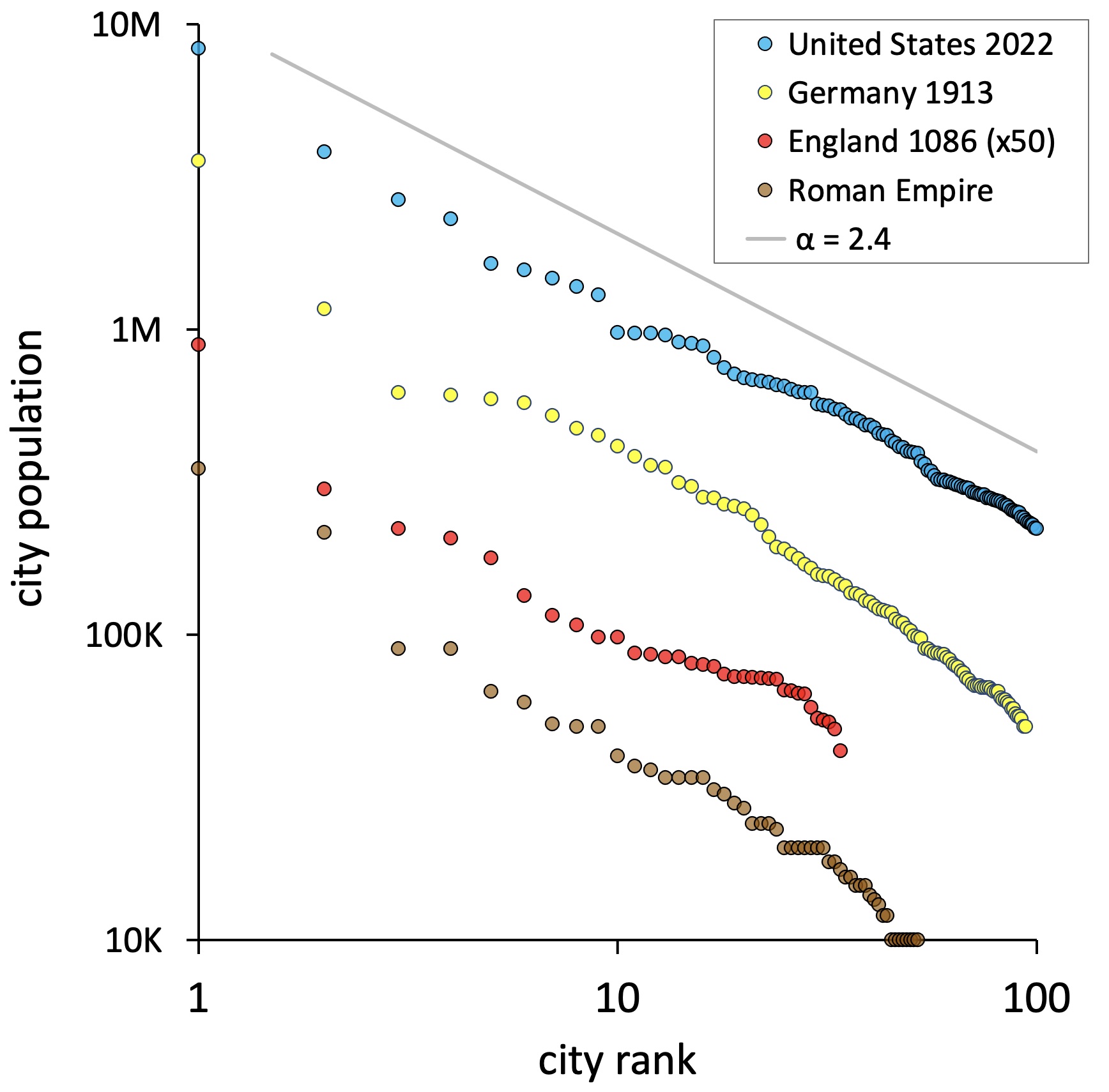} 
        \caption*{Figure 4: Examples of rank-size distributions for city sizes, ancient to modern. The gray line is a power-law function with $\alpha$ = 2.4. See Spencer (2024) and            
            \href{http://scaledinnovation.com/gg/cityGrowth.html}{\xlink} for datasets and sources.}    
        \includegraphics[width=6cm]{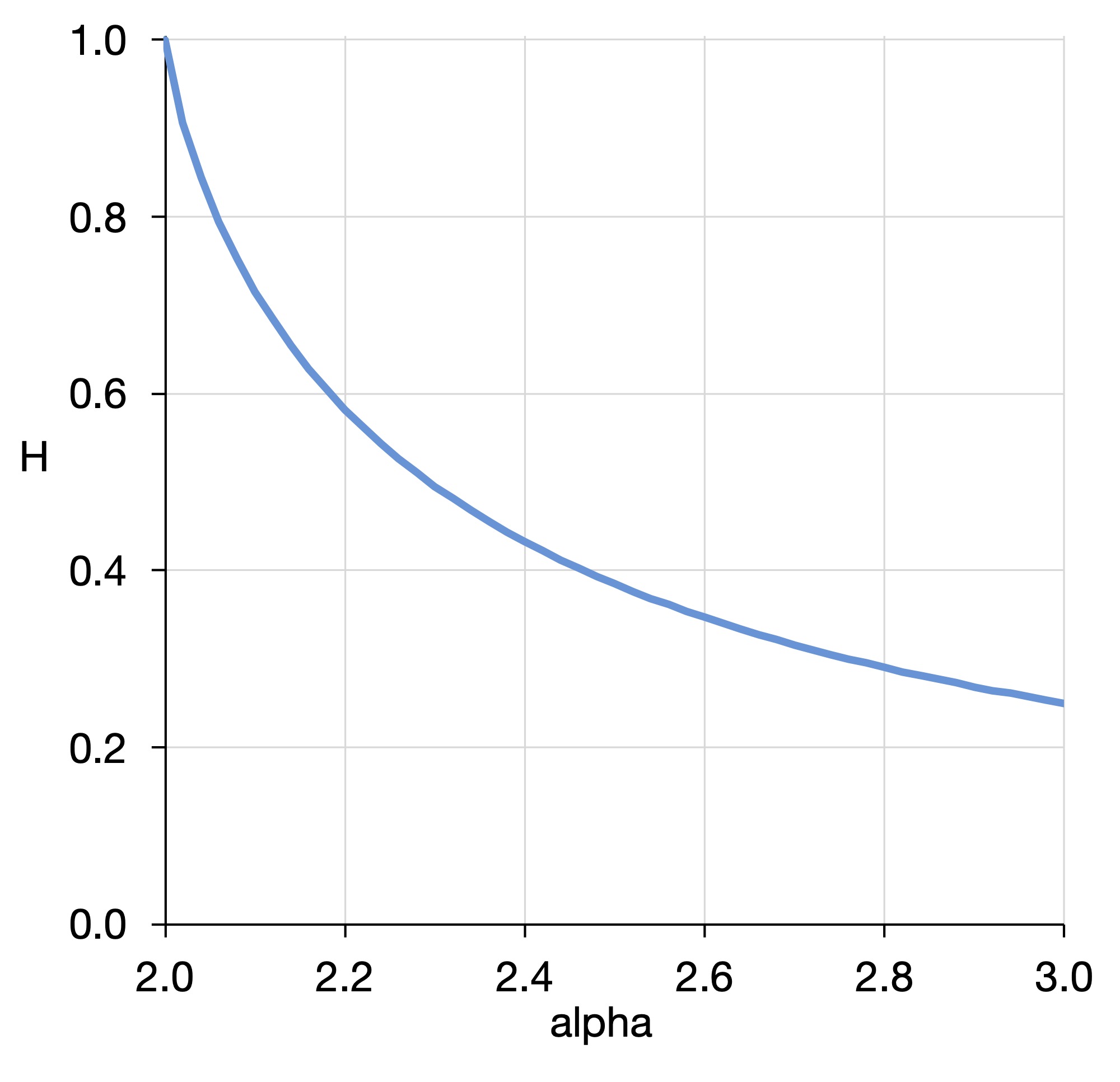} 
        \caption*{Figure 5: Hoover H for a power law distribution as a function of exponent $\alpha$ .}
    \end{wrapfigure}

     The ramifications get more profound when we consider that the substrate for H, namely the sizes of human settlements\footnote{Here I suggest that H may be estimated from settlement populations as well as densities.  Subsequent sections will show that this is valid in the limit of small, equal-area sampling blocks.}, empirically obey a power law distribution which is itself scale-free (Newman 2005).  This is the oldest pattern in demography, first observed by Auerbach (1913), and it applies across centuries and continents with a narrow observed range of its exponent $\alpha$ (Figure 4, Spencer 2024).  Because H is independent of scale and derived from a distribution which is self-similar everywhere, \textit{we should expect to observe a narrow range of its values as a general rule, largely independent of history or situation}.  
     
     The expected value of H can be found by simple algebra.  For power laws, the Gini coefficient $G = 1/(2\alpha - 3)$ and H is a known function of G (Allen 2022): 
     
         $H = \left(\dfrac{1-G}{1+G}\right)^{\dfrac{1-G}{2G}}\left(\dfrac{2G}{1+G}\right)$ 

    \medskip
    Since $\alpha \sim 2.4$ for city sizes (Figure 4, also Clauset et al 2009, Table V), we expect $H \sim 0.4 - 0.6$ as a general rule (Figure 5) and as observed: for 108 international city-size datasets, $H = 0.57 \pm 0.09$ (Spencer 2024, dataset link).  Values of H near zero may be highly egalitarian,  but will be rare and there is no reason to consider them rural or primordial.  If anything, a universal generative model for settlement sizes implies lower $\alpha$ and higher H before the industrial era (Spencer 2024) as discussed below. 

\pagebreak

\subsection*{Population Effects}

    The Lorenz plot is cumulative and normalized to [0..1] on both axes, which makes it universal and unit-independent but also puts distance between the analyst and the data.  Worse yet, tabulated data (perhaps census data) may be plugged into a formula and never examined.  A more direct visualization of the data helps to understand its scaling properties and highlight where and why problems with H may arise.  Figure 6a shows real population density data for the United States, Brazil, and France.   Just as city size distributions are similar over time and space, so are population densities (they would be identical if sampled over fixed-area blocks).  Figure 6b is a sketch of such a distribution which calls out four patterns which may affect the value and interpretation of H.  Three of these: (1) head effects, (2) minimum-value truncation, and (3) saturation, involve the measured population, so I refer to them as \textit{numerator} factors.  Factors involving area, like (4) inclusion of empty spaces, are \textit{denominator} factors.
         
     \begin{figure}[h]
        \centering
        \captionsetup{width=14cm}
        \includegraphics[width=16cm]{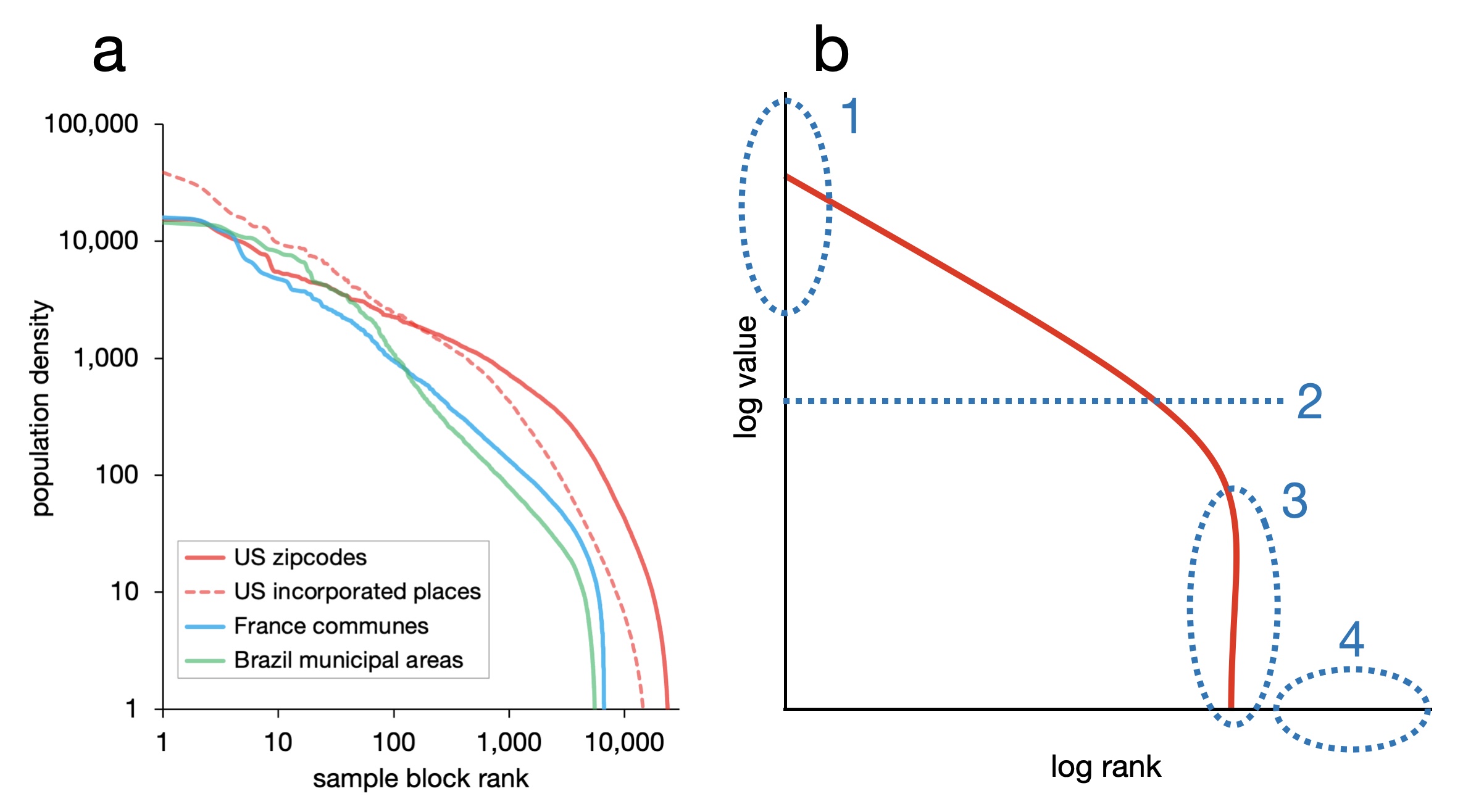}
        \caption*{Figure 6: (a) Rank-value plots of population density.  The most dense block (perhaps central New York City or S\~{a}o Paulo) has rank = 1, and so forth.  US zipcode (2020) and French commune data (2014) treat the raw population data as geolocated point sources which are then sampled on 0.1°x0.1° blocks to give density.  US incorporated place (2020) and Brazilian municipal area (2020) populations are divided by their given areas to give density, in all cases in people per square kilometer.  (b) A sketch of such distributions shows factors which may affect Hoover estimation and interpretation: (1) head effects, (2) minimum-value truncation, (3) saturation, and (4) zero-block effects.}
    \end{figure}
    
    \underline{Head Effects}: Large, dense cities cannot be overlooked, and as they produce and consume most of a country's economic energy tend to be carefully monitored and counted.   There may be debates over where a city's boundary should be or how it is partitioned (is New York City one city or five?), but those do not affect density.  We do not expect, or observe, national-scale effects on H from cities.  Nonetheless, rank-frequency plots may depart from a power law in the head, and since power law models are useful for examining effects on H, these should be examined.
    
     Political or functional blocks often have smaller areas in cities to limit the absolute number of people per block -- there is a practical limit of people per postal code or local political unit.  This manifests in a flattening of the head (upper left) of distributions like Figure 6a.  Alternatively, for historic reasons there are primate cities (Bangkok, London, Vienna) with populations considerably higher than most national distributions would suggest, which raises the head.  But neither of these scenarios has a large effect on H: all developed countries are 80\% or more urban, so that most of the population is already pushed into the far-right of their Lorenz plots.  Squeezing or expanding the right margin of data has little effect on H.  For example, in the US zipcode dataset sampled on 0.1° blocks, artificially doubling the densities of the 100 densest blocks takes H from 0.609 $\pm$ 0.003 to  0.629 $\pm$ 0.004, a 3.3\% increase for an 11\% increase in population.  Halving the same set gives H = 0.602 $\pm$ 0.002, a 1.1\% drop for a 6\% population drop.
    
    \underline{Saturation and Minimum-Value Truncation}: Saturation occurs when a population grows in a finite, bounded space -- at some point people can no longer spread out and the region as a whole must become more dense.  The generative theory is well developed (Bagrow at al 2008) and reproduces observed saturating power laws over both time and space (Spencer 2024).  Since power laws are scale-free, H should be constant.  And since H is scale-independent, it should be unchanged if all samples are multiplied by a constant, which is exactly what saturation of a power law produces. Figure 7 shows results for a minimal model, a function for city sizes which is equivalent to density sampled on unit-area blocks. Figure 7a shows growth over time, as $y_{max}$, the population of the largest city, grows from 10 to 100,000 -- the vertical shift is tantamount to multiplying by a constant.  Figure 7b shows H for these distributions, and above $x \approx 200$ it is constant, as expected.  Saturation \textit{per se} does not affect H.  But at lower x, H falls dramatically which would seem to violate the basic principle of scale-independence.  This illustration with a simple function shows that this effect on H is purely algebraic and not due to any real-world issues of data acquisition or measurement.
        
     \begin{figure}[h]
        \centering
        \captionsetup{width=14cm}
        \includegraphics[width=14cm]{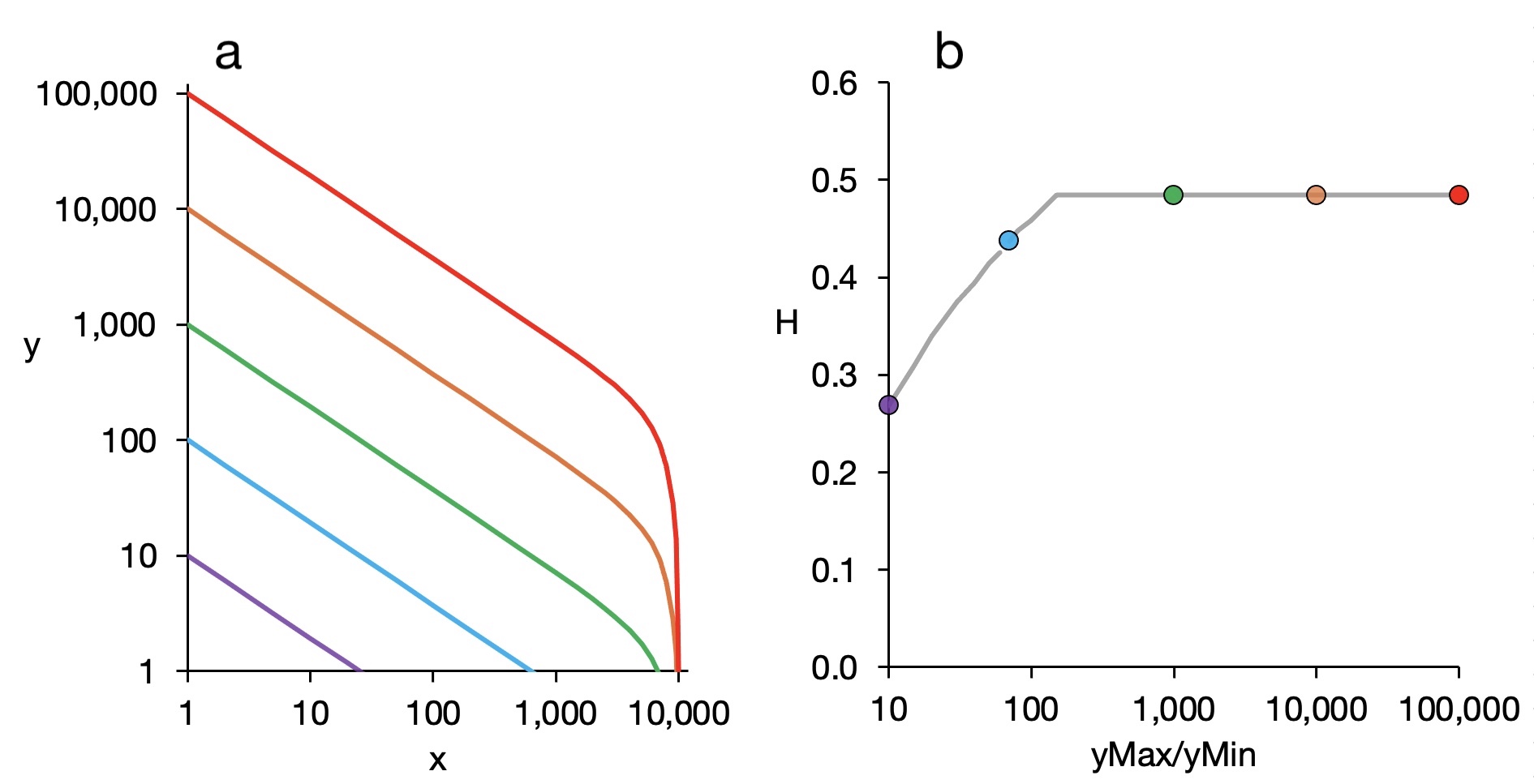}
     \caption*{Figure 7: (a) The function $y = y_{max} \cdot x^{-1/(\alpha - 1)} ( 1 - (x/\beta)^2 ), y \ge y_{min}$ for $\alpha = 2.4$ , $\beta = 10,000$, $y_{min} = 1$, and multiple values of $y_{max}$.  Saturation, the dropoff as $x \rightarrow \beta$, is apparent for $y_{max} > 1000$. (b) H calculated from these distributions, with color dots matching the five distributions of (a) and additional calculated H completing the gray line.}
     \end{figure}
    
        What's going on?  How can H \textit{not} be constant in such a simple system?  The general answer is that the presumption of scale-independence derives from continuous mathematics, while demographic data take on integer (or otherwise truncated) values. The specific answer is that population data (and therefore derived density data) may be truncated.  It is not unusual for census tables to record only settlements above a minimum population, perhaps 100 or 1000 people.  How could it matter?  These uncounted settlements represent a tiny fraction of the whole.  A tiny example illustrates the issue: suppose that we have unit-area blocks with populations 600, 200, 200, 50, 50, 50, 50.  The results for H depend dramatically on our rules for the lowest-valued blocks: if we keep all the data, H = 0.41.  If we truncate values at 100 but keep their areas, our data are 600, 200, 200, 0, 0, 0, 0 which gives H = 0.57.  If we ignore all blocks below 100, both populations and areas, our data are 600, 200, 200 which gives H = 0.27 (this case applies to Figure 7b).  When the data approach the truncation threshold, apparently innocuous rules can swing H in opposite directions by significant amounts.  Both `rules' make sense in different contexts, or worse, may be applied in ignorance, if for example we use density data without knowing that it had been based on truncated population data.
        
        This section closes with a real example that can be directly compared to Figure 7.  Figure 8a shows census-based distributions for US county growth, with the unsurprising signature of power laws approaching saturation.  Figure 8b, in blue, shows H values naively derived from these data (and other years not shown in 8a).  Shortly after the colonial era, H begins near 0.3 and rises smoothly to levels of 0.55 - 0.6 by 1900.  Surely this is another representation of the closing of the frontier and eventual urbanization of the United States?  But the narrative fits too well and the examples above urge caution -- specifically since Figure 8b (blue) has the rise-to-plateau shape of Figure 7b.  There are at least two ways to check the result.  One is to dig carefully into the source data -- were there truncations of small settlements?  Were areas counted fairly or omitted?  But for centuries-old data that would be difficult, so instead we check an independent source, namely massive genealogies in which millions of online users record family history (Spencer \& Otterstrom 2024).  These can serve to estimate H since ancestors' birth dates and locations are easily aggregated into geographic blocks.  The results of 3 million such records are shown in Figure 8b in red: going back to the 1620-1650 birth cohort (well before the first US census in 1790), H is relatively flat at 0.6 - 0.7 up to the 1950-1980 birth cohort (where the genealogy data ends).  This is a more reliable result since genealogy data are not truncated (they are at the indivisible level of the individual) and plentiful (the smallest datapoint in Figure 8b is derived from 10,000 records, and the truncation effect manifests for small datasets).
    
     \begin{figure}[h]
        \centering
        \captionsetup{width=14cm}
        \includegraphics[width=14cm]{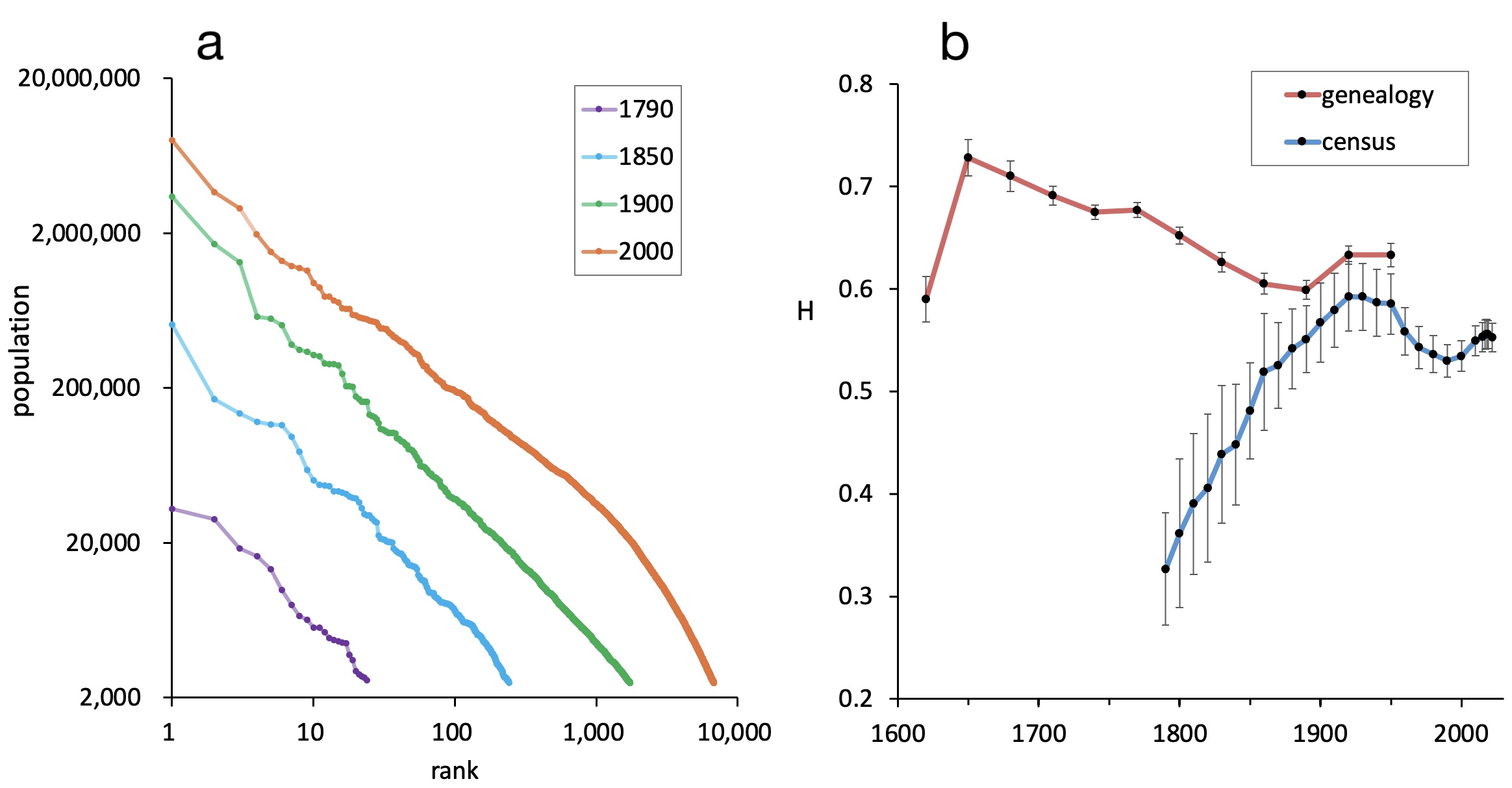}
        \caption*{Figure 8: (a) Population distributions for US counties in the given year. (b) in blue: H calculated from (a), and in red from 3.04 million profiles with birth dates and locations from Kaplanis et al 2018, tallied on 0.1° blocks by 30-year generational time slices. Error bars here and elsewhere are $\pm 1$ standard deviation from 1000 bootstrap cycles.}
     \end{figure}
\pagebreak

\subsection*{Area Choices: the Empty Block Effect}

    \begin{wrapfigure}[18]{r}{3cm}
        \vspace{-12pt}
        \centering
        \captionsetup{width=3cm}
        \includegraphics[width=2.8cm]{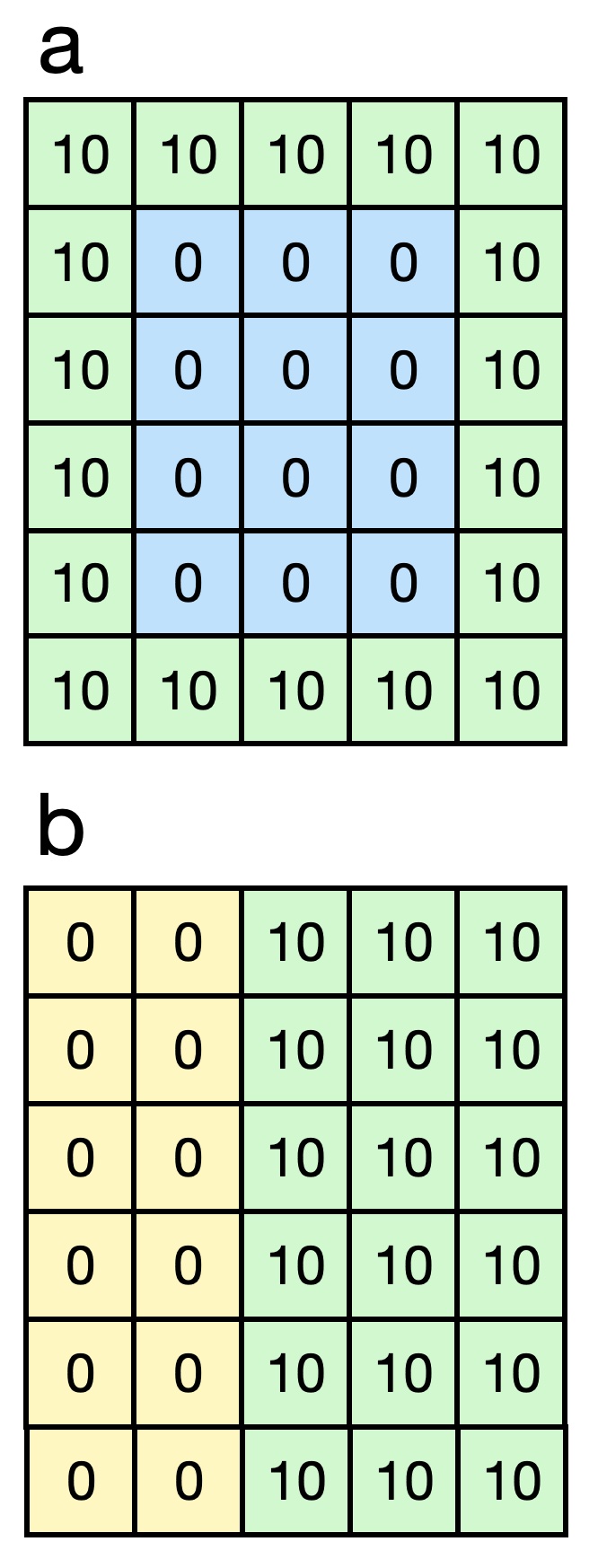} 
        \caption*{Figure 9: Examples to illustrate the effect of empty blocks}
    \end{wrapfigure}
    
    Following the roadmap of Figure 1, the data analyst has two choices about denominator effects on H, namely the areas over which population data are sampled.  Both the scale \& zoning and empty-block choices have profound effects on the value of H; I begin with the latter which is best explained with the simple example of Figure 9.
    
    Figures 9a represents a region with a large central lake.  The lake consists of 12 identical blue blocks where no one lives and is surrounded by 18 same-sized green blocks each with the same population.  Since people can't choose to live in the lake, there's little debate that H should be based only on the occupied blocks, which gives H = 0 (perfect equality for the green blocks).  The ambiguity arises with Figure 9b which has the same numbers and sizes of occupied and empty blocks, but a different narrative:  now the occupied blocks (in green) are the leading edge of a frontier and the empty blocks (in yellow) are just beyond that frontier.  The difficult choice for the analyst is whether the empty yellow blocks should be included in H or not.  If the yellow blocks are severe desert, then probably not, for just like the lake this is not a place where people could reside, so counting only the green gives H = 0.
    
    But maybe we're doing a historic analysis and know, with hindsight, that the yellow blocks are perfectly habitable because they will become occupied in future years.  Figure 9b might represent Kansas in 1850: evenly inhabited to the east and uninhabited (except for the sadly uncounted indigenous people) to the west.  In this case we \textit{should} count the empty blocks, which gives H = 0.4.  Including empty blocks \emph{always} increases the value of H because doing so compresses the Lorenz x-axis to the right.  So in our Kansas frontier model, the frontier scenario has higher H than the later fully settled scenario.  This is opposite to the narrative that H represents urbanization and proceeds in time from lower (primordial, rural) to higher (urban, `concentrated') values.
    
    Whether or not to include unpopulated sample blocks has no clearly correct or incorrect answer; it is a matter of judgment that should be based on separate information like history, climate, or land access.  The choice has substantial impact:  for example 8.4\% Minnesota's surface area is water, and for a model distribution ($\alpha = 2.4, H = 0.33$) adding 8.4\% additional empty space raises H by 7\%.  The considerations are more weighty for Wyoming where nearly half of its area is federally owned, much in national parks and forests which preclude settlement, as if bodies of water.  Choosing to include or exclude that area from Wyoming's calculation changes its H by a factor of two.

    \begin{figure}[htbp]
        \centering
        \captionsetup{width=14cm}
        \includegraphics[width=16cm]{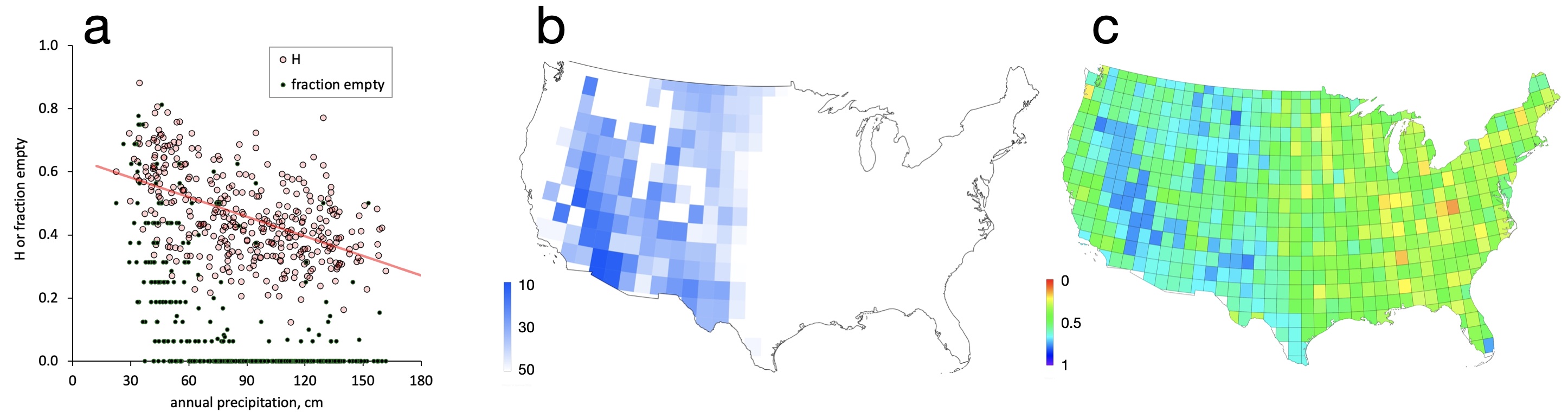}
        \caption*{Figure 10: The empty block effect for 1.2x1.2\textdegree blocks. (a) H and fraction of empty blocks vs average annual precipitation 1901-2000 (b) Precipitation on 1.2\textdegree blocks, scale in cm/year. (c) Same as Figure 2c: H computed with empty sub-blocks, to compare to (b).}
     \end{figure}
     
    In the maps of Figure 2, the value of H for each 1.2°x1.2° colored block is computed from its 16 internal 0.3°x0.3° blocks.  Some of those small interior blocks may have no data, and the question is whether to include them or not in computing H; the consequences are dramatic as Figures 2b vs 2c attest.  With empties included, the blue blocks of Figure 2c at first seem to correspond to very low population density (Figure 2a), though there are differences in the Big Bend of Texas and northern plains.  Instead I suggest a common root cause for the low density and prevalence of empty blocks (and thus very high H), namely extreme aridity: Figure 10a shows the numeric and 10b vs 10c the geographic correlation with average annual precipitation.
    
    If history and geography suggest that empty blocks should be included, then the choice of aggregation scale is critical since $H \rightarrow 1$ as block size $\rightarrow 0$ (Figures 13, 14, 15)\footnote{ $H \approx f_{empty}$ (fraction empty) for $f_{empty} \gtrapprox 0.5$; see Figure 16. In such cases H encodes no information about population distribution, only the fraction of empty blocks.}.  This difficulty can be resolved either by excluding only specifically justified empty areas (which is laborious) or by choosing an aggregation scale that makes the question moot: see Recommendations.

\subsection*{Area Choices: Sample Block Effects}

    The problem of choosing sample blocks for data aggregation, the modifiable areal unit problem or MAUP, is pervasive and problematic.  It has generated strong warnings: ``[MAUP is] unlikely to be solved by geographers who are blinkered by both a statistical perspective and fervent adherence to a paradigm that denies the very existence of the problem" (Openshaw 1983, p 37). ``The results ... are rather depressing” (Fotheringham \& Wong 1991, p 1025). ``[A]lthough MAUP is a statistical and cartographic problem, it has real-world influences. (Buzzelli 2019, p 173). ``Fundamentally, there is little that can be done" (Andresen 2021, p 3). ``The purpose of this paper will have been accomplished ... if it prevents the future computation of meaningless correlations" (Robinson 1950, p 341).  Perhaps another toy  example can convince any remaining ostriches.
    
    \begin{wrapfigure}[9]{r}{3.5cm}
        \vspace{-12pt}
        \centering
        \captionsetup{width=3cm}
        \includegraphics[width=2.8cm]{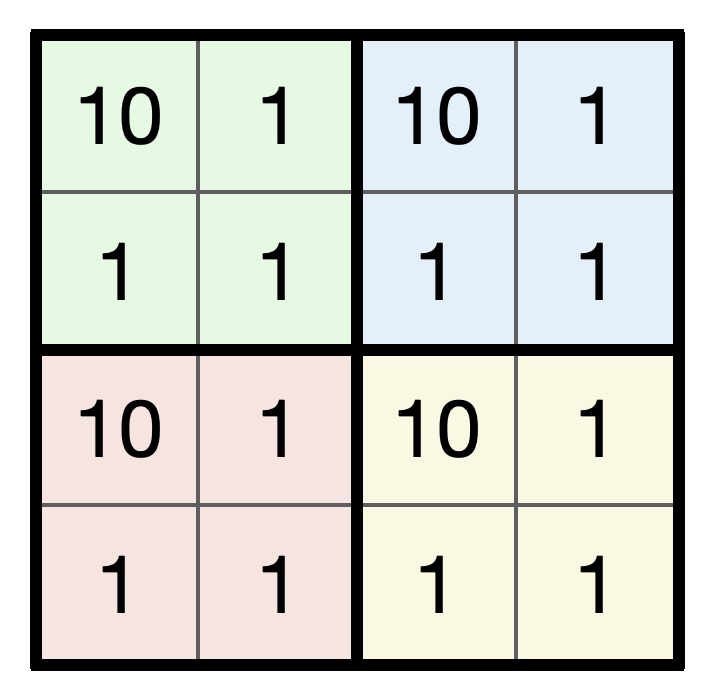} 
        \caption*{Figure 11: Example to illustrate the effect of sample block size}
    \end{wrapfigure}
    
    In Figure 11 the smallest blocks represent townships each with the indicated population (and density on unit area blocks).  The mid-sized colored blocks are counties of four townships each, and the enclosing block is a state with four counties and sixteen townships.  H for each county is computed from [10, 1, 1, 1] = 0.37.  If we choose to aggregate by row, that is two rows each of size [10, 1, 10, 1] and [ 1, 1, 1, 1], then the row H values are 0.41, 0, 0.41, 0 (this is the zoning problem). If the state is considered as four counties, then H = 0 since the counties have identical populations and densities.  But if the state is considered as sixteen townships, then H = 0.37.  This is the scale problem.  Like political gerrymandering, you can get almost any result you wish by redefining boundaries.

    Reactions to the MAUP vary considerably.  Some researchers ignore it, some despair of it, some craft new indices in questionable attempts to evade it, and some, remarkably, view it as a useful indicator of area-specific trends (Vining \& Strauss 1977).  Cohen's recent review of Lorenz-related indices concludes that ``A question remains open: How sensitive are measures of population concentration empirically to different methods of delineating the boundaries and the internal spatial units of cities?" (Cohen 2021, p 1182). It is time to address Cohen's open question with an experimental approach.
    
\subsection*{Experimentally Variable Sampling}

    In the most studies the numerator (population) and denominator (sampled area) of population density are fixed by the data-collecting agency, typically a census department.  Computed H are immediately followed by interpretation; there is no opportunity to explore alternatives since the data are fixed. Occasionally other regions are available, such as commuting zones, but these are also fixed and chosen for their own particular purposes.  Tessellation between datapoints, such as Delaunay triangulation, adapts the sampled areas to the available data, but has significant problems which make it a poor choice (see Supplementary Material).
  
    An experimental alternative and some of its issues are illustrated in Figure 12.  Figures 12a-d show the same settlements (blue dots) with populations and geographic extent suggested by their sizes.  We cannot compute a traditional H for 12a since we have only one sample block.  For 12b, c, and d we can calculate the population densities of each of the 4, 16, and 64 sub-blocks and use those densities to compute H.  Figure 12b poses no particular geographic problem, though with only four values the final value of H will have high variance (which can be estimated by bootstrapping).  But for scenario 12c and especially 12d we must address the empty-blocks problem: do we count them or not?  Given the way that the settlements are distributed in this example, unless we have separate knowledge of lakes or deserts or national parks, we probably should.

    \begin{figure}[!h]
        \centering
        \captionsetup{width=14cm}
        \includegraphics[width=16cm]{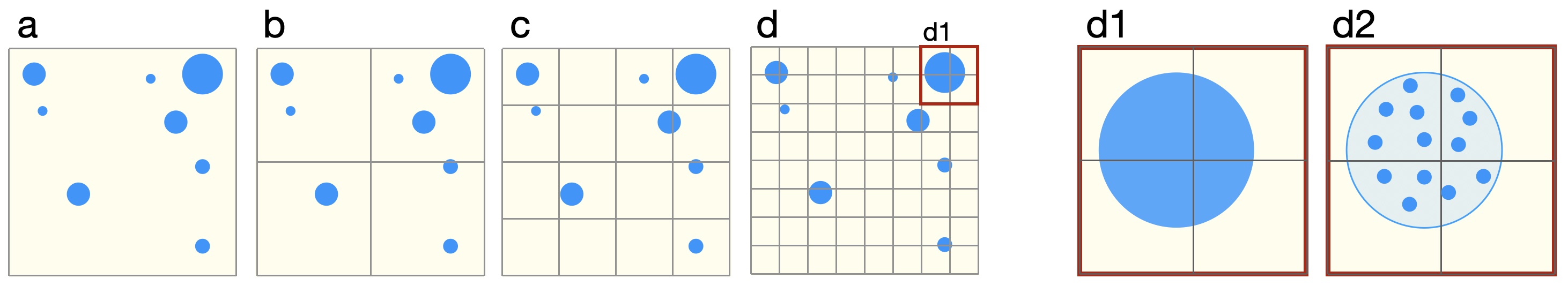}
        \caption*{Figure 12: Fixed settlements (blue dots) are sampled on an experimentally variable sampling grid.}
    \end{figure}

    Figure 12d raises another question in its top-right corner, enlarged in 12d1: How do we count this city that overlaps multiple sample blocks?  The practical answer is suggested by 12d2: use finer-grained data if available.  The variable block method knows only point locations and does not depend on political boundaries (city, county, etc.) so fine-grained population data will give the fairest allocation to each sub-block.  The results below use US census zipcode data, treating each such region as a point source of the given population.  The 2020 US census dataset has 30,323 zipcode datapoints within the contiguous United States -- a ten-fold higher granularity than counties. We can experimentally explore H over three orders of magnitude in sample scale as shown in Figure 13.

    \begin{figure}[!h]
        \centering
        \captionsetup{width=15cm}
        \includegraphics[width=10cm]{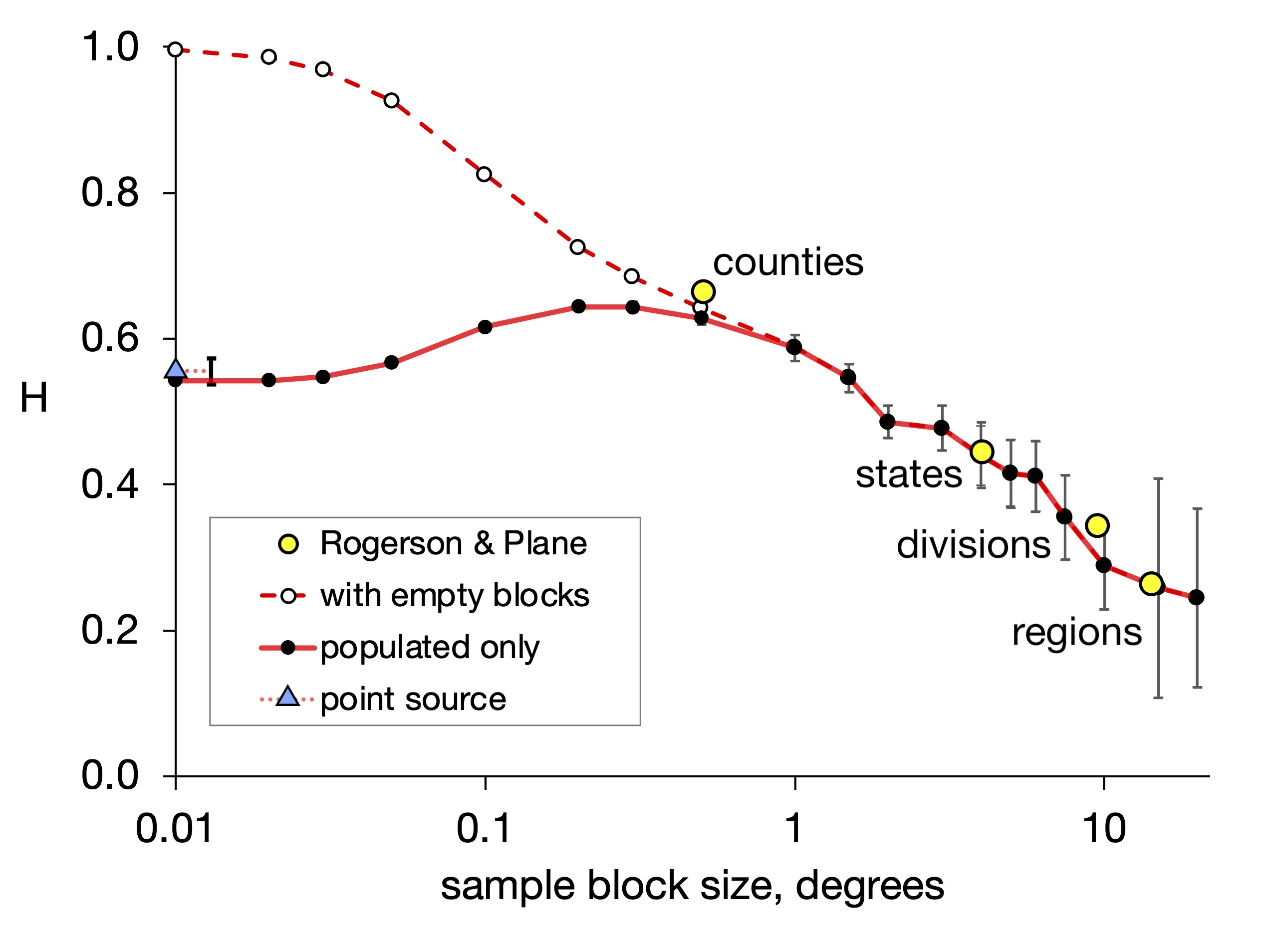}
        \caption*{Figure 13: H for 30,323 contiguous US 2020 zipcode datapoints aggregated on 0.01° to 20°-sided blocks, excluding (solid red) or including (dashed red) empty blocks.  Blue triangle at left axis: H for same datapoints, not aggregated but treated as zero-area point population sources.  Labeled yellow dots are values of H in 2009 from Rogerson \& Plane 2012, Table 1.  Error bars are $\pm$ 1 standard deviation based on 1000 bootstrap cycles.}
    \end{figure}

    In Figure 13 shows that the four wide-ranging values of H for Rogerson \& Plane's census-based blocks (H from 0.66 to 0.26 for counties, states, divisions, and regions)\footnote{The small discrepancies between their four points and my curve is likely due to my sampling only the contiguous 48 states.} are simply points on a continuum.  We also see that including empty sub-blocks has a dramatic effect on H below the scale of counties (about 0.3°x0.3°), such that $H \rightarrow 1$ as block size $\rightarrow 0$.  Also note the small marker at the left axis, which is H calculated solely with zipcode population values as point sources without any sampled area.  As expected, this value is the limit of area-sampled H as block size $\rightarrow 0$ ignoring empty blocks.
    
    Duncan (1957) and Vining \& Strauss (1977) interpret H patterns in terms of the American  frontier narrative of westward expansion.  The same analysis for France and Brazil calls such interpretations into question, since as Figure 14 shows, they have the same pattern as the US yet very different histories -- France hasn't had a frontier since perhaps the early Neolithic.  Inclusion of empty blocks makes H a nearly logarithmic function of block size (i.e. linear on these semilog charts), with $H \rightarrow 1$ as block size $\rightarrow 0$, and $H \rightarrow 0$ as block size approaches the national scale.  Counting populated blocks only, all countries plateau for small blocks to a narrow range of $0.54 \le H \le 0.61$.  

    \begin{figure}[!h]
        \centering
        \captionsetup{width=10cm}
        \includegraphics[width=10cm]{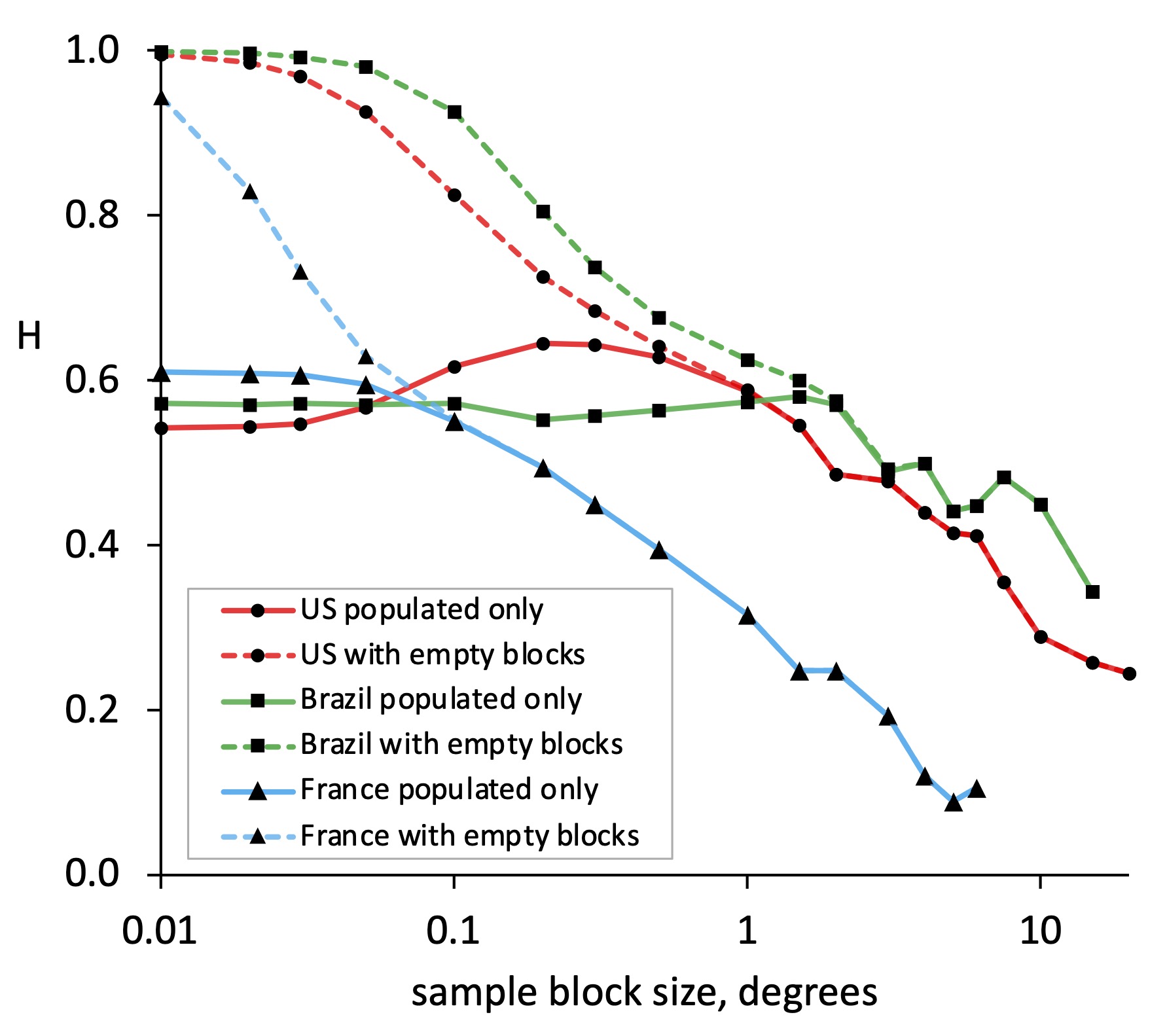} 
        \caption*{Figure 14: As Figure 13, including France (35,812 communes, 2014) and Brazil (5,572 municipal areas, 2020).}
    \end{figure}
       
    These results raise three primary questions about scale effects on H:  
    \begin{enumerateTight}
        \item Why does $H \rightarrow 1$ as block size decreases when empty blocks are included?
        \item Why does $H \rightarrow constant \sim 0.6$ as block size decreases when only populated blocks are included?
        \item Why does $H \rightarrow 0$ for large sample blocks?
    \end{enumerateTight}
    These questions can be addressed by modeling.  In addition, the US data show a slight rise in H to 0.64 around 0.3° sampling which the others do not, a secondary effect that will be examined afterwards with more complex simulation. 
    
\pagebreak
\subsection*{Modeling Aggregation Effects by Multiple Approaches}

    If a complex model generates results that agree with reality, that model might be applicable.  But if a simpler model also agrees with reality, then the simpler model is more likely to apply\footnote{Following Occam and Einstein: ``Everything should be made as simple as possible but not simpler."}.  Thus modeling is more powerful in refutation than confirmation -- it cannot prove what is true but can identify what is false or at least unnecessary.  The three primary questions above are addressed with increasingly abstract data and models:
    
    \begin{enumerateTight}
        \item \emph{Real populations and locations}: US zipcode data as Figures 13, 14.
        
        \item \emph{Population data shuffled}: Same data, with zipcode population values swapped at random with another location.  Locations remain in place, i.e. closer together for urban than rural zipcodes.
        
        \item \emph{Saturated pareto values as point sources}:  Data are 23,600 values drawn at random from a distribution of the form $y = y_{max} \cdot x^{-1/(\alpha - 1)} ( 1 - (x/\beta)^\gamma $ with  $ y_{max} = 30,000, \alpha = 3.5, \beta = 20,000, \gamma = 0.1$ to fit the observed densities of Figure 6a (red).  For this and succeeding cases the numbers of blocks with and without population match those of the real data at each given sample block size.

        \item \emph{Pure pareto values as point sources}:  Data are 23,600 values drawn at random from a distribution of the form $y = y_{max} \cdot x^{-1/(\alpha - 1)} $ with $ y_{max} = 30,000, \alpha = 2.2 $.  The value of $\alpha$ is chosen to give H$_{x \rightarrow 0}$ $\approx$ 0.5; lower $\alpha$ give higher H$_{x \rightarrow 0}$.

        \item \emph{Gaussian values as point sources}: Data are 23,600 values drawn at random from a gaussian distribution with mean 30,000 and standard deviation 10,000.
    \end{enumerateTight}

    \begin{figure}[!h]
        \centering
        \captionsetup{width=12cm}
        \includegraphics[width=12cm]{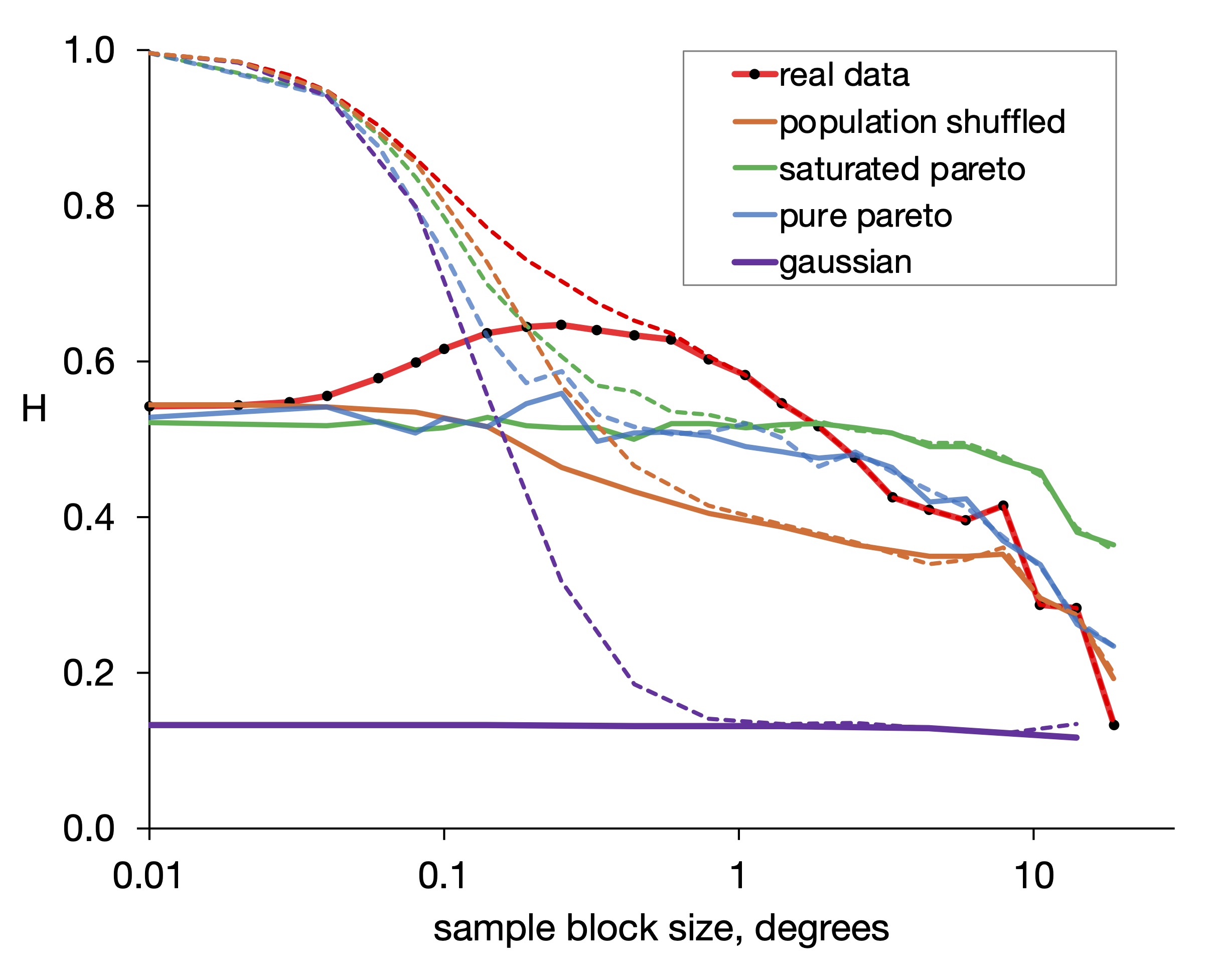} 
        \caption*{Figure 15: Effect of sampling scale on H for real US data and four randomization models.  Solid lines are from sampling only populated aggregation blocks and dashed lines include empty blocks.  Each model is discussed in the text.}
    \end{figure}

    Models 3 and 4 are good matches to the Brazilian data, and French data with a shift in scale.  Model 2 (real data, population values shuffled between locations) inverts the `hump' of the US data.  In all cases the inclusion of unpopulated sample blocks (dashed lines, color-matched to each model) shows the same rise to $H \rightarrow 1$ for small sample blocks.  In all cases for populated blocks only, except gaussian, $H \rightarrow constant \, \sim 0.6$ as block size $\rightarrow 0$; for gaussian data, $H = constant \sim 0.1$, independent of block size.  That H should be near zero for gaussian data is entirely expected, since every block, no matter its size, will have about the same density.  Decreasing the standard deviation in the gaussian model sends H closer to zero. 
    
    \begin{wrapfigure}[15]{r}{5cm}
        \vspace{-12pt}
        \centering
        \captionsetup{width=4.5cm}
        \includegraphics[width=5cm]{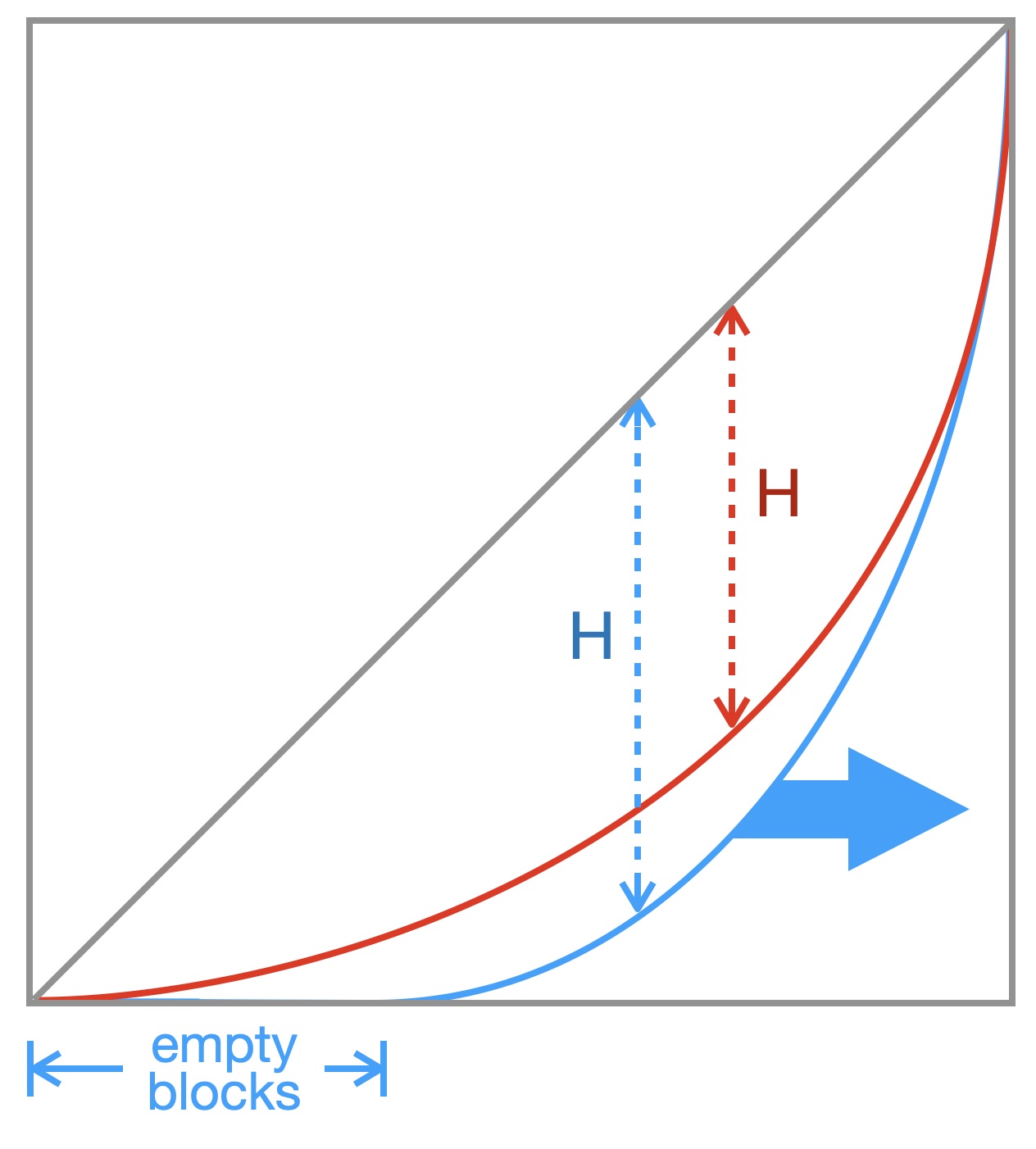} 
        \caption*{Figure 16: Empty sample blocks push the Lorenz curve to the right, increasing H.}
    \end{wrapfigure}

    Addressing the first primary question is straightforward: $H \rightarrow 1$ as block size $\rightarrow 0$ in all datasets and models, which is explained as above for the empty block effect:  including such sample blocks pushes the Lorenz curve to the right, which increases H (Figure 16).  This will be true for all data sources and distributions since the Lorenz curve is monotonic concave-up by construction.
    
    The plateau of H and approach to a constant value as block size decreases is also easily understood.  At some block size, the density of the underlying data (zipcode or town populations, etc.) is such that each non-empty block will contain just one point source (e.g. Figure 12c).  From then on, the distribution by aggregating blocks is the same as the distribution of point sources with its corresponding value of H. 
    
    The last primary question is why $H \rightarrow 0$ as data are aggregated into larger blocks.  This is the most-cited problem of MAUP, and as Fotheringham \& Wong (1991) noted, the answer can be understood intuitively: as the aggregation scale gets larger, each block comes to resemble the whole, i.e. a few cities, many towns, and some rural areas.  Thus the blocks' distributions become self-similar and $H \rightarrow 0$.  Aside from metropolitan Paris, France is relatively homogeneous, which accounts for it reaching this self-similarity threshold at smaller aggregation sizes than the US or Brazil (Figure 14, solid lines). 

    A note about measuring MAUP: while the literature is correct to recognize the problem, most authors do so in misleading terms, namely with correlation coefficients or standard deviations (Openshaw 1983, Fotheringham \& Wong 1991, Wong 2004) which are appropriate for centrosymmetric but not long-tailed distributions.  As Newman (2005) notes, the higher moments of power law distributions with $\alpha \le 3$ diverge, so they have no meaningful variance or standard deviation.
    
    The continuous nature of H vs sample block scale and the common $H \rightarrow 0$ pattern with increasing block scale for different nations as well as random long-tailed distributions are compelling evidence that the county-vs-state, state-vs- division, etc., differences in H observed by Rogerson \& Plane (2012) and others are no more than artifacts of sampling scale.  It is difficult to make a case that MAUP provides interesting information when its dominant pattern is reproduced by arrays of random numbers without any location information.

    The remaining unexplained phenomenon of Figures 14 and 15 is the rise of H in the US dataset for sample blocks of intermediate size.  Its absence in random models 3, 4, and 5 suggests that it reflects spatial correlation, for example a tendency for dense to be near dense, or sparse near sparse, a question that can be approached with simulation.

\subsection*{Simulation}

    I draw a distinction between modeling and simulation:  modeling is the application of mathematics in the abstract, while simulation builds a representation of reality with temporal and spatial dimensions.  Simulation here is based on the checkerboard model of Spencer 2024: the landscape begins empty and adds a person at each time-step.  With probability $p_{stay} = 1/(\alpha - 1)$ (the familiar $\alpha$ of city-size power laws) that person stays at their `birth' location, or with probability $1 - p_{stay}$ jumps (migrates) to another location at a distance chosen at random from a 1/r distribution with given median jump distance.  This is analogous to a Lévy flight across multiple generations (Spencer \& Otterstrom 2024).

    \begin{figure}[!h]
        \centering
        \captionsetup{width=14cm}
        \includegraphics[width=14cm]{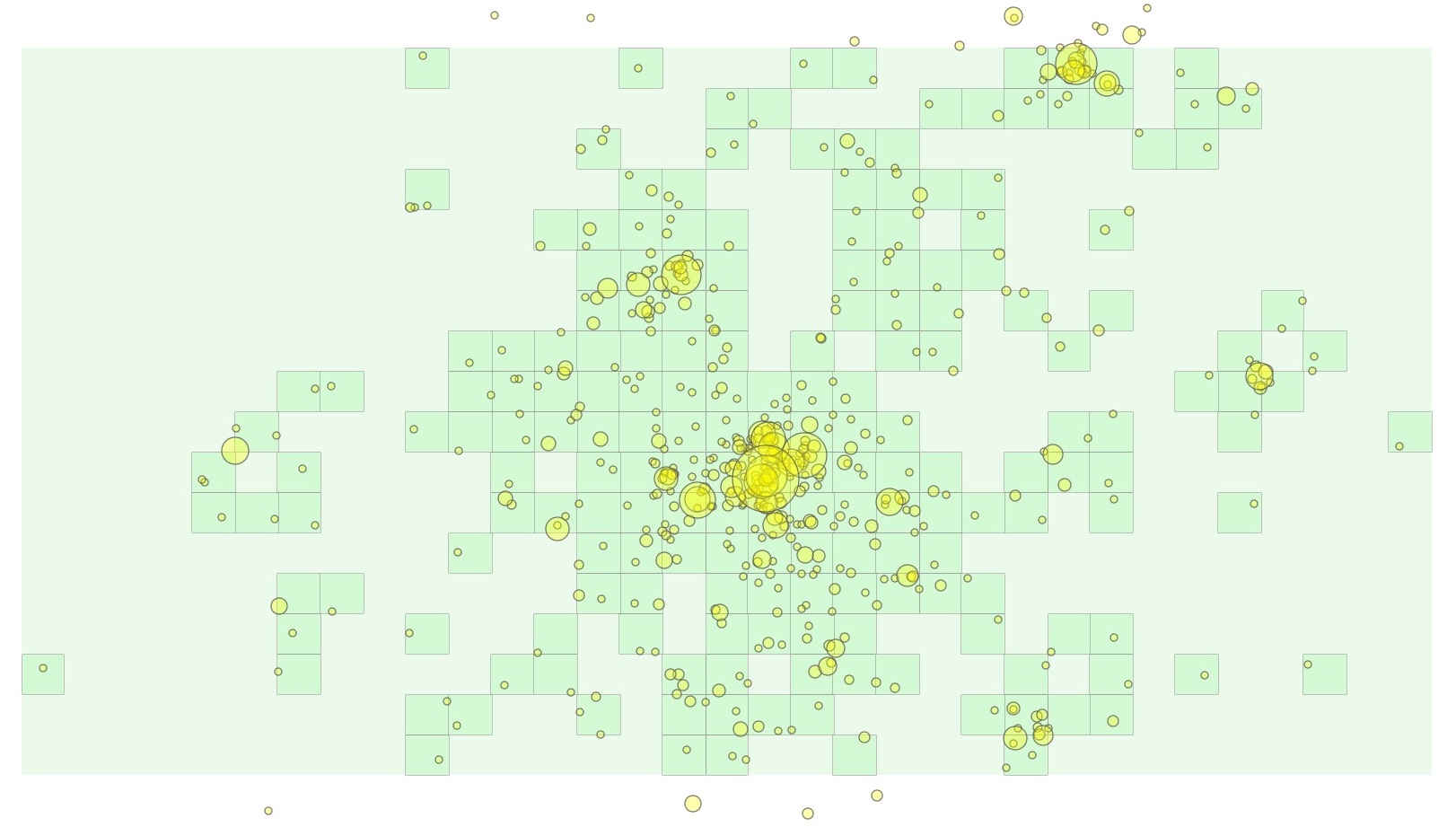} 
        \caption*{Figure 17: Simulation example with landscape 4200 x 2000 scale km, final population = 20000, $\alpha$ = 2.4, 200 scale km median jump, no seeding, and 100 scale km per side sample blocks.  The `national boundary' is in light green, settlements are yellow circles with area proportionate to population, and sample blocks with non-zero population are darker green squares.}
    \end{figure}

    \begin{figure}[!h]
        \centering
        \captionsetup{width=15cm}
        \includegraphics[width=16cm]{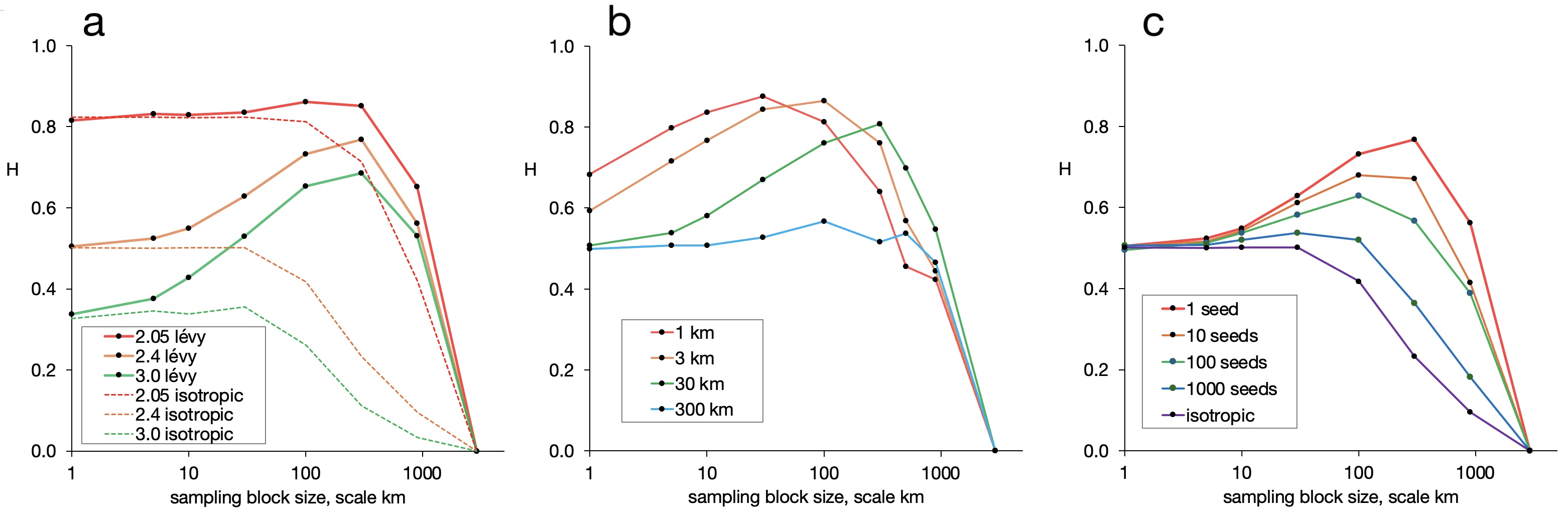} 
        \caption*{Figure 18: Simulation results, all aggregating populated blocks only, final population 10,000. (a) Varying $\alpha$, all 50 km median Lévy (solid) or isotropic (dashed) jumps, one seed. (b) Varying median Lévy jump length, all $\alpha$ = 2.4, one seed. (c) Varying the number of initial random seeds, all $\alpha$ = 2.4 and 50 km Lévy jumps.}
    \end{figure}
    
    The results are straightforward.  Isotropic jumps (Figure 18a, dashed lines) give the monotonic plateau in H of France, Brazil, and the random models of Figure 15, while coordinated jumps (solid lines) give a mid-scale peak in H characteristic of the US data.  As $\alpha \rightarrow 2$, $H_{x \rightarrow 0} \rightarrow 1$, as expected (because at $\alpha = 2$, no one leaves home and the population grows at a single location).  Figure 18b shows that the block scale of the peak in H is a function of Lévy jump distance, with the peak unsurprisingly collapsing for jumps approaching national scale.  Figure 18c shows that additional randomly located `pioneer settlers' (seeds) smoothly transition from the single-pioneer pattern with maximum spatial correlation to fully isotropic with none.
    
    This simulation reproduces the principal features of all datasets.  Notably the simulation has no directionality, no frontier, only spatial correlation via jump selection.  This shows that the effects on H of aggregation block scale are not a legacy of the American frontier, but a general result of spatially correlated suburbs surrounding cities, built on top of a one-parameter model of pareto distributed city growth.
    
\subsection*{The Effect of Population Growth}

    Perhaps the most salient observations with H are those of Figure 19 where Hoover (1941) and others show dramatic changes in H in the United States over time.  Using state-level census data, all authors find that H decreases significantly over time (Figure 19a), while Rogerson \& Plane 2013, like Duncan et al 1961, find that H rises slightly over time for county-level data and simultaneously falls at higher levels of aggregation (Figure 19b).  The trends in H are compelling as are the authors' conclusions: ``The consistent fall in the coefficient, extending over the entire period, indicates the continued spreading of the population more and more evenly over the land area, with more nearly uniform density in different States." (Hoover 1941) and ``We conclude that dispersal [i.e. decreasing H] is more than a statistical artifact" (Vining \& Strauss 1977).

    \begin{figure}[!h]
        \centering
        \captionsetup{width=14cm}
        \includegraphics[width=15cm]{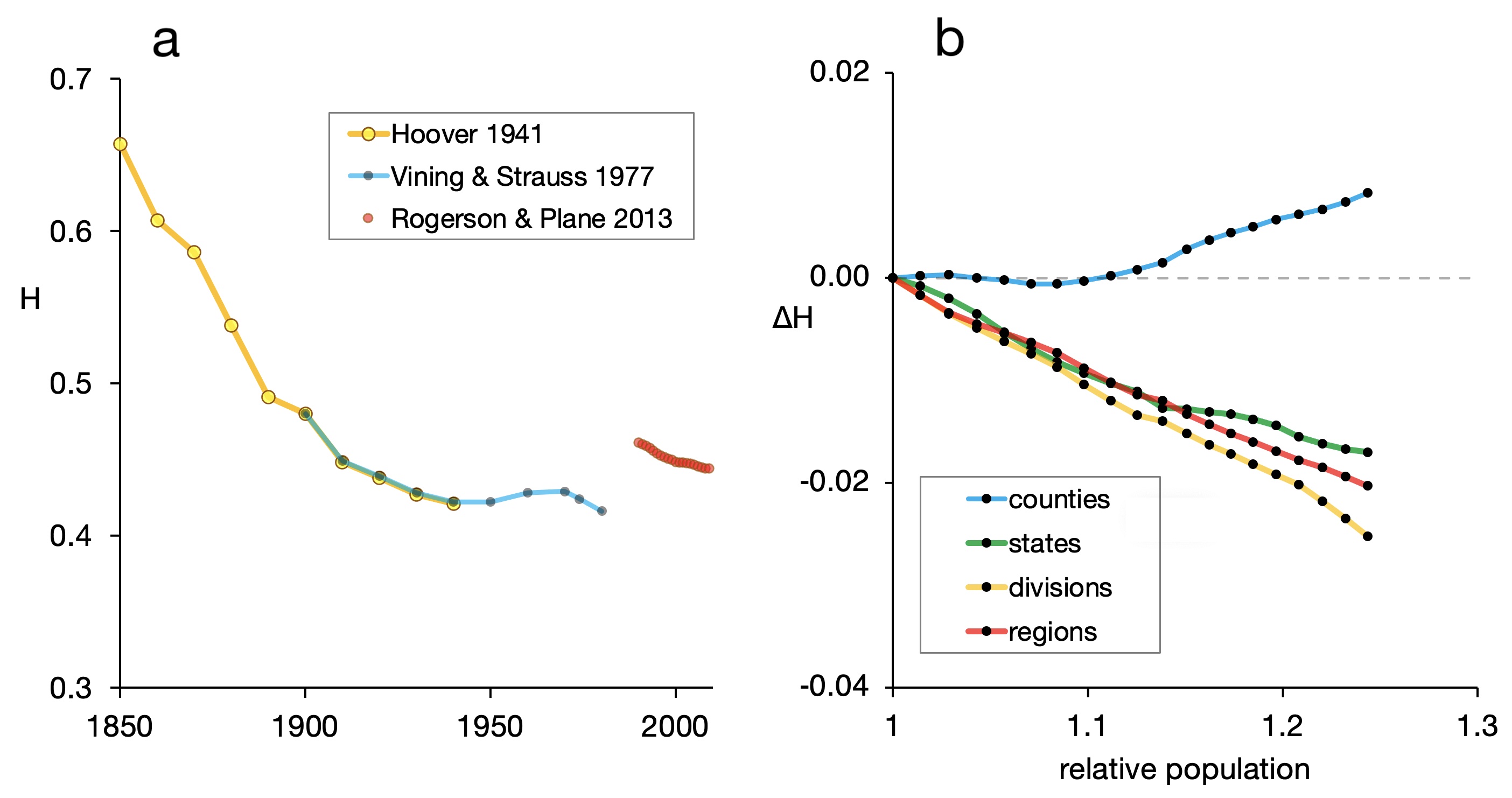} 
        \caption*{Figure 19: (a) H based on state-level census data from Hoover 1941, Vining \& Strauss 1977, and Rogerson \& Plane 2013. (b) The data of Rogerson \& Plane, Table 1, 1990-2009, shown as change in H as a function of relative population (US census, 1990 = 1.0). }
    \end{figure}

    By this point, alarm bells should be ringing.  Hoover's inference that his index's fall indicates ``spreading of the population more and more evenly" is inconsistent with (i) the scale independence of H and (ii) the oldest pattern in demography, the ubiquitous power law distribution of city sizes, and in particular this distribution's consistency in the US from 1790 (the first census) to the present (Figure 8a).  We have seen mathematically that H is constant for a power laws of the same slope, and that H computed in the point-source limit (i.e. the left intercept of Figures 14, 15) is essentially constant in the US after 1900 and after 1630 if computed from massive genealogies (Figure 8b).
    
    The analyses that give constant H are based on unaggregated point sources (city sizes, genealogies) while those that show a decrease (for states and larger) after 1850 are based on state- or higher level aggregation, strongly suggesting that the observed effects are MAUP-related.  The largest demographic change in these time periods wasn't rural-to-urban or urban-to-suburban migration, but simply population growth: the United States grew 12-fold from 1850 to 2000.  Experiments with this factor in the simplest model that replicates observed data, namely pure pareto model of Figure 15, give the results shown in Figure 20.
    
    \begin{figure}[!h]
        \centering
        \captionsetup{width=15cm}
        \includegraphics[width=17cm]{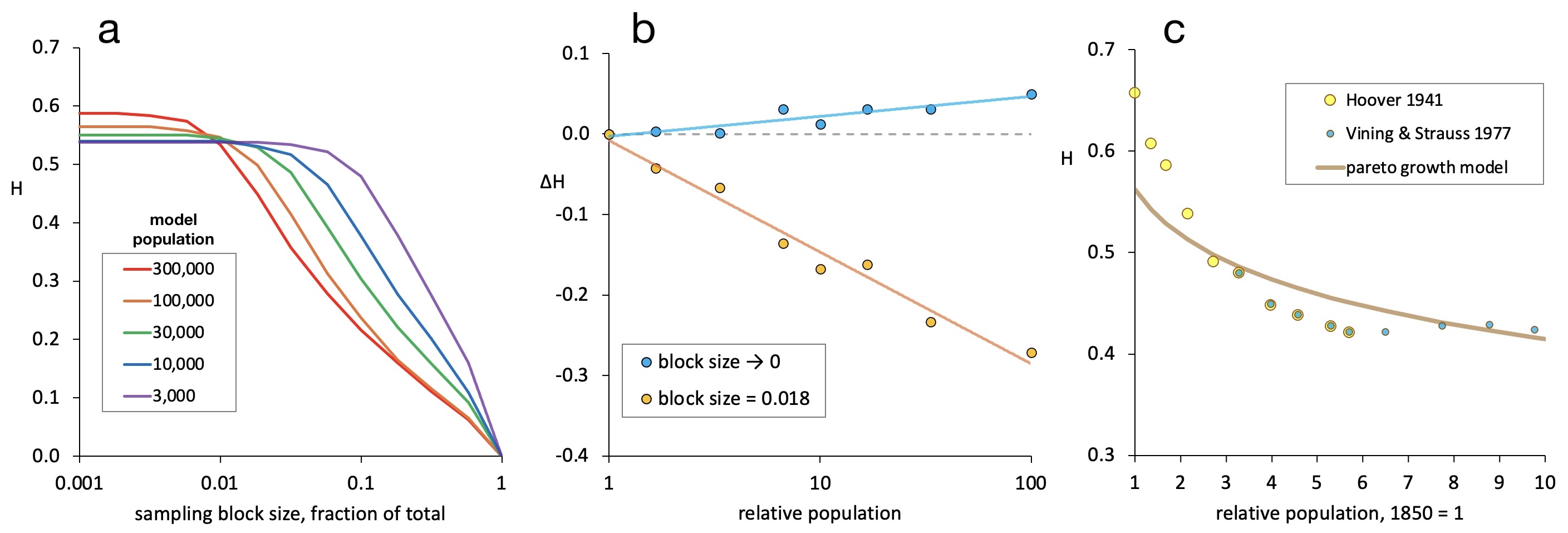} 
        \caption*{Figure 20: (a) Effect of sampling scale for a pure pareto distribution, $\alpha = 2.2$, H values at each block size and population are means of 50 independent iterations. (b)  Vertical slices through (a), for $H_{lim, \, block\, size \rightarrow 0}$ and relative sample block size 0.018.  (c) The logarithmic fit of (b), orange, overlayed on the data of Figure 19a, as a function of relative US population, 1850 = 1.0.}
    \end{figure}

    Figure 20a shows the familiar H as a continuous function of aggregation scale (i.e. like Figures 13, 14, 15, 18) over a hundred-fold range of model population.  At large aggregation block sizes H decreases with population growth, while at small block sizes, H increases slightly, as shown in Figure 20b for two vertical slices through 20a.  The similarity to Figure 19b is remarkable, though a large range of population is used to see a robust effect. The logarithmic fit for block size of 0.018 gives an empirical model which is overlayed on the data of Hoover (1941) and Vining \& Strauss (1977) in Figure 20c.  
    
    \begin{wrapfigure}[17]{r}{6cm}
        \vspace{-12pt}
        \centering
        \captionsetup{width=5.5cm}
        \includegraphics[width=5.5cm]{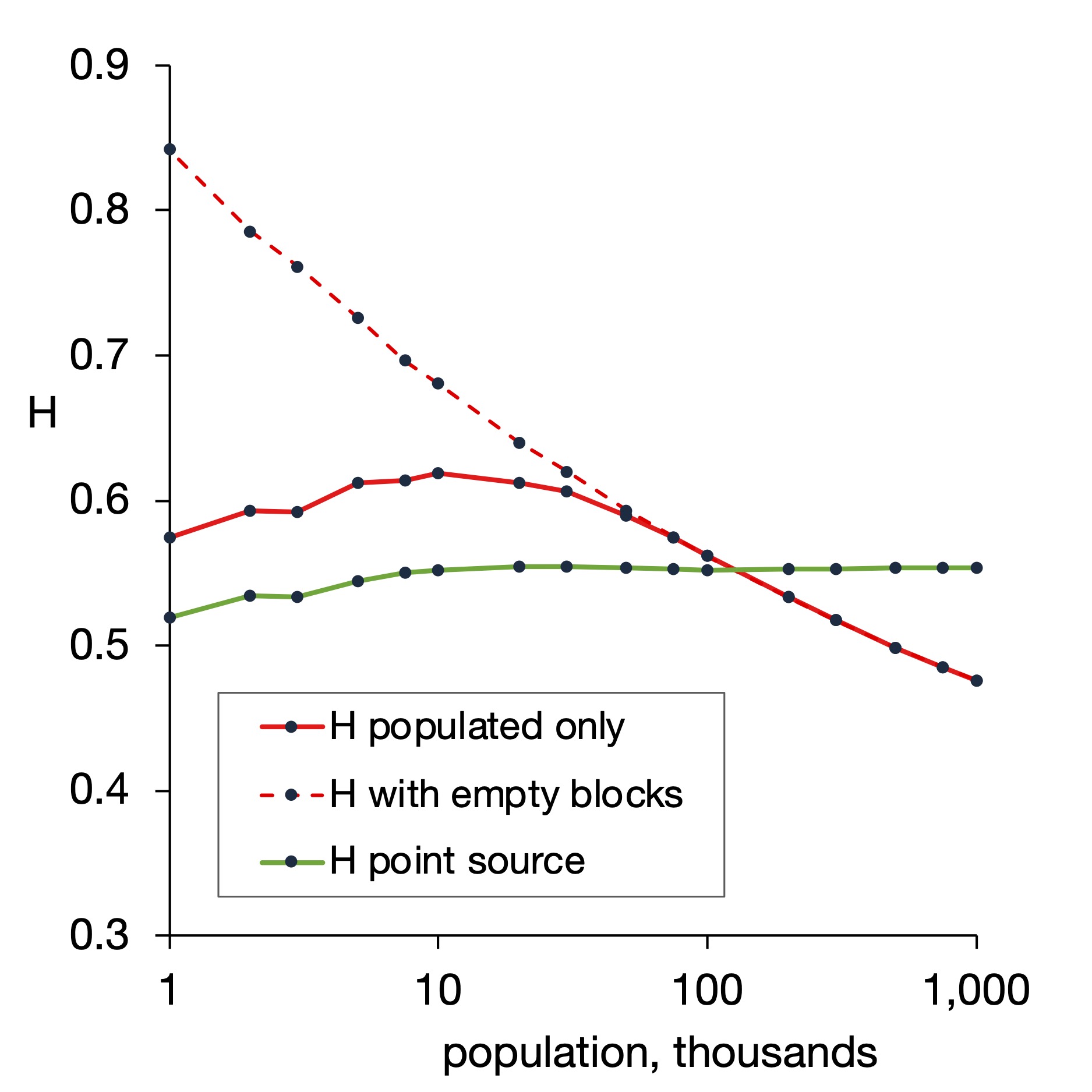} 
        \caption*{Figure 21: Simulation results as Figure 17.  Final population 1 million with data analysis at each intermediate dot, all with $\alpha = 2.3$, 500 km median jump, 1 seed, 100 km aggregation block size.}
    \end{wrapfigure}
    
    This is certainly not a perfect model but it suffices to account for the magnitude and direction of the principal trend in H.  The model consists simply of taking variable-sized subsets of a pareto distribution of random numbers which have no directionality, no spatial or temporal correlation. The results of this algebraic model are mirrored by the results of a more involved simulation (Figure 21) which shows the same shape and magnitude of H decline with growing population.  Simulation additionally explores the effects of including or excluding empty blocks, as well as the value of H determined from settlements as point sources, all in agreement with the common aggregation scale problem.  The population effect on H is not difficult to understand:  as population grows, counties become more like states and states more like census divisions.  The MAUP will arise if the sample block size changes for a static population, or if the population changes for a fixed sample block size.  It is the same phenomenon.
    
    If there is anything to be inferred from the data of Figure 19, it is only what remains after the scale artifacts (and possibly empty block effects) are carefully accounted for.

\subsection*{The Baseline and Primordial Values of H}

	By construction, H = 0 when population density is the same everywhere, and H = 1 when everyone lives in one small region and no one lives anywhere else.  People are dispersed or concentrated \textit{but only relative to each other in the sample}.  Unfortunately ``concentration" is easily misinterpreted in absolute rather than relative terms and so H has become conflated with population density and urbanization -- which toy examples show to be incorrect.  But the confusion persists and some authors suggest that H $\rightarrow$ 0 is a primordial, even idyllic state\footnote{An \textit{idyllic} state is subjective, imagined, and preferred -- the Eloi alone in \underline{The Time Machine}. A \textit{primordial} state is what theory and/or evidence suggest actually existed -- the Eloi and the Morlocks.}.   	We have seen that at small sample scales H will usually be found $\sim 0.4 - 0.7$ (Figures 4, 5, 14), which we can call a \textit{baseline} range, to be expected in the absence of artifacts and exceptional circumstances. Is there also reasonable theory and evidence for a \textit{primordial} range for H, applicable prior to modern urbanization, industrialization, and perhaps far earlier?  

    \begin{wrapfigure}[14]{r}{5.5cm}
        \vspace{-12pt}
        \centering
        \captionsetup{width=5cm}
        \includegraphics[width=5cm]{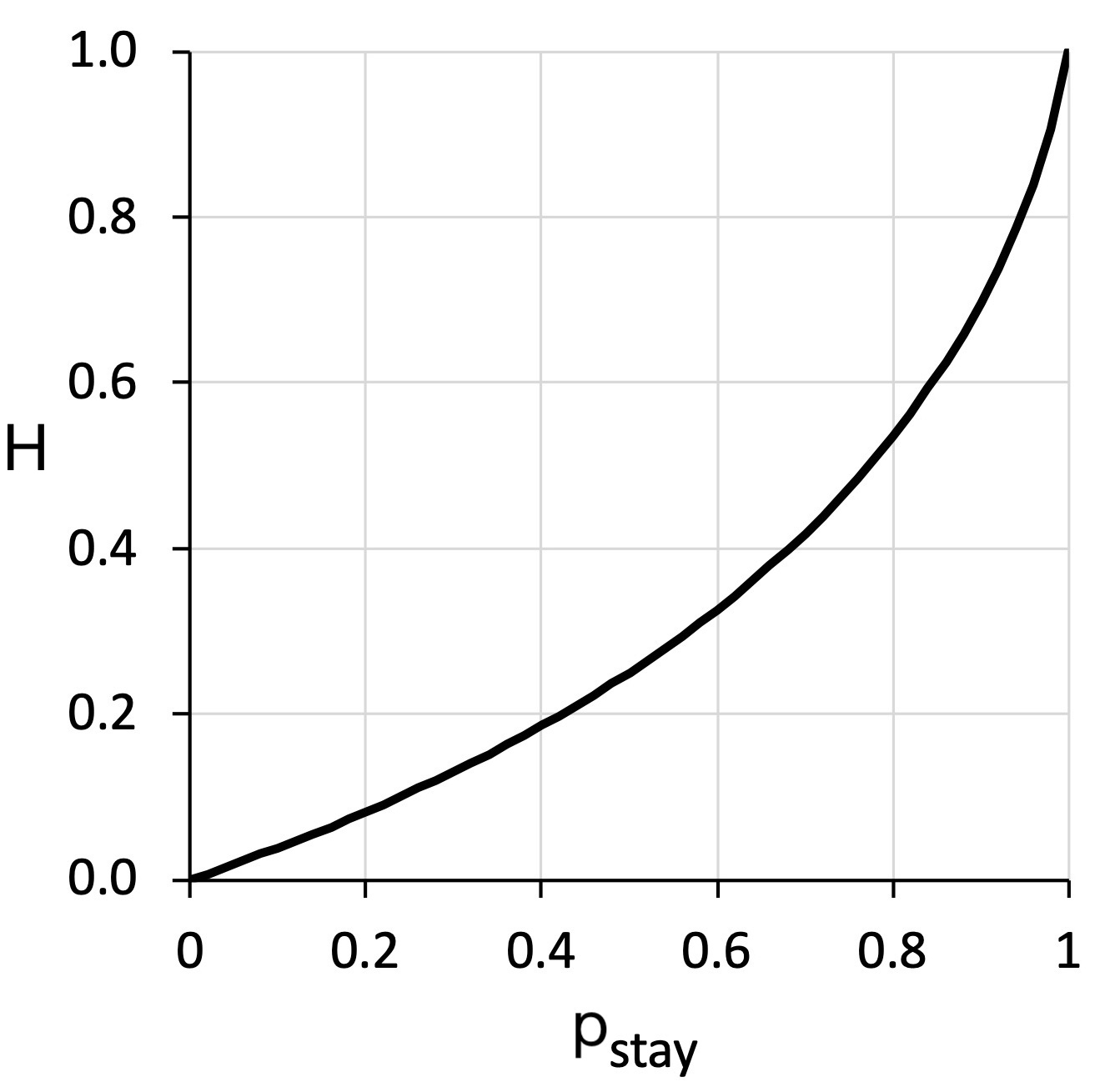} 
        \caption*{Figure 22: H as an algebraic function of the probability that a young adult will stay at home.}
    \end{wrapfigure}
    
    The city-size distribution is ancient and ubiquitous (Figure 4) and connects directly (algebraically) to H.  What makes the city-size distribution primordial is the generative model that produces it\footnote{The stay-vs-leave theory is attractive because of its simplicity, universal applicability, and agreement with observed $\alpha > 2$ (not $\alpha \equiv 2$).  However it does not preclude additional post-industrialization economic forces, though by that time most world cities had long been seeded (Spencer 2024).}.  Each generation of young adults must make a decision to stay near their birthplace or leave and settle elsewhere\footnote{Among other things, avoiding lethal inbreeding requires at least one mating partner to move occasionally, which was recognized in prehistory.  But a more direct and likely cause would be the possibility of inheritance of land or occupation -- facile for a child in a small family, difficult for a child in a large family.}.  This leads directly to a power law distribution of settlement size with exponent $\alpha = 1 + 1/p_{stay}$, where $p_{stay}$ is the probability of staying and $1 - p_{stay}$ of leaving; additionally in a finite landscape this model gives saturated power laws as observed (Bagrow et al 2008, Spencer 2024).  Thus H can be algebraically expressed as a function of $p_{stay}$.  As shown in Figure 22, $H \rightarrow 0$ only if all young adults choose to leave home -- which makes perfect sense since only then will the population become widely dispersed.  Conversely, $H \rightarrow 1$ if everyone stays where they were born: no one moves and exponential growth (however slow) leads to villages outpacing farmsteads with concomitant concentration.

	Barring environmental changes, famine, or warfare, as long as net fertility remains near replacement levels (i.e. two children per couple surviving to adulthood), places will not get crowded and $p_{stay} \sim 1$ is the known, safe choice.  Kaplanis et al (2018), using genealogic datasets extending to the medieval era, show that the median lifetime migration distance in Europe did not exceed 10 km until after 1900 (Figure S18).
	
	There is some irony in this suggestion that if there is a primordial value for H, it is near one and certainly not near zero.  In practice some young adults always do move away, and if 10-20\% do so we get $H \sim 0.6$ as observed across time and geography.

\pagebreak

\subsection*{Does H Have Any Validity?}

    The Hoover index has serious problems.  When its intrinsic scale independence is ignored it can be misconstrued to reflect urbanization.  It is prone to a variety of sampling artifacts -- low-value truncation of datasets, inappropriate empty-block inclusion,  aggregation artifacts due to changes in scale or population growth -- that can send its value high or low by significant amounts.  Most of its trends are mirrored by spatially uncorrelated random number datasets.  Is there anything than can be recovered?  
    
    At a minimum, H must be measured where its principal artifacts are minimal or at least understood and documented, and it must be re-branded not to reflect ``concentration" (whatever exactly that is) but instead heterogeneity, which is what it really measures. The following new patterns for H stand up to experimental resampling and have plausible causation narratives -- though the power of sampling artifacts to move H is so strong that caution is always merited. 

    \begin{enumerate}
        \setlength{\leftskip}{-1.2em}
        \item \underline{Broad areas of lower H}: Decadal census data lend themselves to large tables and line charts  of H over time, as found in most of the literature. But \emph{spatial} trends are easily studied when fine-grained datasets are experimentally aggregated and mapped.  Since $H \sim 0.6$ is a general rule backed by a generative model, settlement-size distribution, and international observation, exceptions are worth consideration.  Figure 23 shows H by geoblock sampling for modern France, Brazil, and the US. The overall green cast is ``baseline" H but all maps have broad yellow/orange regions of $H \sim 0.2-0.5$ which persist over resampling at multiple scales, so they are not MAUP artifacts.  They are certainly not empty-block artifacts because (i) those always lead to higher H, and (ii) in these modern datasets all blocks, except for a few missing squares in the Amazon basis, have tallied populations. 
        
            \begin{figure}[!h]
                \centering
                \captionsetup{width=13cm}
                \includegraphics[width=15cm]{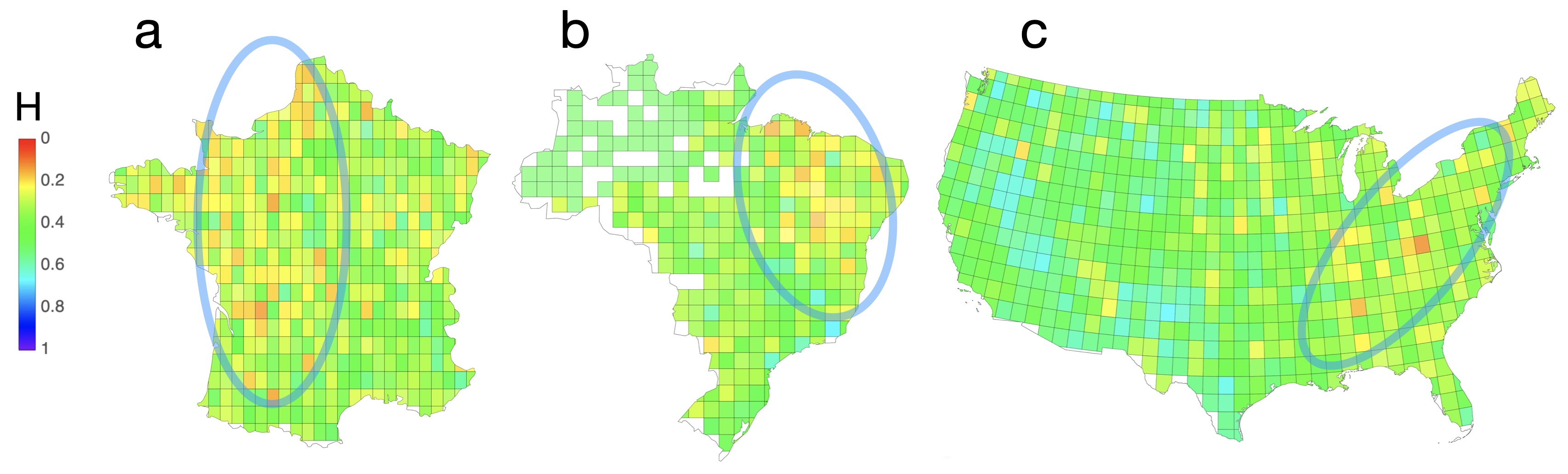} 
                \caption*{Figure 23: H, scale to left, computed on geoblocks of adjustable size and ignoring empty blocks. (a) France, 4x4 sets of 0.1° inner blocks, (b) Brazil, 5x5 sets of 0.3° blocks, US, 4x4 sets of 0.3° blocks.  Blue ovals suggest broad areas of lower H.}
            \end{figure}
            
            Notably these low-H regions are neither highly urban (not near Paris, Marseilles, Lyon; S\~{a}o Paulo, Rio de Janeiro; or New York, Chicago, Los Angeles, Houston), nor are they very low density (upper Amazon basin, arid American west).  What low H tells us is that these blocks are internally self-similar in density.  Constant density does not imply homogeneity; these are mixed-use agricultural regions and will have a mix of similar-sized towns, villages, and farms, with neither large cities nor empty spaces.  Any lingering belief that H is a useful indicator of urbanization or  should be dispelled by these maps.
      
        \item \underline{The mid-scale maximum of H}:  A more subtle pattern in H is seen for the populated-block data of Figures 13, 14, and 15: as aggregation scale increases, H for the US rises from $H_{limit} \approx 0.54$ at its point-source limit to a maximum $H_{max} \approx 0.64$ for 0.2° blocks, forming a broad `hump' until it falls again as the MAUP overrides at larger aggregations.  This hump is notably only in the US data, not French or Brazilian, and not in any of the randomized models of Figure 15.  In Figure 18's simulations the hump is strong for coordinated Lévy jumps starting from a single seed (pioneer), completely absent for isotropic (random, uncorrelated) jumps (Figure 18a), and decreases if the simulation begins with multiple uncorrelated seed locations (Figure 18c).  The absence of a hump for uncorrelated densities is definitive: H is detecting some form of spatial correlation. Two mutually exclusive hypotheses can be considered:
            \begin{enumerate}
            \setlength{\leftskip}{-1.2em}
                \item \underline{Exurban sprawl}: The US hump represents exurban sprawl, consistent with its appearance at aggregation scales of 0.1°-0.5° ($\approx$ 8-40 km) and DeSalvo \& Su's (2019) mean urbanized area scale (Table 3: 85 square miles $\implies$ a square 15x15 km).  Because H is scale-independent, the hump would not occur just with very dense (but size-limited) cities -- significant large-scale sprawl is necessary, on the scale of the Boston-Washington corridor, perhaps enabled by an automotive society.
                \item \underline{Isolated cities}: The US hump is caused by dense cities that \emph{lack} extensive suburbs, transition quickly to low-density surroundings, and are far from other cities. Increasing aggregation scale will pull in otherwise uncounted empty areas, increasing H tantamount to the empty-block effect.  
            \end{enumerate}
            As detailed in the Supplementary Material, difference mapping identifies principal contributors to the hump effect: Las Vegas, Denver, and Phoenix but not Boston, New York City, Chicago, or San Francisco.  While the exurban sprawl narrative sounds compelling it is incorrect and the hump is caused by widely isolated cities which elicit a rise in H via one of its aggregation artifacts.  This is a reminder that there are so many factors affecting the value of H that interpretations must be approached with caution, experimental visualization, and modeling.

        \item \underline{The aggregation scale where the MAUP begins}:  The populated-block data of Figures 13, 14, 15, 18, 20a all show that H follows a fairly stable plateau from small aggregation block sizes and then drops smoothly toward zero as the block size approaches the scale of the outer region (typically a national boundary).  What is notable in the real data of Figure 14 is that the block size where H begins to drop varies from $\sim0.05^{\circ}$ for France, to $\sim1^{\circ}$  for the US, and $\sim2^{\circ}$  for Brazil.
         
         National geographies suggest an interpretation. Much of France, aside from Paris, is self-similar at many scales: rolling farmland and mid-sized towns without many disruptions.  Brazil is the opposite: most of its population clusters along one boundary (the Atlantic coastline) so its heterogeneity will not average out except at the largest scales.  The US is in between: it is heterogeneous but its cities are well distributed, so its blocks will become self-similar when each encompasses one large city, its suburbs, and surrounding towns and land.  It is not the \emph{value} of H that makes the distinction -- these countries have the same small-scale limit of $0.54 \le H \le 0.61$ -- but rather the \emph{scale of aggregation} at which H becomes beset with the MAUP.  Yet as attractive as this interpretation is, it may also be an illusion, especially for more isotropic datasets like France: see the discussion about $S_{maup}$ in the Supplementary Material.

        \item \underline{Increases in H over time}:  None of the artifacts on H would lead to an increase in county-scale or smaller aggregated H values over time\footnote{A famine sufficient to reverse the MAUP effect of population growth would have to be of plague scale.}: minimum-value truncation and inappropriate inclusion of empty blocks can artifactually increase H, but both act at the beginnings of time studies since the artifacts themselves vanish as a population grows.  I exclude all studies of H aggregated above the county level, since Figure 13 shows that such H is on the slippery slope of the MAUP, an artifact of such magnitude that it is possible that subtle changes in sampling could result in small up- or down-shifts in H\footnote{H is often quoted to four significant figures and inferences drawn from $<1\%$ changes (Rogerson \& Plane 2013)}.
        
            What remains are studies at county- or finer-grained analysis that show increases in H over time. Otterstrom (2001) has several examples (Figures 1, 6) sketched in Figure 24a.  Whether or not to include empty sample blocks is a legitimate choice, and Otterstrom's choice to include them is generally valid\footnote{Except perhaps for the Mountain region, known to have significant areas uninhabitable by aridity or legal prohibition.} because we know in hindsight that most of the areas were subsequently inhabited.  However, by including empty blocks, H is subject to two separate forces: empty-blocks' upward effect on H abates as those blocks become occupied, driving H down, and the isolated city effect drives H up. 
            
            Both effects on H are valid, but perhaps the interpretation is clearer if the effects are separated by ignoring empty blocks (Figure 24a, solid line).  It is not the timing of the minimum of Figure 24a that matters, but rather the timing of the early down-slope (the filling-in of empty spaces, arguably the closing of the frontier) and the timing of the later up-slope (probably the growth of isolated cities).
        
            \begin{figure}[!h]
                \centering
                \captionsetup{width=14cm}
                \includegraphics[width=10cm]{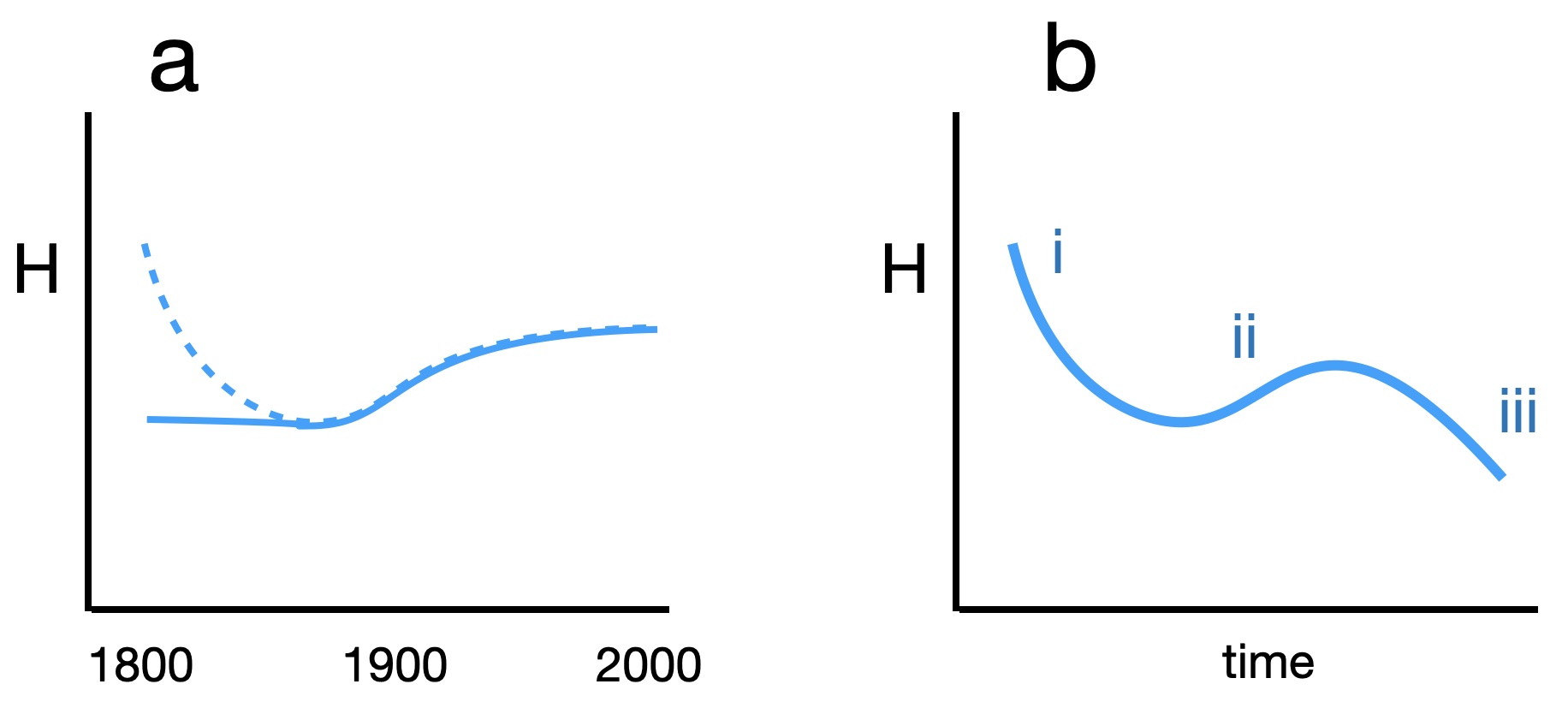} 
                \caption*{Figure 24: Independent effects pull H down and up over time.  (a) Dashed line, as in Otterstrom 2001, Figures 1 and 6, with empty blocks included; solid line, tallying populated blocks only. (b) Potential effects on H: (i) H decreases as empty blocks become populated, a frontier scenario. (ii) H rises as isolated cities exaggerate the distinction vs rural areas, and (iii) above county-level aggregation, the MAUP decreases H as the population continues to grow. }
            \end{figure}
            
            If H had been sampled at a higher level of aggregation (states instead of counties) we would have the complex story of Figure 24b, with an initial drop in H as empty blocks filled in, followed by a rise, followed again by a steep drop as population growth caused the temporal MAUP.  The timing of the minima and maxima have no particular meaning but reflect when one factor or another dominates H.
                  
    \end{enumerate}
    
\subsection*{How \textit{Should} We Measure Urbanization?} 

    The Hoover index, like the phrase \textit{``80\% of the people live on 20\% of the land"} reflects distribution but not urbanization, and applies equally to a city with adjacent suburbs or to a rural town with adjacent hamlets and farms.  Is there way to measure urbanization that reflects the essential difference between urban and rural, is immune to sampling artifacts, and has sufficient usage and history to be readily understood?
    
    Other indices have been proposed (e.g. Fotheringham \& Wong 1991, Cowell 2000, Cohen 2021), often justified with elegant mathematics.  But toy examples like Figures 9 and 11 show that Lorenz- (Gini, Hoover) and entropy-based (Theil, Atkinson) indices all suffer from the zero-block inclusion effect and the MAUP.  No mathematics can evade bias in the input data.
    
    Arriaga (1970) notes that an index based on percents of population gives odd results, for example that Hawaii is as urban as England.  The toy examples show that such oddities are inevitable when the underlying city-size distribution is scale-free.  The absolute size of cities must be taken into account, and Arriaga proposes a new index of urbanization that equals the size of the settlement in which a randomly chosen person lives.  This has the desirable properties of an index (Lemelin et al 2016), namely that it increases with increased population density, conforms to the transfer principle, and is consistent in aggregation.  When sampled in equal-area blocks, Arriaga's index is proportionate to population density.

    There are advantages to keeping things simple and the obvious metric is simply population density.  It can be used in absolute units (people per unit area) or cast as fractional population above or below a threshold in absolute units (which is necessary to break the scale-free implications of settlement growth).  One may quibble that population density doesn't measure exactly what is wanted in a particular study, but at least it's clear, readily calculated, and directly interpretable.
  
\subsection*{Recommendations}

    This criticism of H is ultimately intended to be constructive, and so I offer recommendations for future research.
    
    \begin{itemize}
    \setlength{\leftskip}{-1.2em}

    \item \underline{Take an experimental approach}: I strongly endorse an experimental approach beginning with the finest-grained data available -- postal regions or geolocated settlements at the smallest possible sizes.  Compute H with and without empty sample blocks.  Compute by sampling at many scales, from point-source (as if all datapoints have zero area) to within a factor of 2 or 3 of the scale of the dataset (typically a nation).   If the data have an obvious distribution, try extending the data along that distribution to the ultimate sampling limit (the individual) to gauge any minimum-value truncation problems.  Bootstrap to estimate a standard deviation for H\footnote{A thousand bootstrap cycles are used throughout this work, though sampling statistics generally converge after only ten.}.  Recompute all spatial datasets (i.e. not point-source) after random shuffling of populations and locations; such randomized data are still prone to the MAUP (Figures 15, 18a isotropic) and the comparison will reveal the extent of the MAUP with the real data as well as the existence of any real spatial correlation.
    
    \item \underline{Find the benchmark values for H}:  Figure 25 sketches the four notable values the H can have, depending on analyst choices about empty blocks and aggregation scale.  All results (Figures 13, 14, 15, 18, 20a) demonstrate that H has two limiting values at small sample scale:  if empty blocks are included, $H \rightarrow 1$, and if excluded, H approaches a stable value $H_{limit}$, the point-source limit, typically 0.5 - 0.6.  The empty-block $H \rightarrow 1$ is well understood and uninteresting but $H_{limit}$ is a quotable value of H for a given dataset.  It is readily computed from a simple array of settlement populations which are thereby treated as zero-area point sources.  $H_{limit}$ is immune from the MAUP since no aggregation is involved.  Assuming a pareto or saturated pareto distribution of settlement sizes (the most pervasive distribution in demography), $H_{limit}$ is simply an algebraic function of the pareto exponent $\alpha$, and like $\alpha$ can be expected to have a narrow range across all places and eras (like Brazil, France, and the US from the colonial era to the present, Figure 8b).  A value of $H_{limit}$ outside its typical range conveys exceptional circumstances that merit close examination of the raw data.
        
        \vspace{-6pt}
        \begin{figure}[!h]
            \centering
            \captionsetup{width=6cm}
            \includegraphics[width=4cm]{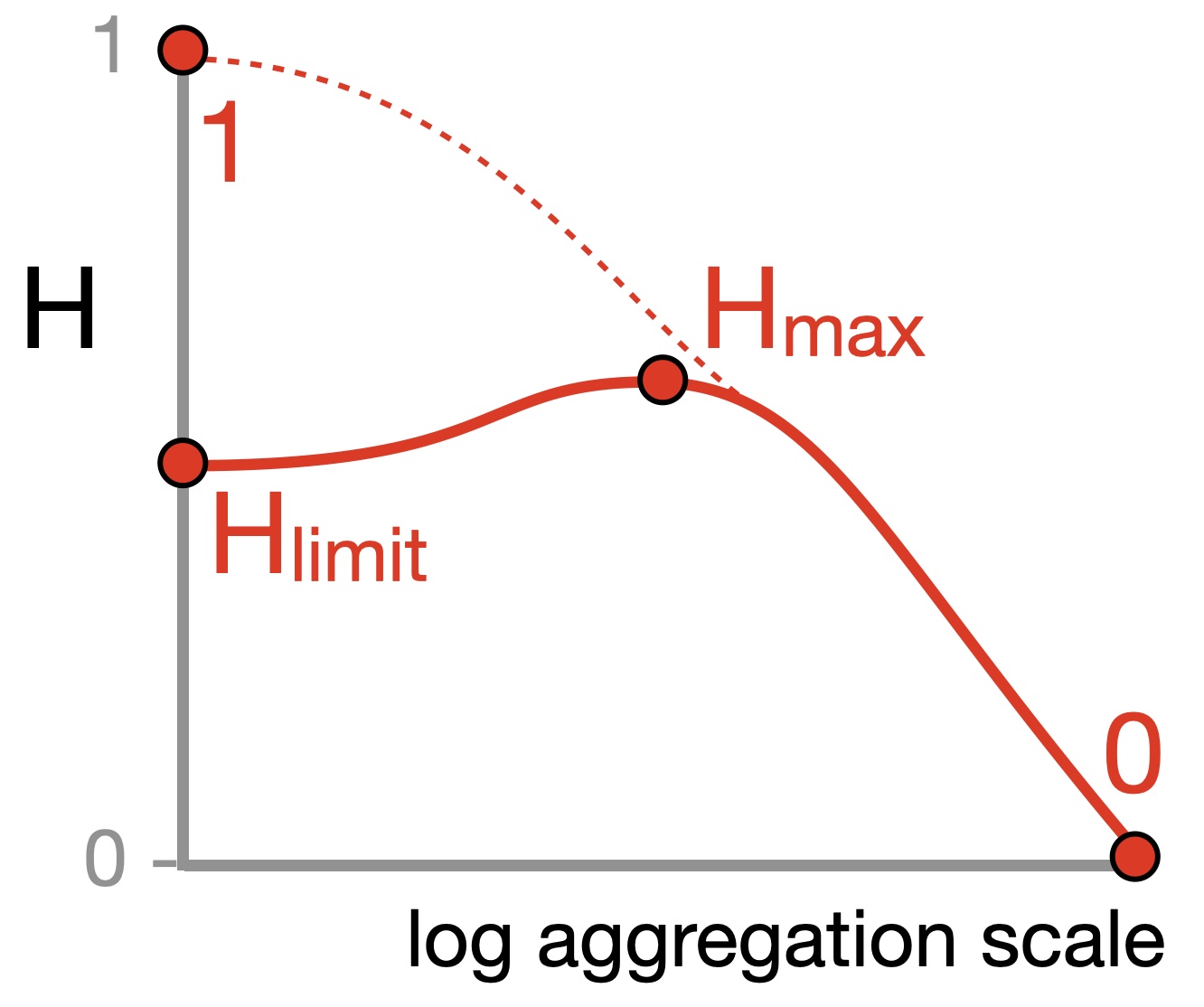} 
            \caption*{Figure 25: Benchmark values of H}
        \end{figure}
        \vspace{-6pt}

        But $H_{limit}$ is not the whole story, particularly because by construction it cannot encode spatial correlation.  $H_{max}$, the highest value of H as a function of aggregation scale, is arguably the best choice if one must choose a single value to represent a dataset.  $H_{max}$ encodes spatial correlation, specifically of isolated cities, and it occurs at the smallest aggregation scale that renders the empty-block question moot (see Supplementary Material).  Importantly $H_{max}$ also occurs just as the MAUP begins, thereby avoiding the most problematic artifact.  If a dataset does not have significant isolated cities, H plateaus from  $H_{limit}$ to the MAUP threshold, thus $H_{max} = H_{limit}$.
        
    \item \underline{Avoid the MAUP}: Examine H at all aggregation scales (e.g. Figure 13) and avoid further analysis at any scale down the steep slope of the MAUP; for the US this means aggregate only at $\sim0.5-1^{\circ}$ (counties) or smaller, or for France, $\sim0.03^{\circ} \sim3$ km or smaller.  Above this scale any variations in H will be dominated by aggregation artifacts. 
    
    \item \underline{Visualize the raw data}: Cast the data into its original units and examine as log-log rank-frequency (Figure 6a).  Compare to expected distributions (pareto or saturated pareto) to be aware of possible truncation bias.  Use log scales for population and scale axes to see the full range of the data.
    
    \item \underline{Visualize in maps}: One of the best ways to avoid Mark Twain's imprecation about statistics is simply to visualize the data and results in choropleth maps.  Technical problems are easily recognized (i.e. if sample blocks include large bodies of water) as are interpretive problems (anyone who knows where American cities are located cannot conflate H with urbanization after seeing Figure 2b).  Mapping may also suggest causation -- for example high-H blocks in the US coincide with high aridity rather than with arable (but low population density) prairies.  Difference maps, which show where a localized change in the data (typically randomization) for one block may affect the overall H, can readily distinguish hypotheses.
    
    \item \underline{Consider empty sample blocks}: Assess whether empty areas should or should not be included based on external knowledge.  They probably should for historic data if future data show areas to be habitable and should not if areas are known to be uninhabitable (water, severe desert, legally excluded).  However if empty areas are included, H becomes a steep function of aggregation scale across its entire range (Figures 13, 14, 15, 21), approaching one at small aggregations\footnote{Consider the contiguous United States, area $7.7 \cdot 10^{12}\, m^2$ and population $3.4\cdot10^8$, giving density $\sim 1/20,000\, m^{-2}$.  Allocating each person 1 square meter gives 20,000 empty blocks for each occupied block, so $H \sim 1 - 1/(20,000) > 0.9999$.} which makes H independent of the data and useless.  The choice of aggregation scale is therefore critical if empty areas are included, and must be stated and justified.  A better option would be to include specific empty areas for specific reasons, otherwise tally only populated blocks at the smallest feasible aggregation scale.

    \item \underline{Separate the factors that affect H}: Figure 26 sketches the most significant factors that affect the value of H, and shows that different circumstances and analyst choices can give a wide range of values and trends in H.  If H is to have any value, these factors must be separated, documented, and artifacts identified as such.
    
        \begin{figure}[!h]
            \centering
            \captionsetup{width=13cm}
            \includegraphics[width=13cm]{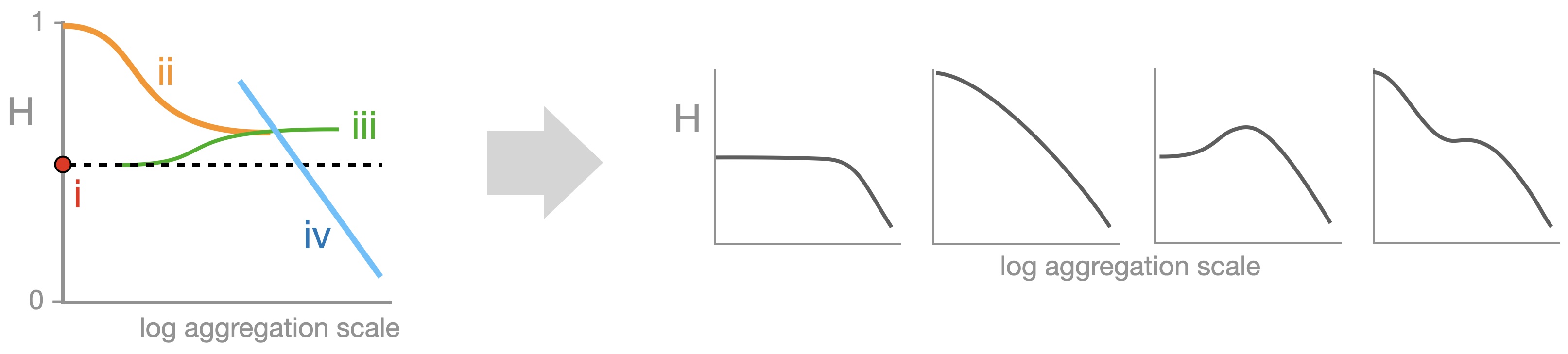} 
            \caption*{Figure 26: Separate factors affecting H can, in different combinations, give a wide array of apparent trends and values.  Factors are (left) (i) Red dot and dashed line: the point-source value = $H_{lim \, x\rightarrow 0}$, (ii) empty block inclusion, (iii) isolated city increase, (iv) sampling scale (MAUP) artifact.  For a growing population, each of the H vs scale patterns at the right gives rise similar patterns of H over time, all highly dependent on the analyst's choice of sampling scale.}
        \end{figure}

    \item \underline{Examine other datasets to distinguish common from unique causes}: Analyses for Brazil, France, and the US make it clear that the decrease of H with sample scale is an artifact (MAUP) and does not have a unique national narrative.  On the other hand, robust differences can lead to interesting historic interpretations.
    
    \item  \underline{Tell the story in the data, not the story you anticipated}:  Confusion of ``concentration" with density has led to much discussion of $H \rightarrow 0$ in terms of a frontier or idyllic, pre-urban state.  But here the data show, for example, that regions of low H correspond to large self-similar mixed areas of towns and farms, neither very low density nor urban/suburban (Figure 23).  That is a less romantic but completely sensible interpretation of low H.
    \end{itemize}
    
\subsection*{Conclusions}

    Rogerson (2019) has the appealing title ``I dream of Gini".  I considered calling this paper ``The Hoover, Damned", but the results are too serious for flippancy.  There is very little salvageable about the Hoover index; it is beset by artifacts not only of data aggregation scale (the MAUP), but also of empty-block inclusion, data truncation, population growth, dataset size and national area (see Supplementary Material for scale-axis measurable). Ingenious sampling can make things worse (Supplementary Material on Tessellation).  Other indices (Gini, Theil, Atkinson) fare no better.  H is frequently misinterpreted as measuring urbanization instead of heterogeneity.  Low values of H are sometimes presumed to indicate a primordial or ideal state, when in fact its natural value is 0.4-0.7 because it is a close cousin of the universal settlement-size pareto distribution which may truly have prehistoric roots (Spencer 2024).
    
    Any future studies that rely on H should, at the very least, show results of aggregation at multiple scales and empty areas should be treated sensibly and overtly.  Better yet, leave H behind and base conclusions on something clear and direct like population density.

\subsection*{Sources, Methods, and Data Availability}

    All population and map data are from readily accessed national sources: Brazil municipal areas 2020 \href{https://www.ibge.gov.br/en/geosciences/territorial-organization/territorial-meshes/18890-municipal-mesh.html?edicao=24069&t=downloads}{\xlink}, France communes 2014  \href{https://www.data.gouv.fr/fr/datasets/decoupage-administratif-communal-francais-issu-d-openstreetmap/}{\xlink},  United States zip code populations 2020 \href{https://data.census.gov/table/DECENNIALDHC2020.P1}{\xlink}, zip code geography \href{https://www.census.gov/cgi-bin/geo/shapefiles/index.php?year=2016&layergroup=ZIP+Code+Tabulation+Areas}{\xlink}, and US climate data \href{https://www.ncei.noaa.gov/access/monitoring/climate-at-a-glance/county/mapping/110/tavg/202005/12/value}{\xlink}.
    
    Population aggregation is carried out on degree-based blocks for computational efficiency, with density then corrected for the dependence of east-west distance on latitude.  Spatial correlation effects on H can be visualized with difference maps, in which H for the entire region (nation) is recomputed with sample blocks randomized one at a time, and plotting the resulting difference in H.  This is equivalent to recomputing Figure 2b several hundred times.
    
    The simulation of Figure 17 is an extension of the checkerboard city-growth model of Spencer (2024).  Like that model, at each time-step a person is added to a finite two dimensional space.  With probability $1 - p_{stay}$, that person migrates to another location, chosen in a random direction and at a distance taken from a 1/r distribution (like all human migrations, Spencer \& Otterstrom 2024).  The jump distance is characterized by a single adjustable parameter, the median distance, thus $d_{jump} = d_{median}\cdot(\frac{1}{random} - 1)$ where \textit{random} is a uniform random number on [0...1].  Simulations begin with one or more `pioneer settlers' seeded at random locations.  Spatial correlation is generated by analogy to human migration: each new person is assigned a `parent' chosen at random from the current population, and the new person's migratory jump begins at the parent's location.
    
    The software for all computations and simulations is open-source and available, with working demonstrations, at \url{https://scaledinnovation.com/gg/reports/hoover/hooverAnalysis.html}.  All software and datasets for city-size distribution analysis is similarly available at \url{http://scaledinnovation.com/gg/cityGrowth.html}.  All software is original by the author without frameworks or libraries.

\subsection*{Acknowledgement}
    The author is very grateful for the encouragement, advice, and critical reading of this research and manuscript by Samuel M. Otterstrom of the Department of Geography, Brigham Young University, Provo, UT 84602.
    

\subsection*{References}
\setlength{\parskip}{0.5\baselineskip}
\setstretch{0.95}

    Allen, E. (2022). Relation Between Two Income Inequality Measures: The Gini coefficient and the Robin Hood Index. WSEAS Transactions on Business and Economics, 19, 760-770.
    
    Andresen, M. A. (2021). Modifiable areal unit problem. The Encyclopedia of Research Methods in Criminology and Criminal Justice, 2, 854-855.

    Arriaga, E. E. (1970). A new approach to the measurements of urbanization. Economic Development and Cultural Change, 18(2), 206-218.

    Auerbach, F. (1913). Das gesetz der bevölkerungskonzentration [The law of population concentration] Petermanns Geographische Mitteilungen 59, 73-76. 
    
    Bagrow, J. , Sun, J. \& ben-Avraham, D. (2008). Phase transition in the rich-get-richer mechanism due to finite-size effects. Journal of Physics A: Mathematical and Theoretical, 41(18), p.185001.
     
    Buzzelli, M. (2019). Modifiable areal unit problem. International encyclopedia of human geography, 169.
    
    Clark, C. (1951). Urban population densities. Journal of the Royal Statistical Society. Series A (General), 114(4), 490-496.

    Clauset, A., Shalizi, C. \& M.E.J. Newman, M. (2009) Power-law distributions in empirical data, SIAM Review 51(4), 661-703 (2009)
    
    Cohen, J. E. (2021). Measuring the concentration of urban population in the negative exponential model using the Lorenz curve, Gini coefficient, Hoover dissimilarity index, and relative entropy. Demographic Research, 44, 1165-1184.
    
    Costa, R. N., \& Pérez-Duarte, S. (2019). Not all inequality measures were created equal: The measurement of wealth inequality, its decompositions, and an application to European household wealth (No. 31). ECB Statistics Paper.
   
    Cowell, F. A. (2000). Measurement of inequality. Handbook of income distribution, 1, 87-166.
    
    DeSalvo, J. S., \& Su, Q. (2019). The determinants of urban sprawl: Theory and estimation. International Journal of Urban Sciences, 23(1), 88-104.
    
    Duncan, O. D. (1957). The measurement of population distribution. Population Studies, 11(1), 27-45.
    
    Duncan, 0. D., Cuzzort, R. P. \& Duncan, B. (1961). Statistical Geography:Problems in Analysing Areal Data. Glencoe: Free Press.
    
    Elledge, J. (2024). A history of the world in 47 borders: the stories behind the lines on our maps. Hachette UK.
 
    Farrell, H., Gopnik, A., Shalizi, C., \& Evans, J. (2025). Large AI models are cultural and social technologies. Science, 387(6739), 1153-1156.

    Fotheringham, A. S., \& Wong, D. W. (1991). The modifiable areal unit problem in multivariate statistical analysis. Environment and planning A, 23(7), 1025-1044.
 
    Gehlke, C. E., \& Biehl, K. (1934). Certain effects of grouping upon the size of the correlation coefficient in census tract material. Journal of the American Statistical Association, 29(185A), 169-170.

    Hoover Jr, E. M. (1941). Interstate redistribution of population, 1850–1940. The Journal of Economic History, 1(2), 199-205.
   
    Hayek, F. A. (1945). The Use of Knowledge in Society. The American Economic Review, 35(4), 519-530.
    
    Hayek, F. A. (1975). The pretence of knowledge. The Swedish Journal of Economics, 77(4), 433-442.

    Kaplanis, J., Gordon, A., Shor, T., Weissbrod, O., Geiger, D., Wahl, M., ... \& Erlich, Y. (2018). Quantitative analysis of population-scale family trees with millions of relatives. Science, 360(6385), 171-175.   Mirror data site \url{https://osf.io/fd25c/} 
    
    Lemelin, A., Rubiera-Morollón, F., \& Gómez-Loscos, A. (2016). Measuring urban agglomeration: A refoundation of the mean city-population size index. Social Indicators Research, 125, 589-612.

    Newman, M.E., (2005). Power laws, Pareto distributions and Zipf's law. Contemporary physics, 46(5), pp.323-351.  

    Openshaw, S. (1984). The modifiable areal unit problem. Concepts and Techniques in Modern Geography 48. Geo Books, Norwich, UK.

    Otterstrom, S. M. (2001). Trends in national and regional population concentration in the United States from 1790 to 1990: from the frontier to the urban transformation. The Social Science Journal, 38(3), 393-407.

    Robinson, W. S. (1950). Ecological Correlations and the Behavior of Individuals. American Sociological Review, 15(3).

    Rogerson, P. A., \& Plane, D. A. (2013). The Hoover index of population concentration and the demographic components of change: an article in memory of Andy Isserman. International Regional Science Review, 36(1), 97-114.

    Rogerson, P. A. (2019). I dream of Gini: measures of population concentration and their application to US population distribution. Population, place, and spatial interaction: essays in honor of david plane (pp. 1-17). Singapore: Springer Singapore.

    Spencer, R. (2024). City size distributions are driven by each generation’s stay-vs-leave decision, https://arxiv.org/abs/2405.02129 [physics.soc-ph]   See also            
            \url{http://scaledinnovation.com/gg/cityGrowth.html} for datasets, sources, and simulations.
    
    Spencer, R. \& Otterstrom, S. (2024) Massive Genealogies Distinguish Frontier from Steady-State Internal Migration,	arXiv:2410.18235 [physics.soc-ph] See also \url{https://scaledinnovation.com/gg/migration/migration.html} for data sources and maps.

    Vining Jr, D. R., \& Strauss, A. (1977). A demonstration that the current deconcentration of population in the United States is a clean break with the past. Environment and planning A, 9(7), 751-758.

    Wong, D. W. (2004). The modifiable areal unit problem (MAUP). In WorldMinds: geographical perspectives on 100 problems: commemorating the 100th anniversary of the association of American geographers 1904–2004 (pp. 571-575). Dordrecht: Springer Netherlands.
    
    Ye, X., \& Rogerson, P. (2022). The impacts of the modifiable areal unit problem (MAUP) on omission error. Geographical Analysis, 54(1), 32-57.

\pagebreak
\hspace{0pt}
\vspace{6cm}
\begin{center}
\fontsize{2cm}{2.4cm}\selectfont Supplementary Material
\break
\break
\normalsize 
\end{center}
\setcounter{footnote}{0}
\setstretch{1.125} 
\pagebreak

\section*{The Rise of H with Aggregation Scale}

    In the literature H is generally derived from fixed aggregations like county or state census tables.  No one has previously examined H as a continuous function of aggregation scale, and so the maximum `hump' of H for the United States at intermediate aggregation scales (Figures 13, 14) is a new observation.  This hump of H is interesting for several reasons: it is new and unexplained, it is absent in the French and Brazilian datasets, absent in all randomization models (Figure 15) but strongly present in correlated jump simulations (Figure 18), and known artifacts (the MAUP) lead to a decrease in H with aggregation scale but never an increase.  Given the French and Brazilian results it would be tempting assert some distinction based on different national histories, but the many artifacts on H urge caution:  for example, it is plausible that France would show such a hump in H, but its local-scale self-similarity causes the MAUP artifact to appear at small scales and suppress the effect.  The absence of the hump in all randomized datasets and models strongly suggests that spatial correlation must be involved.   
    
    \begin{wrapfigure}[12]{r}{5cm}
        \vspace{-12pt}
        \centering
        \captionsetup{width=4cm}
        \includegraphics[width=4cm]{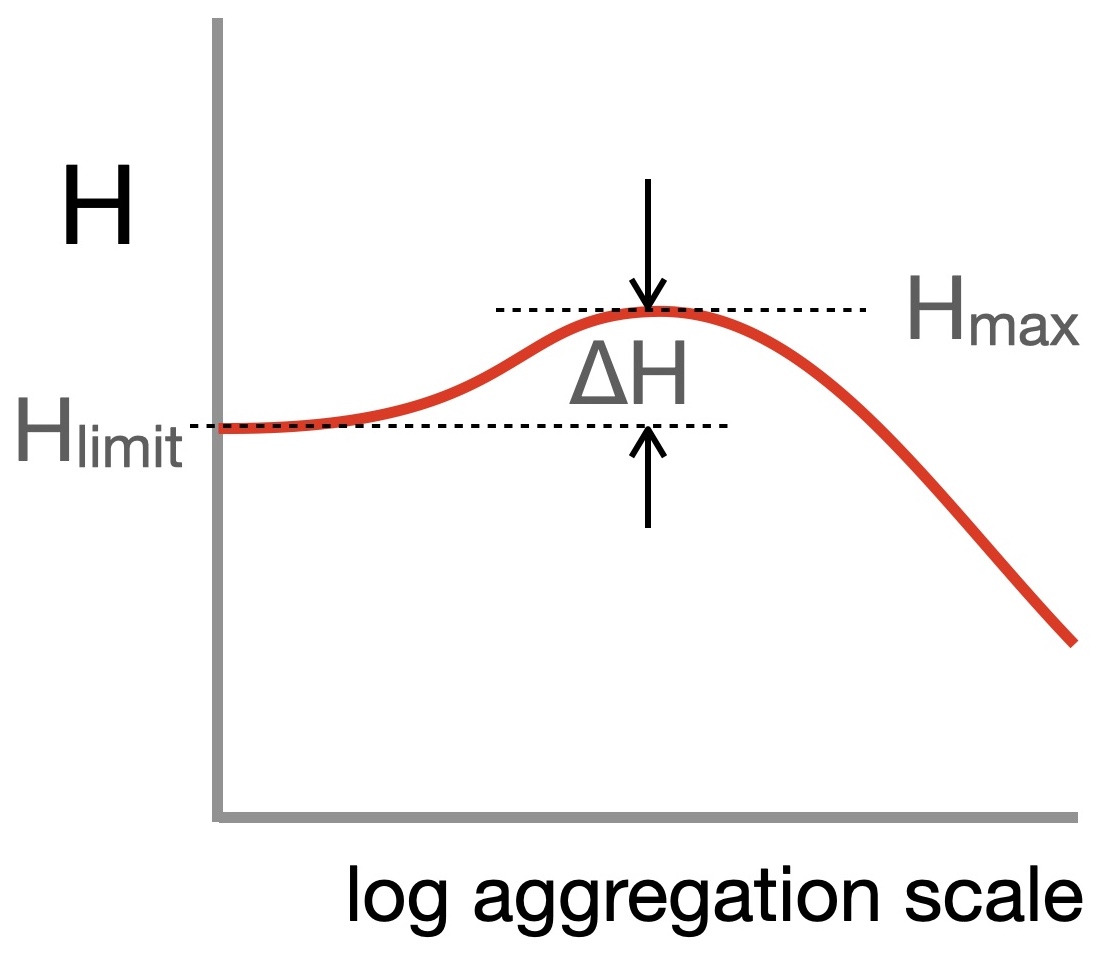} 
        \caption*{Figure S1: Definitions of $H_{limit}, H_{max}$ and $\Delta H$.}
    \end{wrapfigure}
    
    But taking the recommended advice here we do more analysis and visually map the results before suggesting interpretation. With plentiful fine-grained data the analyses applied at a national level can also be applied at a state level; Figure S1 provides useful definitions.
    
    The results in Figure S2a are startling: The aggregation profiles for Colorado and West Virginia have the same $H_{limit}$ as the US ($\sim0.55$), but then Colorado has a very strong rise to $H_{max}\sim$ 0.76 at a scale of 0.4°, while West Virginia has no hump at all and by a scale of $\sim$0.04° has begun its MAUP-driven decline.   Adding in more states shows that the picture is still more complex: in Figure S2b not only do various states display or lack a peak of H, but their small-aggregation $H_{limit}$ values range from 0.3 to nearly 0.7 which is the full gamut seen with international city-size datasets. 

    \begin{figure}[!h]
        \centering
        \captionsetup{width=12cm}
        \includegraphics[width=14cm]{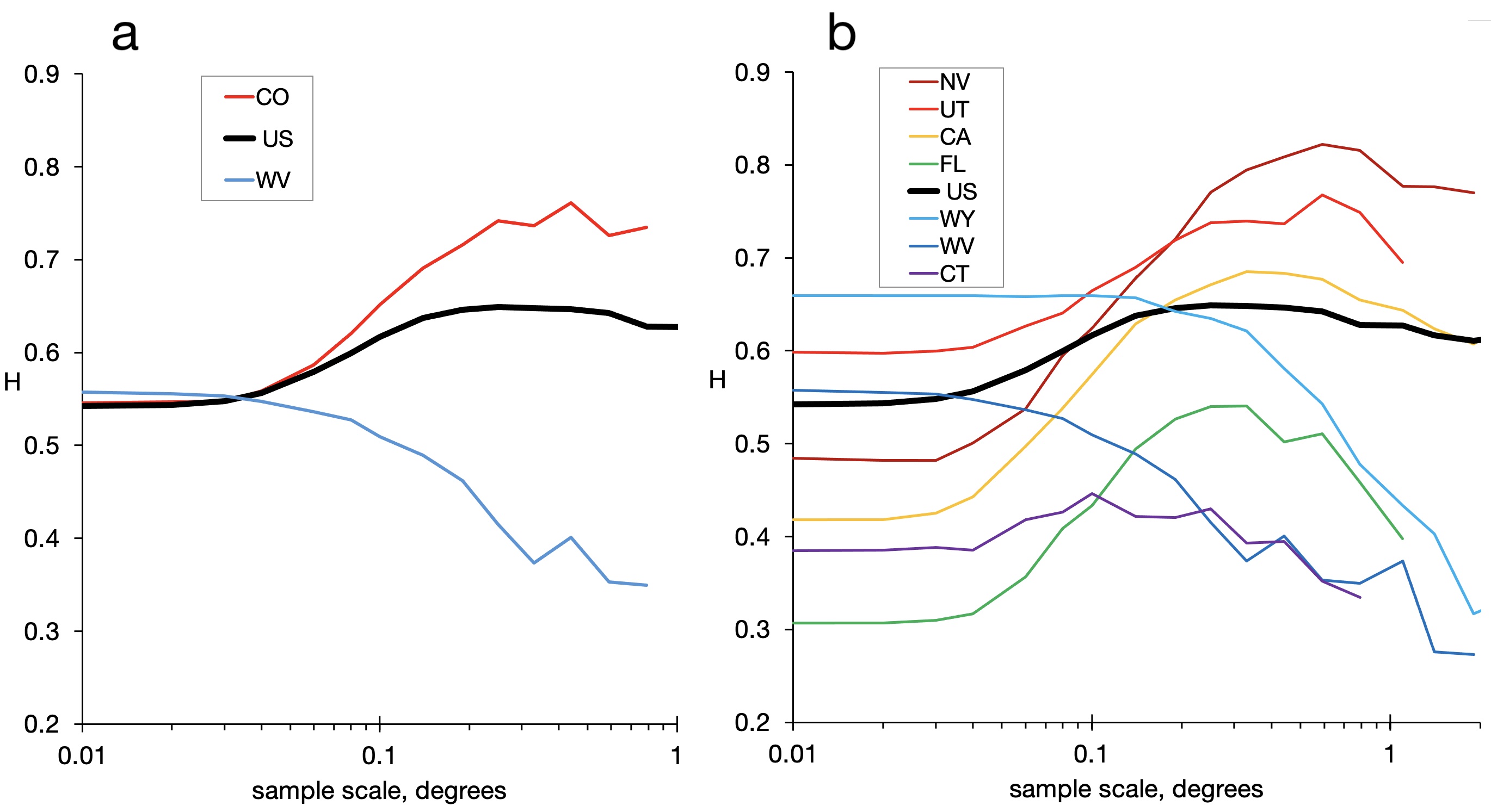}
        \caption*{Figure S2: H for states as a function of aggregation scale, as in Figure 13, counting only populated blocks. The thick black line recapitulates the analysis for the contiguous US. }
    \end{figure}
        
\pagebreak

    The 48 contiguous US states are plotted in Figure S3.  Their values of $H_{limit}$ and $H_{max}$ each span a range of 0.4, and since they are uncorrelated any interpretive model will need at least two degrees of freedom: no one factor can suffice.  
    
    It is instructive to look for geographic, demographic, historic, or perhaps economic patterns in the data.  Those with the greatest ``hump" in H , along the dashed blue line, seem to be large Western states (California, Arizona, Nevada, Colorado, Utah) -- but so does Florida, while Massachusetts and Texas (which could not be more different) have about the same $\Delta H$ as Utah.  Montana, large, sparse, and Western, has $\Delta H = 0$.  East coast states have the lowest $H_{limit}$, except for Virginia and Maine.
    
    \begin{figure}[!h]
        \centering
        \captionsetup{width=13cm}
        \includegraphics[width=12cm]{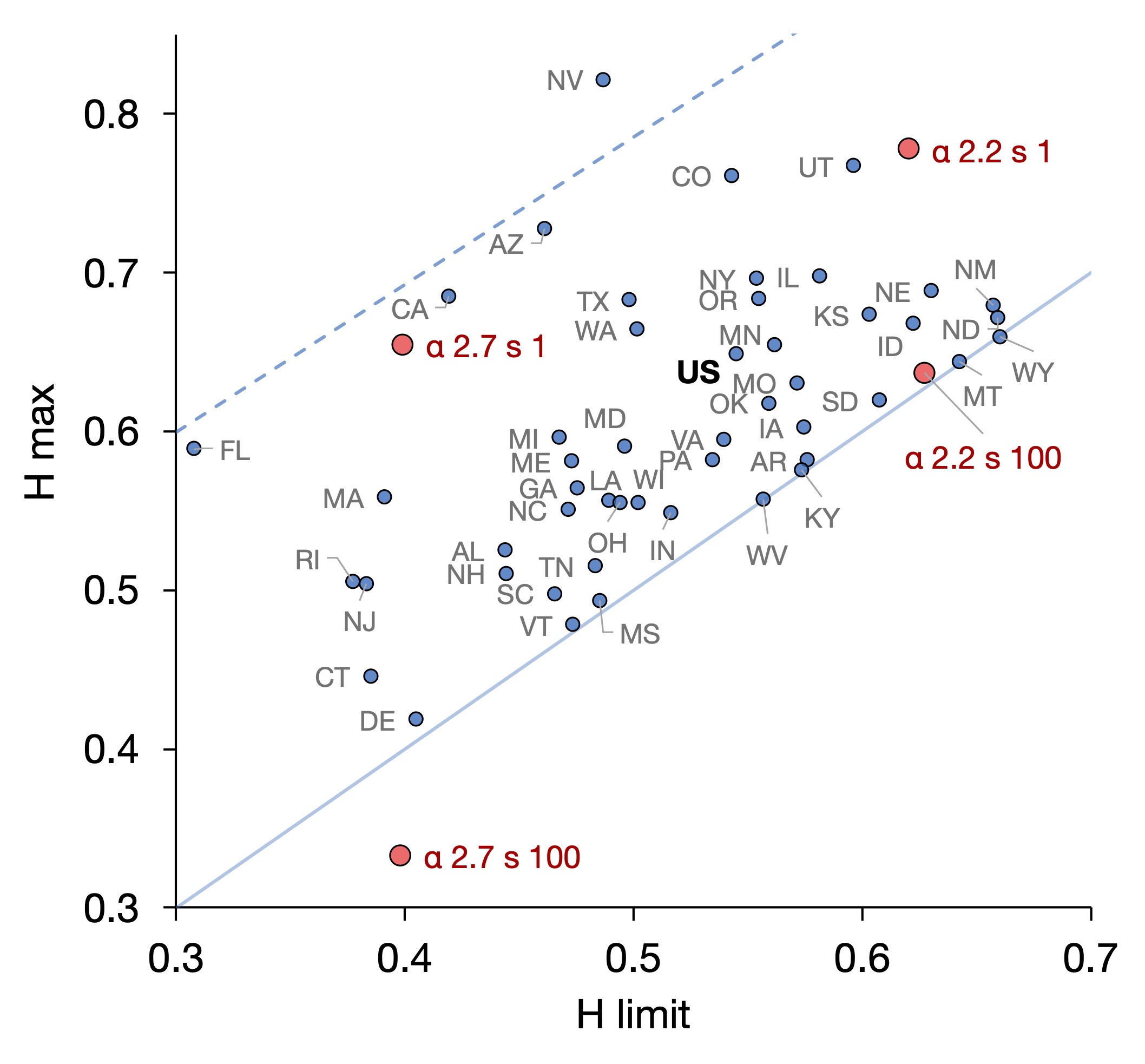} 
        \caption*{Figure S3: $H_{max}$ vs $H_{limit}$ for US states (blue dots), determined by analyzing zipcode data by state in 0.01° to 8° aggregation blocks, ten per log. ``US" shows the states treated as a single region.  Red dots show the results for simulations with the indicated values of $\alpha$ and s = number of seeds, each the mean of 20 independent simulations.  The solid blue line indicates $H_{max} = H_{limit}$ and the dashed blue line $H_{max} = H_{limit} + 0.3$. }
    \end{figure}
    
    Several states have $\Delta H \approx 0$ (blue line), but what do Delaware, Vermont, Mississippi, West Virginia, Kentucky, Arizona, Montana, and Wyoming have in common?  Sparse prairie and Western states (Nebraska, Kansas, North and South Dakota, Montana, Idaho, New Mexico) have high $H_{limit}$, but so does Illinois which includes Chicago.  For every hint of a pattern there is a contradiction; familiar descriptors -- dense, sparse, rural, urban, mountain, farmland -- do not seem to apply.

\pagebreak

    Fortunately the simulations of Figure 18, originally designed to demonstrate the MAUP, also span this wide range of $H_{limit}$ and $H_{max}$. A different way of thinking -- namely visual inspection of the simulations' output -- is suggestive, and for that purpose the simulations of Figure S4 are arranged in the same pattern as their red dots in Figure S3.   Two independent factors in the simulation, $\alpha$ and the number of seeds, suffice to reproduce the US states' range of $H_{max}$ and $H_{limit}$\footnote{The median Lévy jump length can be used instead of the number of seeds.  Increasing either will result in multiple settlements far from the first founder, thereby decreasing H.  Multiple seeds are placed completely at random (instead of just far away) and therefore any resulting `cities' will have no spatial correlation to each other.}.
    
    Two visual patterns are apparent: with only one seed (i.e. the simulation begins with a single founder), one large city predominates; with more seeds there are more cities.  Secondly, for $\alpha = 2.2$ the empty spaces are `emptier' -- they have very little background of tiny settlements, while for $\alpha = 2.7$ there are more small settlements in the sparse spaces.  Note that the empty-block problem does not apply here: H is calculated with final populations of 100,000 or more where there are no empty blocks, and above that threshold H is independent of population (Figure 7b).   The reader may reproduce these and other scenarios at \url{https://scaledinnovation.com/gg/reports/hoover/hooverAnalysis.html}.

    \begin{figure}[!h]
        \centering
        \captionsetup{width=14cm}
        \includegraphics[width=16cm]{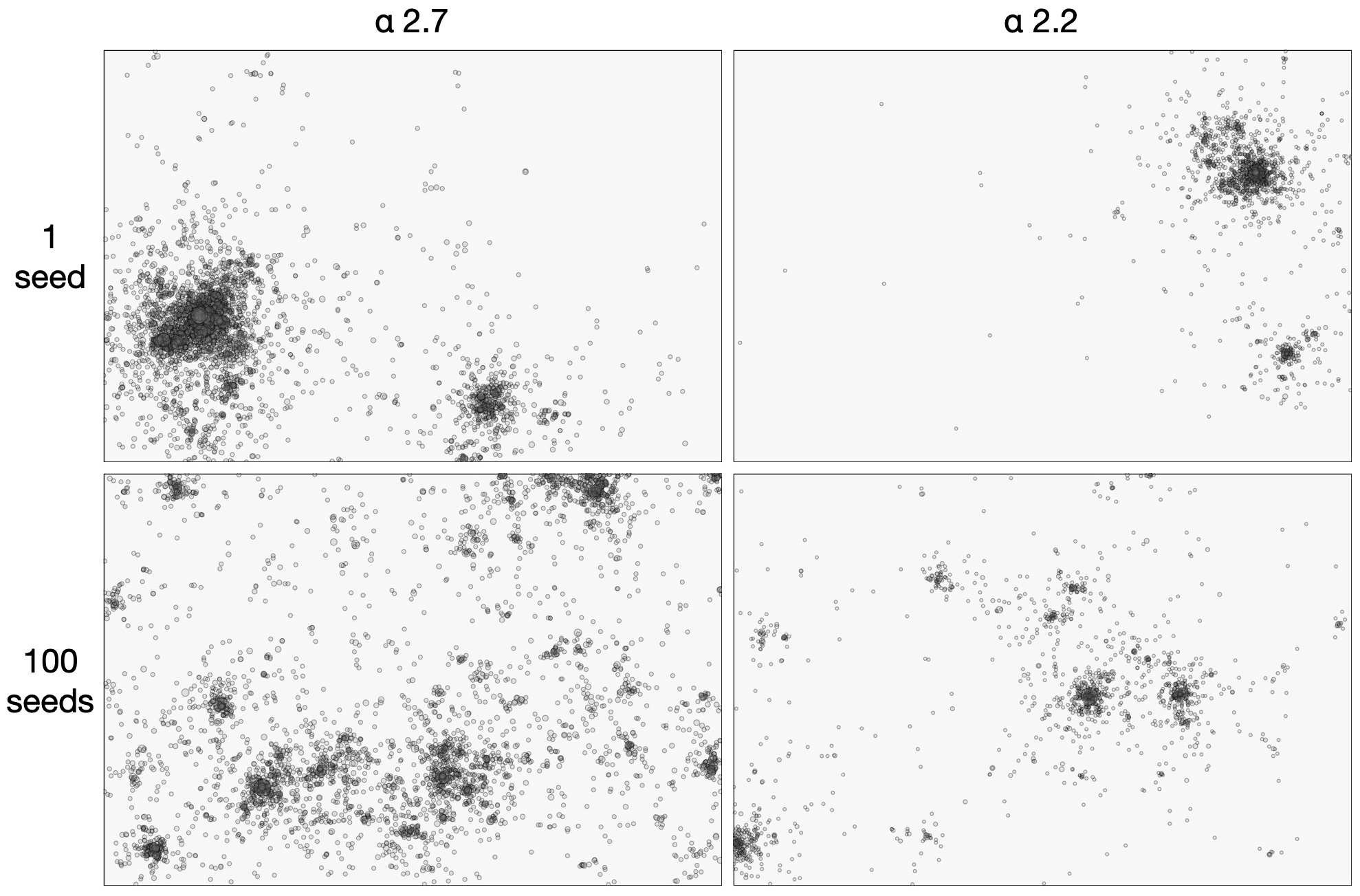}
        \caption*{Figure S4:  Simulations as Figure 17, all with 10,000 final population, boundaries 2000 x 3000 scale km, 50 scale km median jump, 300 scale km sample block size (at $H_{max}$, Figures 18a,c),  $\alpha$ and number of seeds (founders) as indicated.  Circle areas are proportionate to population.}
    \end{figure}
    
\pagebreak

    Figure S3 suggests that a few states -- Florida, California, Arizona, Nevada, Colorado, Utah -- contribute most to the overall H hump for the US.   Finer-grained mapping reveals what they have in common that other states lack:  Figure S5 uses the technique used to create Figures 2b and 2c where each 1°x1° block is subdivided into 0.2°x0.2° inner blocks and population density calculated for each inner block.  This suffices to calculate $H_{limit}$ and $H_{1 \,degree}$ (a surrogate for $H_{max}$), and thus $\Delta H$, which is mapped.

    \begin{figure}[!h]
        \centering
        \captionsetup{width=14cm}
        \includegraphics[width=13cm]{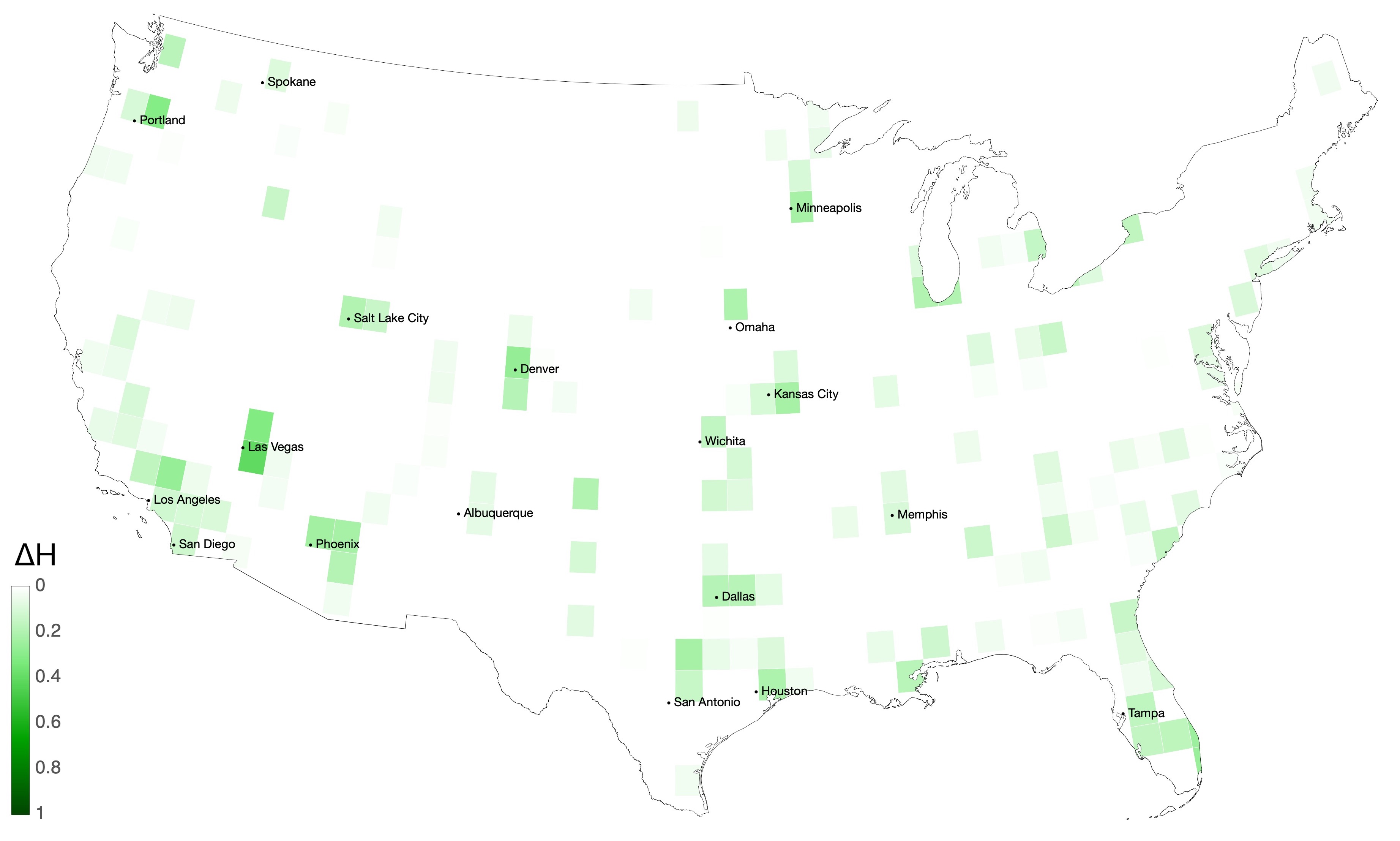}
        \caption*{Figure S5: Map of $\Delta H = H - H_{limit}$ for 1x1° blocks.  Each 1°x1° block is sampled as 25 0.2°x0.2° inner blocks, from which H (from the 25 internal densities) and $H_{limit}$ (from all internal zipcodes as point sources) are calculated. While $H_{max}$ for the US occurs at $\sim$0.5° aggregation (Figure 13), 1° aggregation is used here to be visible in illustration.  The results are the same when sampled at 0.5°. }
    \end{figure}
    
    This is an intriguing result.  The implicit question of Figure S3 -- what makes states similar or different? -- is now redirected to a specific set of cities.  Firstly note what cities do \emph{not} have $\Delta H > 0$: New York City, Boston, San Francisco, Baltimore, and dozens of others.  The cities in green blocks with $0.2 < \Delta H < 0.5$,  Las Vegas, Phoenix, Denver, Portland, Salt Lake City, Kansas City, are mid-sized.  They would not seem to be similar, or dissimilar, to many others not highlighted ($\Delta H < 0.05$).  Comparable maps of France and Brazil lack blocks with $\Delta H > 0.1$, consistent with their flat plateaus in Figure 14.
    
    \begin{wrapfigure}[14]{r}{9cm}
        \vspace{-8pt}
        \centering
        \captionsetup{width=7cm}
        \includegraphics[width=9cm]{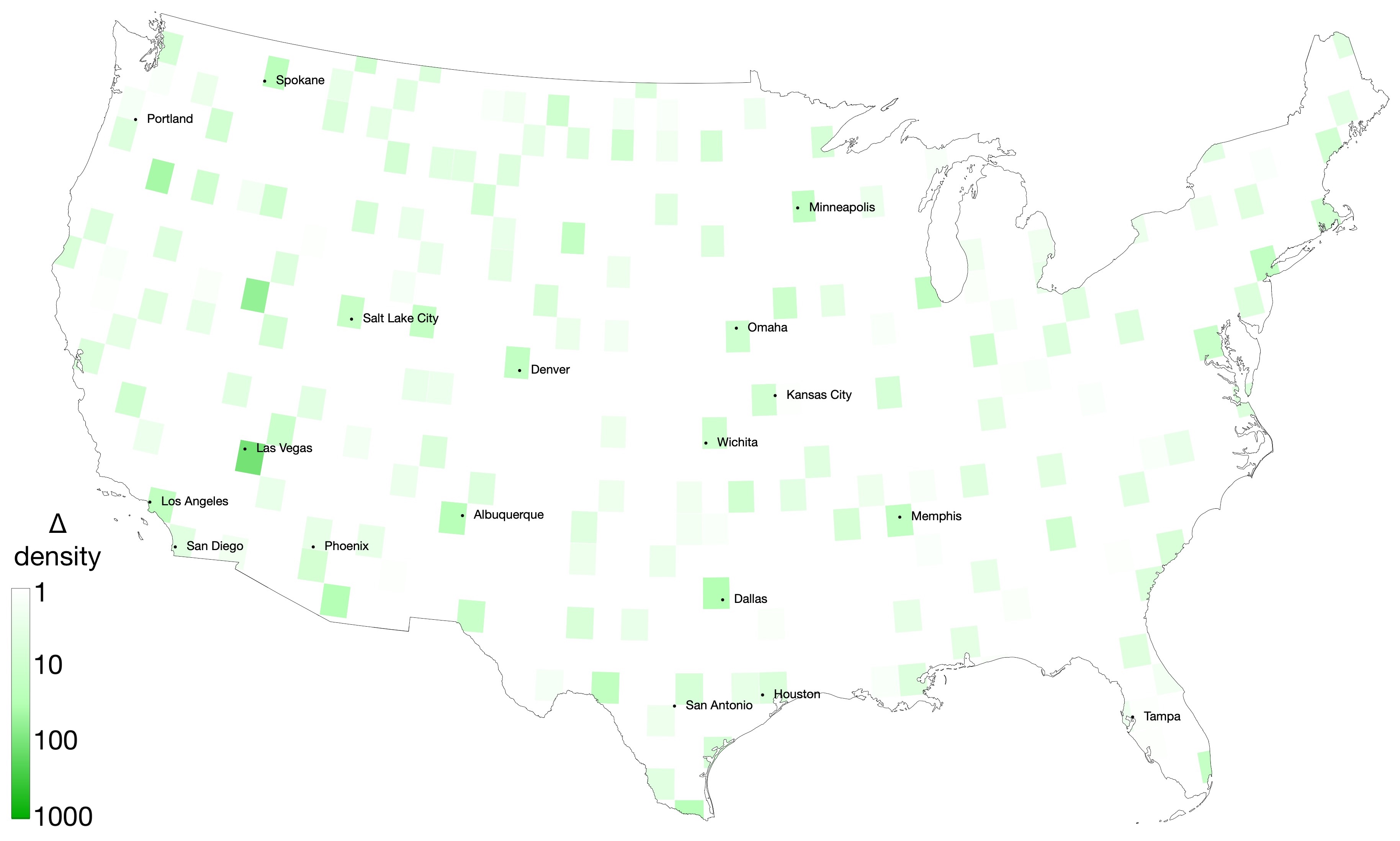} 
        \caption*{Figure S6: Population density difference of 1° blocks relative to their orthogonal neighbors}
    \end{wrapfigure}
    
    One characteristic is consistent with the simulations and highlighted cities:  these are \emph{isolated} cities.  They rise from relatively sparse surroundings, due to inhospitable land, water barriers, or historic factors. 
    
    This is confirmed in Figure S6, which shows the relative population density for every 1° block, calculated as $\Delta\rho = (\rho_{center} - \bar{\rho}_{NSEW})/\bar{\rho}_{NSEW}$ where $\rho_{center}$ is the density of a given block and $\bar{\rho}_{NSEW}$ is the mean density of its immediate north-south-east-west neighbors.  The correspondence between effect on $\Delta H$ and $\Delta\rho$ is compelling:  Las Vegas is the standout in both cases, an exceptionally isolated city with population density (at 1° sampling) 500-fold that of its neighboring blocks and largest $\Delta H$.  Phoenix, Spokane, Memphis, Denver, and Dallas appear in both, and New York, Boston, Baltimore, San Francisco in neither.
    
    Why would an isolated city elicit a rise in H at intermediate aggregation scales?  The effect certainly has to do with spatial correlation, since it vanishes upon randomization.  But now we see that it is a particular form of spatial correlation, namely a dense core region with sparse surroundings.  Dense-near-dense does not apply since the Boston-Washington corridor is clear of the effect, nor does sparse-near-sparse since the prairies are also clear.  But more to the point, how does such an arrangement lead to an increase in H?  
    
    \begin{wrapfigure}[13]{r}{10cm}
        \vspace{-4pt}
        \centering
        \captionsetup{width=9cm}
        \includegraphics[width=10cm]{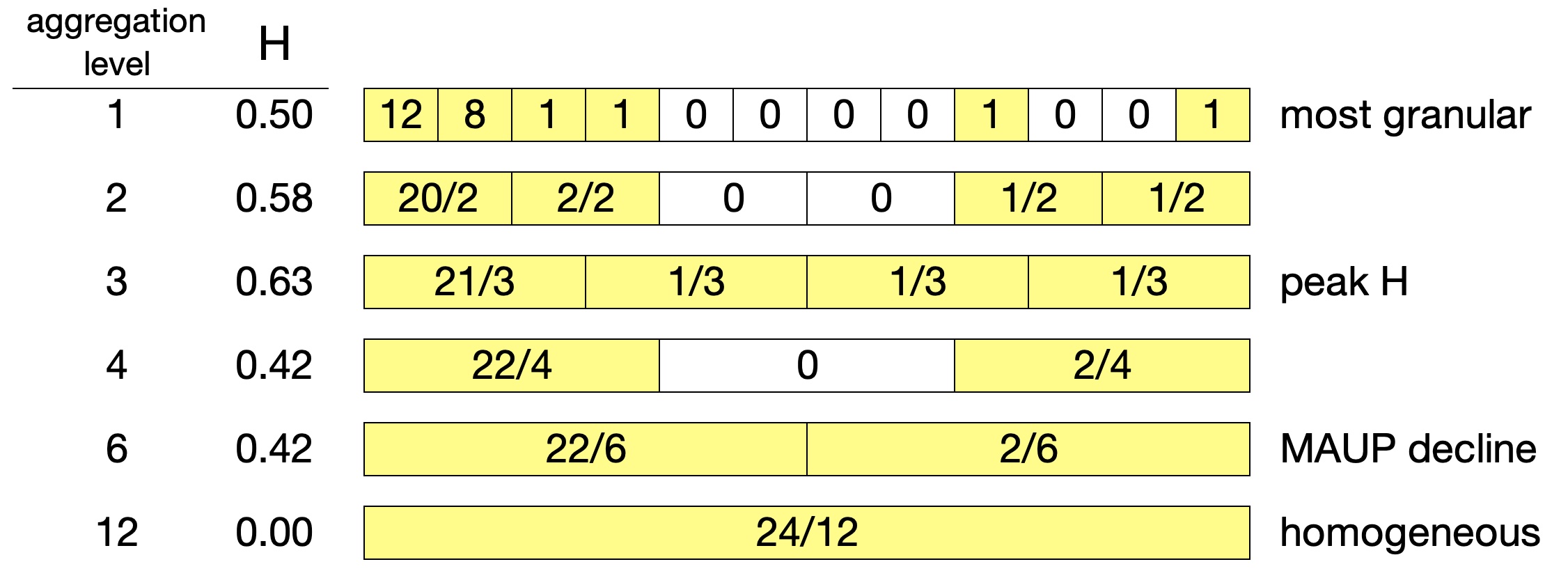} 
        \caption*{Figure S7: One-dimensional aggregation example with populated blocks in yellow and empty in white.  Values are population density in arbitrary units.}
    \end{wrapfigure}
    
    The answer is that the geography triggers the empty-block effect upon aggregation, as illustrated by the one-dimensional landscape of Figure S7.  The rows are copies of the same 12-block landscape at increasing levels of aggregation; there is a dense settlement at the left, then considerable empty space and a couple of distant, sparse settlements.  Because in these analyses we do not count empty blocks (the only situation in which the hump of H is seen -- note Figure 13), more or less area is included in the denominator as aggregation increases, and the maximum H occurs when the numerator is the least averaged while the denominator is maximal -- the third row in Figure S7.  The example also shows that some distant settlements are necessary to rope in a larger area as aggregation increases, and how far those sparse settlements are from the core (the four empty squares in the top row) will determine the scale at which H is maximal.  This is visible in the state data of Figure S2b: Nevada, which is large and empty, reaches $H_{max}$ at $\sim0.9^\circ\approx90$ km aggregation, Florida at $\sim30$ km, and Connecticut at $\sim13$ km.  For the same reason this is the scale at H values computed with and without empty blocks converge.
    
    How do such isolated cities appear?  The appearance of dense cities in isolated areas may be a post-industrial phenomenon -- something that would not be expected in Europe where most cities were seeded millennia ago.  Absent natural barriers, an old city will have had time to accrue suburbs and exurbs as many generations made their 1/r local migrations.  An isolated city is unlikely to be sited at a natural crossroads, a river junction or natural harbor, but instead since its founding to have had post-industrial access like the railroad to facilitate its growth in otherwise inaccessible surroundings.  A large city founded in the modern era also implies growth by (internal) immigration, since natural growth is too slow to account for its size.  All these are attributes of the cities highlighted in Figures S5 and S6.
    
    Pursuit of a pattern in H has led to a new topic in population geography, the recently-seeded isolated city.  Further studies should, however, not use H or $\Delta$H as metrics because these are complex, scale dependent, and prone to multiple artifacts -- indeed the hump of H is essentially an artifact of the empty-block effect.  Better to follow-up with population density and its various differences and ratios, to keep things simple and directly interpretable.

\pagebreak
\section*{Measurable Scale Parameters in Aggregation Analysis}

    \begin{wrapfigure}[14]{r}{5.5cm}
        \vspace{-12pt}
        \centering
        \captionsetup{width=5cm}
        \includegraphics[width=5cm]{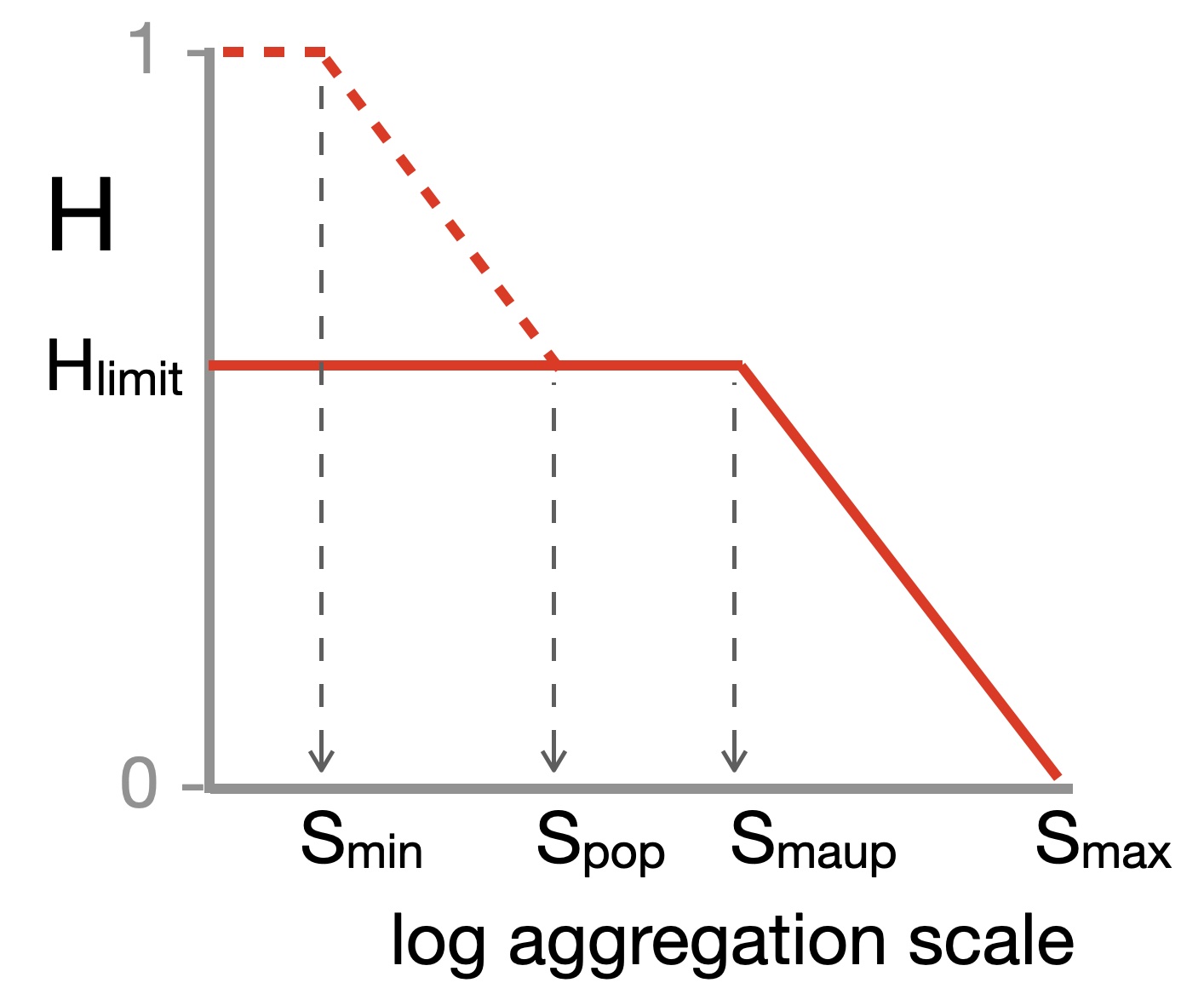} 
        \caption*{Figure S8: Sketch of Figures 13, 14, 15 with definitions of measurables on the scale axis. Solid red line: H from populated aggregation blocks only. Dashed red: including unpopulated blocks.}
    \end{wrapfigure}
    
    A major theme in this work is the variation in H as a function of aggregation scale.  $H_{limit}$ and $H_{max}$ are important measurables from such analysis and are detailed above.  In addition there are four measurable values on the scale axis, defined in Figure S8 and detailed below.
    
    \underline{$S_{max}$} is the maximum relevant scale for any given dataset and may be estimated as $\sqrt{area}$ of the overall region, typically a nation.  As the aggregation scale approaches $S_{max}$, $H \rightarrow 0$, since the aggregated area becomes unified and homogeneous.  This is an intrinsic result and simply sets the maximum scale endpoint.

    \underline{$S_{min}$} is the scale at which H (including empty blocks) plateaus to 1.0.  Its value is a function of the number of points in the dataset and may be estimated by a balls-in-urns statistical model:  if $M$ balls (datapoints) are tossed at random into $N$ urns (sampled geographic blocks), the expected fraction of empty urns $f_{empty} = (1 - 1/N)^M$, graphed in Figure S9.  Consideration of the Lorenz curve, especially for x-axis values above $\sim0.5$, shows that $H \approx f_{empty}$ (Figure 16).  The cusp point occurs at $S_{min} \approx 1/(2\sqrt{2M})$ in units of fractional scale and the calculations for the national datasets (Table S1) are in good agreement with Figures 14 and shown as triangles in Figure S10.  Thus France is left-shifted relative to the US simply because it's smaller in area (therefore in linear scale) while Brazil is right-shifted because its dataset is smaller.
    \vspace{6pt}
    \begin{center}
        \begin{tabular}{ c c c c c c c }
        \, & datapoints = M & area, $km^2$ & $S_{max}$ & $S_{min}$ & $S_{pop}$ & $S_{maup}$ \\
        \hline
        United States & 32,800 & 8,080,000 & $31.6^\circ$ & $0.062^\circ$ & $0.17^\circ$  & $1^\circ$ \\ 
        France        & 35,800 &   552,000 & $9.46^\circ$ & $0.018^\circ$ & $0.050^\circ$ & $0.05^\circ$\\ 
        Brazil        &   5570 & 8,510,000 & $26.3^\circ$ & $0.12^\circ$  & $0.35^\circ$  & $2^\circ$\\ 
        \hline
        \end{tabular}
        \captionsetup{width=12cm}
        \captionof*{table}{Table S1: Particulars of the national datasets.  $S_{max} = \sqrt{area}$, converted to degrees at mid-latitude. $S_{min} = S_{max}/(2\sqrt{2M})$, $S_{pop} = S_{max}/\sqrt{M}$, and $S_{maup}$ estimated from Fig 14.}
    \end{center}
    \vspace{6pt}

    \begin{wrapfigure}[16]{r}{6cm}
        \vspace{-12pt}
        \centering
        \captionsetup{width=5.5cm}
        \includegraphics[width=6cm]{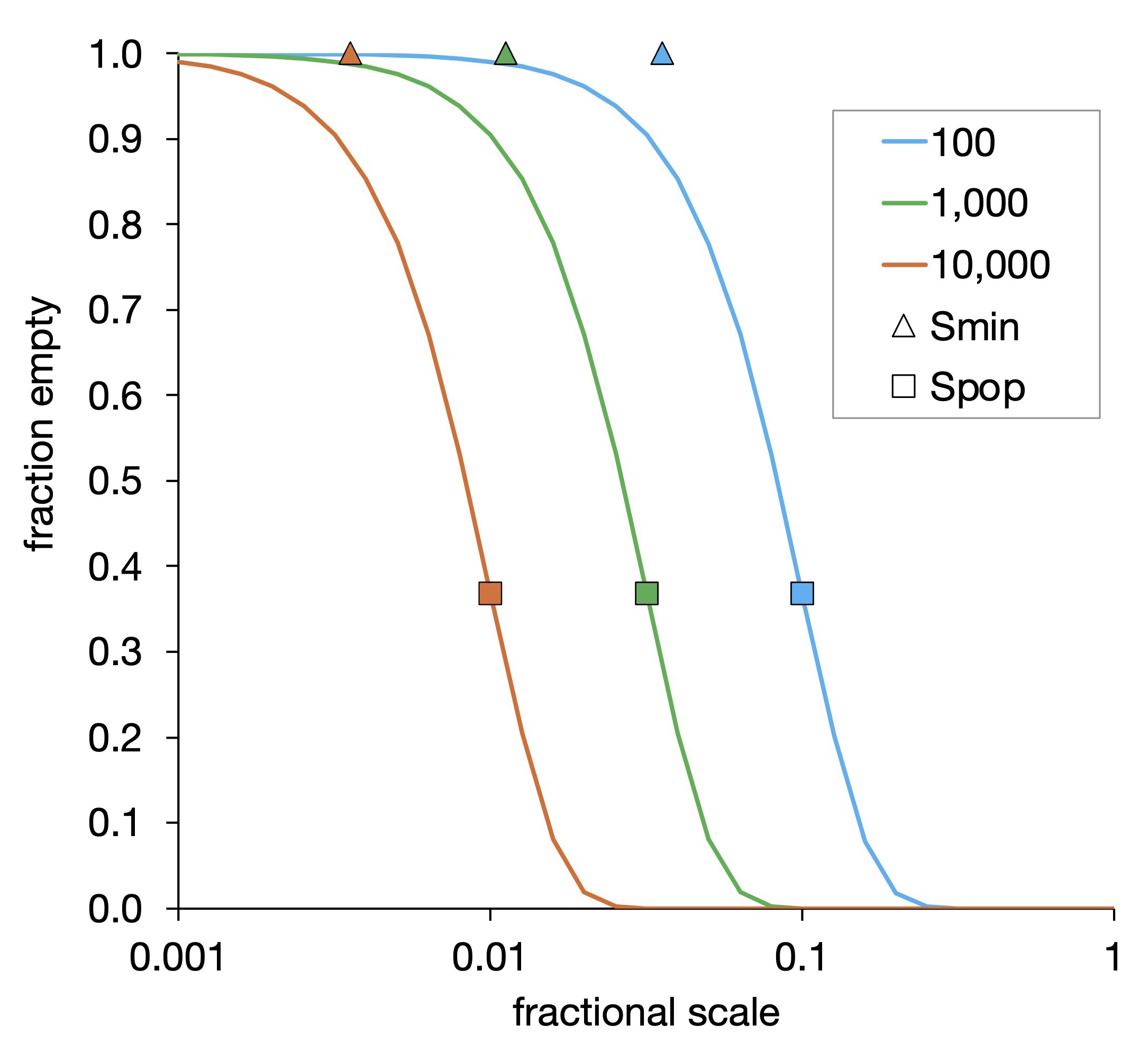} 
        \caption*{Figure S9: The balls-in-urns model, $H \approx f_{empty} = (1 - 1/N)^M$ where M = number of datapoints (100, 1000, 10,000) and N = number of sampling blocks = $1/(fractional \, scale)^2$ .}
    \end{wrapfigure}

    Equivalently $S_{min}$ may be estimated from inter-datapoint distances, for example as $\sqrt{median\,area}$ of Delaunay triangles, which from Table S2 is 7.8 km for the US, 2.4 km for France, and 15 km for Brazil.  $S_{min}$  can also be found as the median nearest-neighbor distance for all datapoints (9 km US, 3 km France, 21 km Brazil). $S_{min}$ is a property of the density of the dataset and the area it covers; it sets the lower range of scale that need be considered.  The balls-in-urns value is in excellent agreement with the real data (Figure S10, triangles; compare to Figure S9), because at small sampling scales there is minimal aggregation and the data are closest to isotropic and independent. $S_{min}$ tells us nothing about history or geography.
    
    \underline{$S_{pop}$} is the scale at which every aggregation block contains as least one datapoint, above which there are no empty aggregation blocks.  In the balls-in-urns model this is guaranteed only as $N \rightarrow \infty$, so we must choose a probability; a convenient point is the probability of no-empty-urns is 0.368 = 1/e which occurs when M = N, i.e. when the number of sampled blocks equals the number of datapoints (Figures S9, S10, squares).  Again $S_{pop}$ is a property of the dataset and the area it covers.  The balls-in-urns model is a poorer fit since non-random spatial correlations (like the hump in H) and the MAUP appear at larger scales.

    \underline{$S_{maup}$} is critical for the analyst to recognize since is the scale above which the MAUP begins, where aggregation blocks become self-similar and the value of H begins falls toward zero, dominating all other effectors. 
     Like $S_{min}$ and $S_{pop}$, $S_{maup}$ is also a function of dataset size.  Figure 20a shows that the H-vs-scale curves shift right ($S_{maup}$ increases) with decreasing size of a synthetic dataset.  Figure S10 shows that this holds true for real data: random subsets of 10\% of each dataset shift the US and French curves to the right. 
     
    \begin{figure}[!h]
        \centering
        \captionsetup{width=15cm}
        \includegraphics[width=15cm]{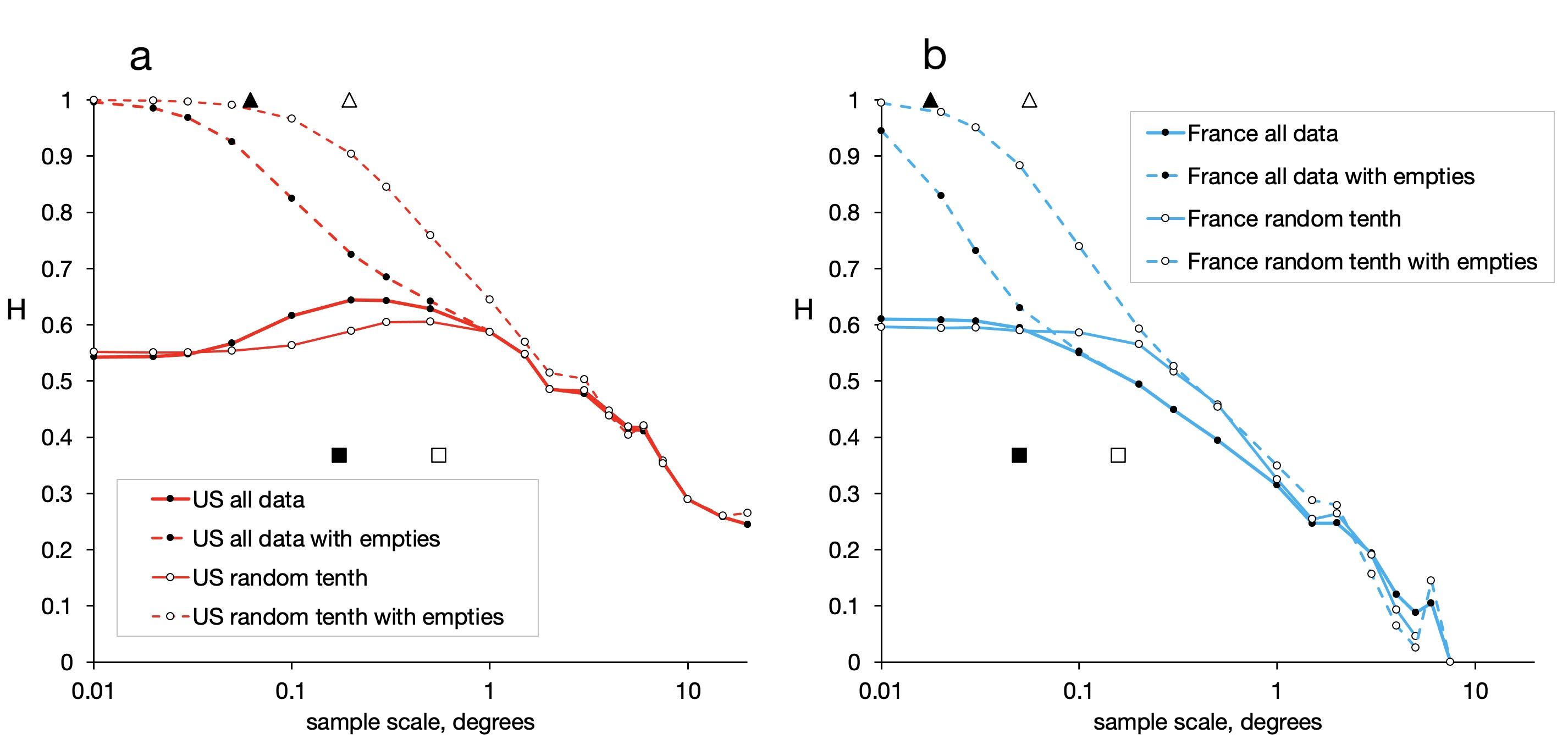}
        \caption*{Figure S10: The effect of dataset size: (a) Filled dots, US data as in Figure 14.  Open dots, same analysis for a 10\% random subset of the data (thus 3280 datapoints). (b) Same for France. Triangles denote $S_{min}$ and squares $S_{pop}$ as calculated from a balls-in-urns model, black for full dataset and white for 10\% subset.}
    \end{figure}
    
     But $S_{maup}$ is more complex since the empty block effect ($S_{pop}$), spatial correlation ($\Delta$H, Figure S1), and the MAUP all affect H at this scale.  Simulations may help to dissect these factors (Figure S11).

    \begin{figure}[!h]
        \centering
        \captionsetup{width=15cm}
        \includegraphics[width=16cm]{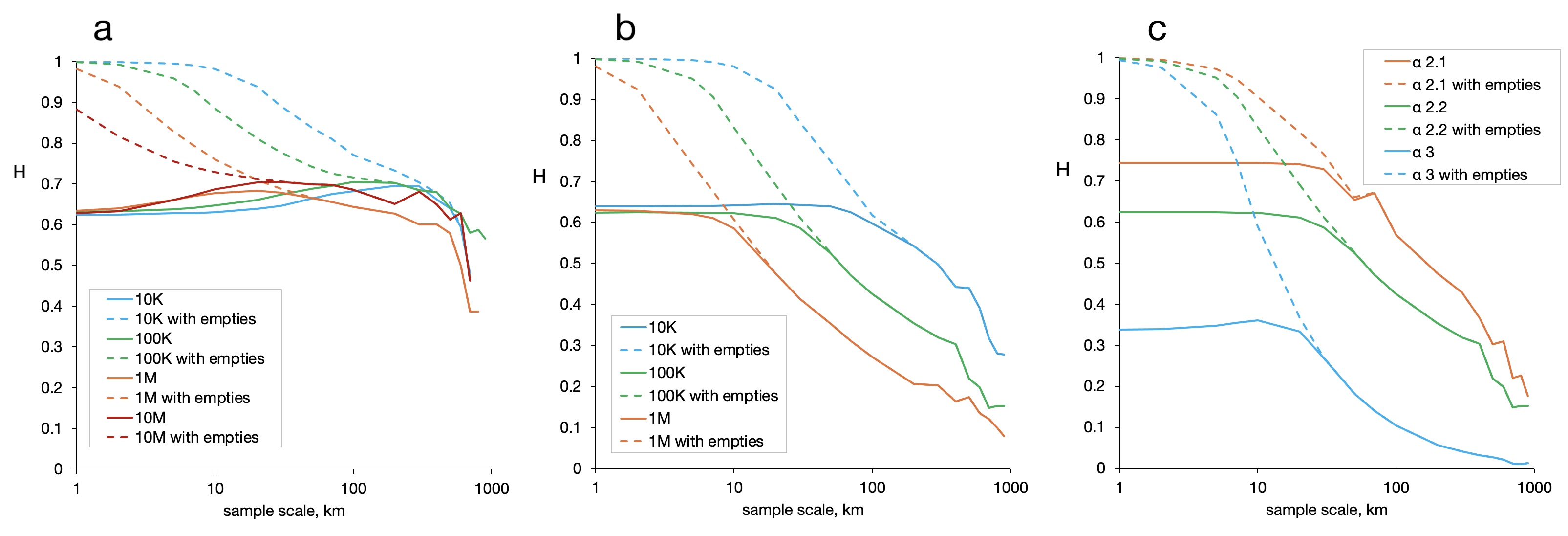}
        \caption*{Figure S11: Simulations as in Figure 17 and S4. (a) Locations derived from median 200 km Lévy jumps from parent location, pareto city size distribution with $\alpha = 2.2$, varying final population 10K to 10M.  (b) Same as (a) but isotropic random jumps. (c) Same as (b) but varying $\alpha = 2.1, 2.2, 3.0$, all 100K final population.}
    \end{figure}
    
    Figures S11b and S11c show that $S_{maup}$ may coincide with $S_{pop}$, as it does for France in Figure S10b.  But this is apparently a property of isotropic datasets, since $S_{maup}$ in Figure S11a is a log order larger than $S_{pop}$ for populations over one million.  With $\alpha = 2.2$ (a very steep city-size distribution) and parent-child location based on 1/r Lévy jumps, the Figure S11a dataset is highly spatially correlated, as also evidenced by its $\Delta$H humps (which isotropic datasets lack).
    
    Figure S11c shows that the slope $\alpha$ of the underlying city-size distribution can affect $S_{maup}$, but since $\alpha$ determines $H_{lim}$, if $H_{lim}$ values are the same (as they are for the US, France, and Brazil), then $\alpha$ cannot be a determinant of $S_{maup}$ -- though this may be true for some of the US state datasets (Figure S2b).
    
    So for isotropic datasets $S_{maup} \approx S_{pop}$ and is therefore just a function of dataset size and area covered.  Its value does not say anything interesting and the MAUP is apparently a continuation of the same averaging process that underlies the empty block effect on H.  For spatially correlated data, $S_{maup}$ may be partially decoupled from $S_{pop}$ but then becomes conflated with the $\Delta$H hump.  We are left with $S_{maup}$ as an important measurable without a clean interpretation.


\section*{Spatial Sampling by Tessellation}

    There are multiple ways to partition populations and areas.  Most of the literature uses fixed aggregations like counties or states, while this work principally aggregates data in latitude/longitude blocks in order to allow experimentation with scale.  H may also be computed from space-filling tessellations such as Delaunay triangulation or Voronoi tiling, which are the graph theory duals of each other.
    
    Figure S12 is a detail of a Delaunay triangle US population density map based on the 2020 zipcode dataset; zipcodes have specific locations which are taken as vertices of triangles, with population of the enclosed area calculated as the mean of the vertices' values.  The results in Table S2 include the number and median areas of the triangles, as well as point-source $H_{limit}$ values for comparison.

    \begin{figure}[!h]
        \centering
        \captionsetup{width=13cm}
        \includegraphics[width=14cm]{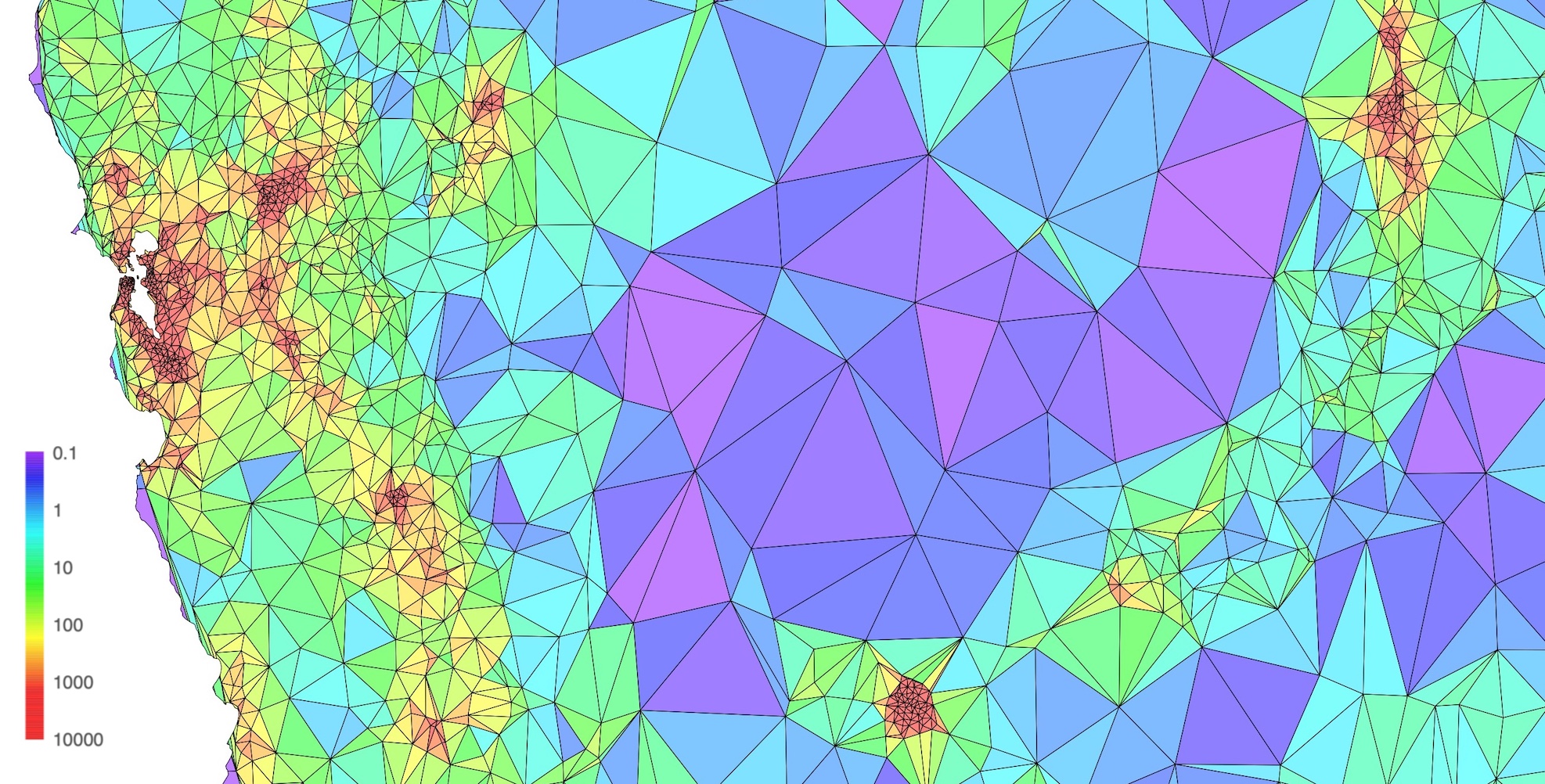}
        \caption*{Figure S12: Detail of Delaunay tessellation based on US zipcode dataset.  Scale in population per square kilometer.  San Francisco, left; Las Vegas, center-bottom; Salt Lake City, top-right. }
    \end{figure}
    
    \vspace{0.3cm}
    \begin{center}
        \begin{tabular}{ c c c c c c }
        \, & triangles & median area, km$^2$ & H$_{delaunay}$ & H$_{limit}$ & $\Delta$H \\
        \hline
        United States & 65,401 & 62  & $0.723 \pm 0.003$ & $0.545 \pm 0.002$ & 0.18  \\ 
        France        & 71,604 & 6.0 & $0.633 \pm 0.004$ & $0.613 \pm 0.007$ & 0.02  \\ 
        Brazil        & 11,127 & 230 & $0.694 \pm 0.020$ & $0.574 \pm 0.019$ & 0.12  \\ 
        \hline
        \end{tabular}
        \captionof*{table}{Table S2: Results for H via Delaunay triangle tessellation}
    \end{center}

    Tessellation is a poor choice for the estimation of H:
    \begin{itemize}
    \setlength{\leftskip}{-1.2em}

    \item Tessellation intrinsically adapts to the granularity of the dataset, giving the smallest possible local sample regions.  But their areas may vary widely within and between datasets (note in Table S2 the median areas range from 6 to 230 km$^2$).  Given the artifacts of H with sampling scale this precludes comparison of H between datasets.
    
        The 65,000 areas of the US tessellation cover five orders of magnitude in a saturated power law distribution with $\alpha$ = 2.1.  It would be difficult to find a less uniform distribution.
        
    \item By construction there are no empty regions in tessellation; this removes a legitimate choice for the analyst, so that uninhabitable areas (lakes, severe deserts: the purple triangles in Figure S12 across Nevada and Utah) are always included and can send H artificially high.  This is why the US has $H_{delaunay}$ 0.18 higher than $H_{limit}$ while for France they differ by only 0.02.  Essentially tessellation bakes-in the empty block problem.
    
    \item Like county and state census data, tessellations are fixed and not amenable to experiment.  Just as any empty-block problems are masked by tessellation, so are any MAUP problems if the tessellations are large enough to average out the interesting heterogeneity of the data.
    
    \item It is not clear how to interpret H from tessellation.  For the 48 contiguous US states, $H_{delaunay}$ is not strongly correlated to $H_{limit}$ or $H_{max}$ ($H_{delaunay}$ vs $H_{limit}$ has $r^2 = 0.23$ while $H_{delaunay}$ vs $H_{max}$ has $r^2 = 0.57$).   Because both the sample scale and effect of empty areas vary across tessellations, $H_{delaunay}$ can take on any value between the populated-only (solid red) and empty-block included (dashed red) lines of Figure 13.
    \end{itemize}

\end{document}